%% file: main.tex
\documentclass{article}

\usepackage[final, main, nonatbib]{neurips_2026}

\usepackage[numbers,compress]{natbib}

\usepackage[utf8]{inputenc}
\usepackage[T1]{fontenc}
\usepackage{hyperref}
\hypersetup{colorlinks=true, linkcolor=blue!50!black, citecolor=blue!50!black, urlcolor=blue!50!black}
\usepackage{url}
\usepackage{amsmath}
\usepackage{amsfonts}
\usepackage{booktabs}
\usepackage{tabularx}
\usepackage{nicefrac}
\usepackage{microtype}
\usepackage{xcolor}
\usepackage{float}
\usepackage{graphicx}
\usepackage{multirow}
\usepackage{enumitem}

\IfFileExists{siunitx.sty}{
  \usepackage{siunitx}
}{
  \newcommand{\num}[1]{##1}
  \newcommand{\SI}[2]{##1\,##2}
}

\title{Scaling Storm-Resolving Atmospheric AI Simulation to the Entire Planet}

\author{%
  Zeyuan Hu\thanks{Equal contribution.} \\
  NVIDIA
  \And
  Akshay Subramaniam \\
  NVIDIA
  \And
  Noel Keen \\
  LBNL
  \And
  Tao Ge \\
  NVIDIA
  \And
  Jaideep Pathak \\
  NVIDIA
  \AND
  Mohammad Shoaib Abbas \\
  NVIDIA
  \And
  Suman Ravuri \\
  NVIDIA
  \And
  Karthik Kashinath \\
  NVIDIA
  \AND
  Naser Mahfouz \\
  PNNL
  \And
  Peter Caldwell \\
  LLNL
  \And
  Mike Pritchard \\
  NVIDIA
  \And
  Noah Brenowitz\footnotemark[1] \\
  NVIDIA
}

\begin{document}

\maketitle

\newcommand{\blfootnote}[1]{%
  \begingroup
  \renewcommand{\thefootnote}{}\footnotetext{#1}%
  \addtocounter{footnote}{-1}%
  \endgroup
}
\blfootnote{Corresponding author(s): \texttt{zeyuanh@nvidia.com}, \texttt{nbrenowitz@nvidia.com}.}

\begin{abstract}
    Kilometer-scale convection shapes precipitation extremes, tropical
    organization, and cloud feedbacks, but most global atmospheric models
    approximate these processes at 25--100\,km resolution. 
    Global storm-resolving physics models resolve convective systems explicitly, but at a cost (roughly one MWh per simulated day on exascale supercomputers) that limits long-duration simulation.
    We introduce STRATA (Storm-resolving Tile-based autoRegressive Atmosphere Transformer Architecture), the first autoregressive AI emulator for global storm-resolving atmospheric dynamics.
    STRATA is trained on the highest-resolution atmospheric dataset yet used for global AI emulation:
    17 days of SCREAM physics-model output at 4.9-km resolution (${\sim}25$ million grid cells) sampled every 10 minutes.
    Our central premise is that on 10-minute timescales atmospheric dynamics are predominantly local, so training on small spatial tiles trades scarce global temporal samples for abundant local spatial samples and enables global rollout via overlapping-tile blending.
    STRATA combines 3D patch embedding and local 3D neighborhood attention, a novel Stereographic Rotary Position Embedding
    (StereoRoPE) for grid-invariant encoding, and a pixel-space
    de-aliasing decoder that suppresses patch-scale rollout artifacts.
    An iso-FLOP scaling study reveals that km-scale emulation requires ${\sim}10\times$ more FLOPs per grid point than coarse-resolution AI weather models, consistent with the higher information density of convective-scale dynamics.
    Trained on only 17 days of data, STRATA produces stable 24-hour global rollouts with realistic km-scale dynamics across diverse regimes, though large-scale biases develop with lead time.
    It achieves 48 simulated days per megawatt-hour (more than 50 times better energy efficiency than the SCREAM physics model) and 741 simulated days per wall-clock day at 512 H100 GPUs.
    Code, model weights and training data are publicly released; see \url{https://github.com/NVlabs/STRATA}.
    
\end{abstract}

\begin{figure}[t!]
  \centering
  \includegraphics[width=0.99\linewidth]{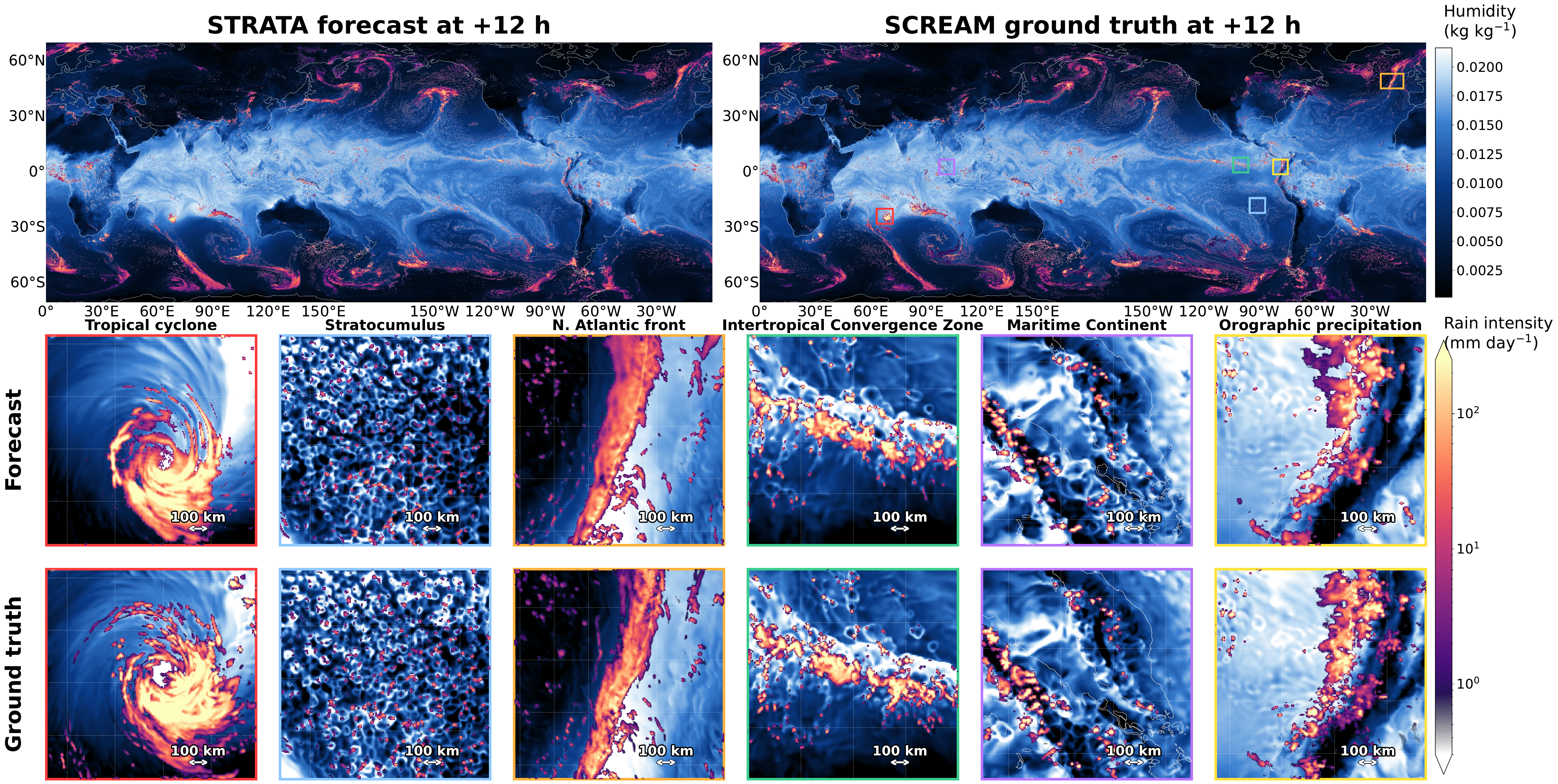}
  \caption{%
  \textbf{Twelve-hour STRATA rollout compared with SCREAM.}
  Top: near-surface specific humidity and rain intensity at +12\,h from STRATA (left) and the
  corresponding SCREAM reference simulation (right) on the native 4.9-km output
  grid ($6\times2048^2\approx25$\,M horizontal cells). The colored
  boxes in the global SCREAM panel mark the zoom locations.
  Bottom: zoomed comparisons across six representative regimes. Row~1: STRATA forecast. Row~2: SCREAM ground truth at native 4.9-km
  resolution. Humidity colors in the zoom panels are independently
  rescaled for each location to emphasize local structure. Scale bars are
  100\,km.
  Supplementary videos of the full 24-hour rollout are available at
  \url{https://research.nvidia.com/labs/climate/strata/}.
 }
  \label{fig:big-qv}
\end{figure}

\section{Introduction}
\label{sec:intro}

Long-range atmospheric simulation is 
among the most challenging computational workflows in science.
Weather and climate depend on processes spanning more than ten orders of magnitude in space, from microphysics interactions between cloud droplets and raindrops to planetary-scale circulation, and physical fidelity requires accurately representing all of them.

In practice, the finest scales are typically approximated by closed-form parameterizations rather than simulated explicitly, introducing substantial uncertainty in long-range projections \citep{bonyCloudsCirculationClimate2015,schneiderClimateGoalsComputing2017}. To reduce this uncertainty, state-of-the-art physics models push to ever-finer grids that explicitly resolve convective storms rather than approximating them---currently reaching ${\sim}1$\,km globally. These models, known as global storm-resolving models (GSRMs;  \citealp{caldwellConvectionPermittingSimulationsE3SM2021,seguraNextGEMSEnteringEra2025}; Appendix~\ref{sec:app_gsrm_background}) are correspondingly expensive: the leading GSRMs achieve a throughput of roughly 1 simulated day (SD) per MWh on exascale supercomputers \citep{taylorSimpleCloudResolvingE3SM2023,donahueExascaleSimpleCloudResolving2024}, roughly 2 SD per ton of CO$_2$ emitted or 1 SD per passenger for a one-way coach flight from Seattle to New York.

AI emulators could offer similar fidelity at  lower cost because their core computations are more arithmetically-intense and suited to modern hardware accelerators. While traditional atmospheric solvers are memory-bandwidth bound---spending most of their time moving data rather than computing---transformer-based architectures \citep{vaswaniAttentionAllYou2023} operate on dense, low-precision matrix multiplications that saturate GPU compute units. Coarse-resolution global atmospheric emulators and limited-domain km-scale emulators have realized this efficiency gain in their respective regimes (Section~\ref{sec:related}) but achieving autoregressive rollout at both \textit{global and km-scale}, the only regime that can replace a full-resolution GSRM, remains an unsolved challenge.

Reaching this milestone is hindered by several challenges that do not arise when using AI to simulate weather at coarse resolution. Convective dynamics are 3D, have elevated information density, and require 10-min (36$\times$ more) sampling in time and 4.9 km (25$\times$ more) sampling in space to constrain.
This raises acute computational challenges.
A given day of GSRM data occupies thousands of times more memory than a day of the ERA5 dataset. Such high resolution data is difficult to store on disk and activation memory-use during training becomes prohibitive. 
The upshot is a problem that demands especially scalable training strategies working backwards from an ambitious global inference goal---rapidly evolving 25M atmospheric horizontal grid cells globally.

We introduce STRATA (Storm-resolving Tile-based autoRegressive Atmosphere Transformer Architecture), to our knowledge the first AI emulator to demonstrate successful autoregressive rollout at global km-scale. Its central premise is that atmospheric phenomena on 10-minute timescales are predominantly local, so a tile-based training strategy can convert temporally scarce global simulations into spatially abundant samples, with overlapping tile blending reassembling a coherent global state at inference.

Our key technical solutions to the challenges of training a high-throughput global high resolution model include:
\begin{itemize}[topsep=3pt,itemsep=0pt,parsep=0pt]
\item\textbf{Tile-based training of global auto-regressive models.} The activation memory bottleneck is overcome by training on crops. Only inference is global.
  \item \textbf{3D local-attention backbone.} 3D patch tokenization and neighborhood attention further exploit atmospheric locality for efficient roll-out.
  \item \textbf{Stereographic Rotary Position Embedding (StereoRoPE).} Grid-invariant position encoding via stereographic projection, enabling consistent spatial representation across tile locations and transfer to unseen grids.
  \item \textbf{Patch de-aliasing.} A spectral stability analysis of patch-embedded architectures and a pixel-space de-aliasing decoder that greatly reduces the resulting checkerboard artifacts.
\end{itemize}

Additional accomplishments are as follows. STRATA produces stable 24-hour rollouts reproducing realistic km-scale dynamics across diverse weather regimes. Scaling analysis reveals ${\sim}10\times$ more FLOPs are needed per grid point than coarse-resolution forecasting, with an information-theoretic explanation based on the higher entropy of convective-scale data. Most importantly, inference runs at a throughput of 48 simulated days per MWh---more than 50 times better energy efficiency than SCREAM---and scales to 741 simulated days per wall-clock day at 512 H100 GPUs (Table~\ref{tab:hero-compute}a). 

\section{Related Work}
\label{sec:related}

\textbf{Coarse-resolution emulators}
Global AI weather models now rival operational forecasts at ${\sim}0.25^\circ$ resolution and 6-hour timesteps \citep{pathakFourCastNetGlobalDatadriven2022,biAccurateMediumrangeGlobal2023,lamGraphCastLearningSkillful2023,chenFuXiCascadeMachine2023,bodnarFoundationModelEarth2024,nguyenScalingTransformerNeural2024,priceGenCastDiffusionbasedEnsemble2024,langAIFSCRPSEnsembleForecasting2024a,bonevFourCastNet3Geometric2025a,kossaifiDemystifyingDataDrivenProbabilistic2026,yuScalingLawsGlobal2026}, and several coarse-resolution atmosphere emulators have produced stable multi-decadal to century-long rollouts \citep{kochkovNeuralGeneralCirculation2024,cresswell-clayDeepLearningEarth2025,watt-meyerACE2AccuratelyLearning2024,cachayProbabilisticEmulationGlobal2024,chapmanCAMulatorFastEmulation2025}. However, these systems operate at resolutions at least an order of magnitude coarser than kilometer scale and therefore cannot simulate explicit convective-scale phenomena.

\textbf{Regional km-scale emulators.} \citet{floraWoFSCastMachineLearning2025} show that realistic convection can be generated with a deterministic model when the resolution is fine enough in space (3-km) and time (10-minute). However, their simulations are limited to spatial extents of $(\SI{900}{\km})^2$ and are only rolled out to two hours. Other regional emulators have been trained on 1-hour time resolution data, but either become blurry \citep{abdiHRRRCastDatadrivenEmulator2025} or require expensive diffusion sampling \citep{pathakKilometerScaleConvectionAllowing2024}.

\textbf{Super-resolution.}
Generative diffusion models offer a complementary path to km-scale output by conditioning on coarse atmospheric states rather than rolling out autoregressively---producing km-scale details diagnostically as a function of the coarse input \citep{mardaniResidualCorrectiveDiffusion2023,brenowitzClimateBottleGenerative2025,perkinsHiROACEFastSkillful2026}. Inspiring our approach, global coverage is made tractable via patch-based training \citep{Bar-Tal2023-sg, brenowitzClimateBottleGenerative2025}. While practical for inpainting realistic details when trained on undersampled datasets, diffusion-based generation is purely downscale, so cannot model the two-way feedback between fine and large scales that we are after.

\textbf{Spherical architectures.}
Faithfully representing data on the sphere is a core challenge for global atmospheric emulators, motivating a range of geometry-aware
architectures mostly targeting resolutions of ${\geq}0.25^\circ$; scaling to km-scale grids remains an open problem. For example, Spherical Fourier Neural Operators~\citep{bonevSphericalFourierNeural2023} replace standard FFTs with spherical harmonic transforms for globally equivariant weather modeling ${\sim}0.25^\circ$, but by using global basis functions they preclude our memory-saving local training strategy.
Alternately, MeshGraphNet-based architectures~\citep{keislerForecastingGlobalWeather2022,lamGraphCastLearningSkillful2023,floraWoFSCastMachineLearning2025} avoid polar geometric distortions and can be run globally or locally, but suffer from low throughput because they are based on sparse message passing operations.

\section{Proposed Method}

\begin{figure}[t!]
\centering
\includegraphics[width=0.99\textwidth]{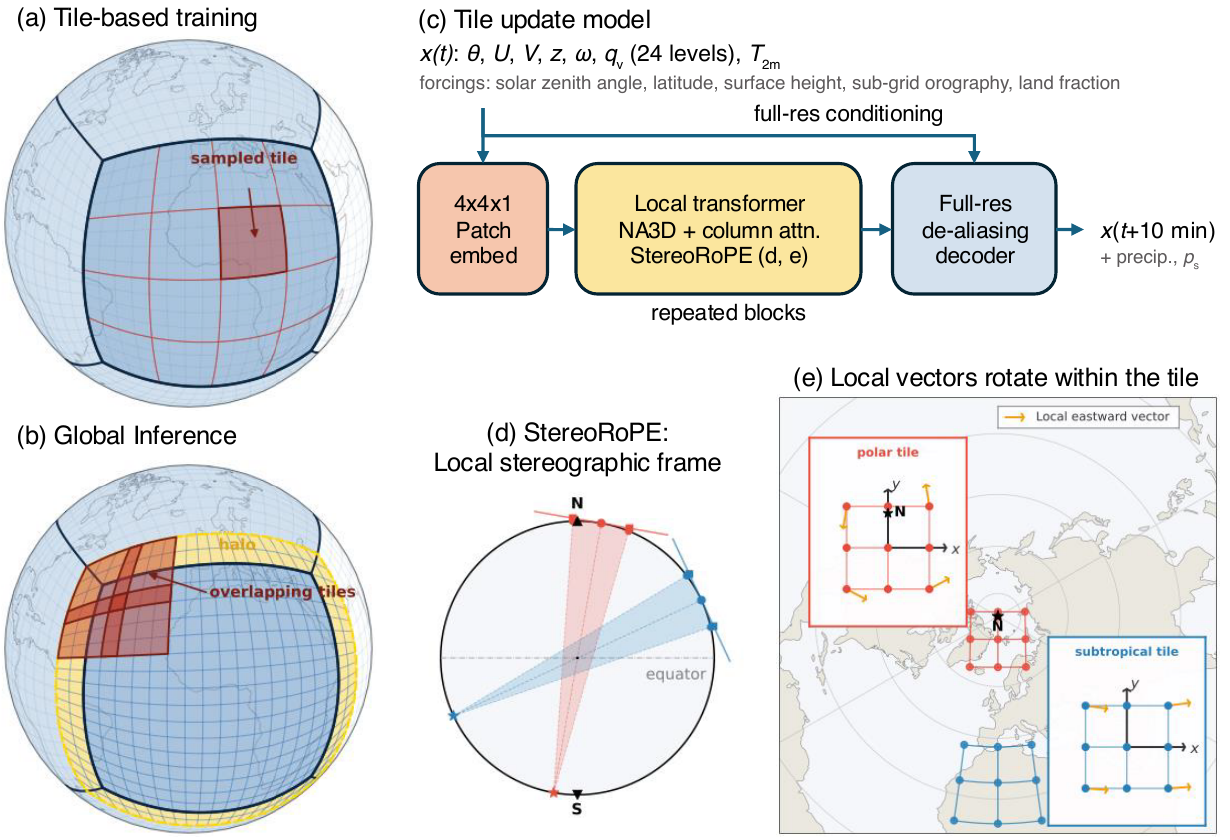}
\caption{%
\textbf{STRATA method overview.}
\textbf{(a,b)} STRATA learns 10-minute updates from local SCREAM tiles and
applies the same model globally using overlapping tiles that are blended into a
continuous rollout; Duo-grid halo padding supplies geometrically consistent
context near cubed-sphere face boundaries.
\textbf{(c)} The tile update model combines 3D patch embedding, local attention
mechanisms, and a pixel-space de-aliasing decoder adapted from PixelDiT.
\textbf{(d,e)} Stereographic Rotary Position Embedding (StereoRoPE) projects
each tile to a local stereographic frame and rotates vector inputs into the
same tile-centered frame, providing grid-invariant position information for
the transformer.
\label{fig:method-overview}
\label{fig:stereographic-rope-schematic}
}
\end{figure}

\subsection{Tile-based training and inference}

We train on independent $128\times128$ tiles (${\sim}576$\,km across)---sufficient to capture all physically relevant signals in a
10-minute step, since even acoustic waves propagate only ${\sim}200$\,km per step---using the many spatial tiles in each
global snapshot as training samples to trade scarce temporal coverage for abundant spatial samples (Figure~\ref{fig:method-overview}a).

At inference, STRATA is applied globally by decomposing each cubed-sphere face into overlapping tiles, processing them independently, and blending the overlap regions into a continuous global state (Figure~\ref{fig:method-overview}b). For tiles near cube-face boundaries, accurately predicting state changes requires atmospheric context from the adjacent face. These cross-face neighbors cannot be used directly, however, because the cubed-sphere grid orientation changes discontinuously at face boundaries, and at corners where three faces meet, naive padding is geometrically ill-defined. Duo-Grid halo padding~\citep{mouallemImplementationNovelDuoGrid2023} addresses this by extending each face's equiangular grid past its boundary and interpolating neighboring-face values onto this extended grid (yellow shading in Figure~\ref{fig:method-overview}b), giving boundary tiles a locally coherent neighborhood that the model processes identically to interior tiles. This tile-based inference protocol also enables scalability to a multi-GPU distributed environment, yielding strong scaling efficiency of up to 98.4\% from 8 to 512 GPUs and a throughput of 741 simulated days per wall-clock day at 512 GPUs. Tile-based training and inference do not depend on the network architecture. More inference details are in Appendix~\ref{sec:app_inference_details}.

\subsection{Architecture}
\label{sec:app_arch}

Km-scale atmospheric dynamics have two properties that shape the
architecture: they are intrinsically 3D and predominantly local on
10-minute timescales.
STRATA addresses these with three components
(Figure~\ref{fig:method-overview}c): a 3D patch embedding layer, a
local-attention DiT transformer backbone, and a pixel-space de-aliasing
decoder.
Full architecture specifications are provided in
Appendix~\ref{sec:app_arch_spec}.

\textbf{Transformer backbone.}
The STRATA backbone follows the Diffusion Transformer (DiT)~\cite{peeblesScalableDiffusionModels2023} architecture (Appendix~\ref{sec:app_arch_spec}, Figure~\ref{fig:dit-schematic}a), adapted for deterministic atmospheric emulation rather than diffusion sampling. All input fields are concatenated and treated as a single volumetric tensor spanning the horizontal tile and all vertical levels. At a 10-minute timestep, atmospheric interactions are largely local, so we use 3D Neighborhood Attention (NATTEN~\cite{hassaniNeighborhoodAttentionTransformer2023a,hassaniGeneralizedNeighborhoodAttention2025}), with stereographic rotary position embeddings (see \S\ref{sec:stereorope}) inside local $(x,y,\mathrm{lev})$ windows. With a horizontal patch embedding size of 4 and a NATTEN kernel size of 9, each attention window covers $36\times36$ horizontal native grid cells (${\approx}160\times160$\,km). We alternate these blocks with column-wise attention along the vertical dimension to capture processes that are computed column-by-column in SCREAM with no horizontal coupling, such as vertical radiation transfer and cloud microphysics.

\textbf{3D Patch-embedding and computational efficiency.} 
Like the original DiT~\cite{peeblesScalableDiffusionModels2023}, STRATA uses patch embedding to tokenize the input, grouping $p_h \times p_h \times p_v$ 3D grid cells into a single token before the transformer sees them. Unlike coarse-resolution atmospheric transformers, we treat the vertical dimension as a spatial axis ($p_v=1$) rather than a channel dimension ($p_v=\text{\# of levels})$. This decision is motivated on both physical and computational grounds---it allows modeling the 3D dynamics of atmospheric convection at the token level and acts as an independent lever for scaling computational complexity, and therefore accuracy \citep{hoffmannTrainingComputeOptimalLarge2022,kaplanScalingLawsNeural2020,yuScalingLawsGlobal2026,peeblesScalableDiffusionModels2023}.

The computational complexity of our architecture, measured in terms of FLOPs, is approximated by $2NP$ per forward pass, where $P$ is the number of parameters and $N$ the number of patches $N \propto (p_v p_h^2)^{-1}$. The number of parameters $P\propto d^2$ where $d$ is the embedding dimension of the network. From this we see that reducing the vertical patch size to $p_v=1$ while holding the embedding dimension fixed increases $N$ and the computational complexity of the model overall.

\begin{figure}[t]
    \centering
    \includegraphics[width=0.9\textwidth]{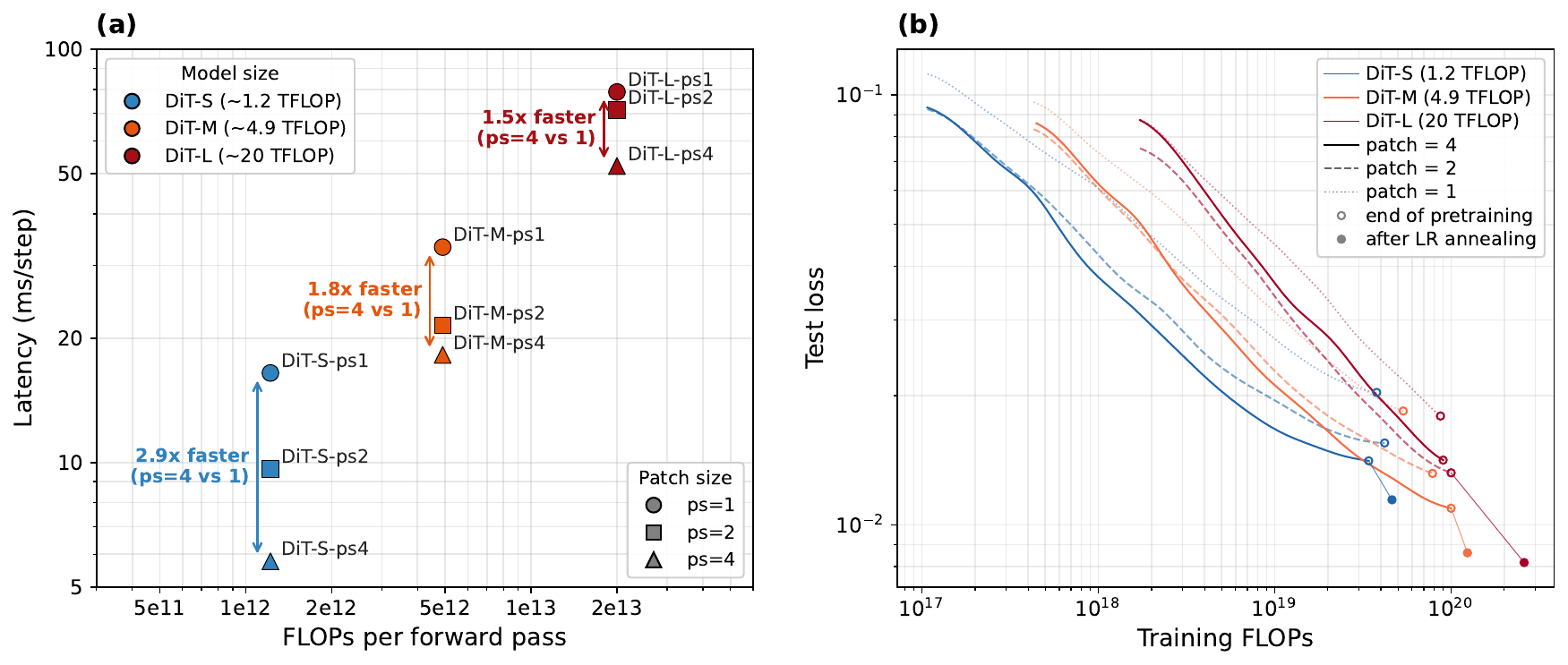}
    \caption{\textbf{Patch-size and model-size scaling.}
    \textbf{(a)} inference latency vs.\ FLOPs per forward pass (estimated as
    $2NP$, where $N$ is the token sequence
    length and $P$ the number of parameters) for DiT-S/M/L models with horizontal patch sizes $ps \in \{1,2,4\}$;
    at equal FLOPs, $ps=4$ is $1.5$--$2.9\times$ faster than $ps=1$.
    \textbf{(b)} test loss vs.\ training FLOPs for the same sweep, counted as $2NP$ per
    training sample, i.e.\ the forward pass only. The common convention of $6NP$, which
    also counts the backward pass, would multiply every point by 3 and leave the comparison
    between models unchanged.
    Patch size 1 is worse than patch sizes 2 and 4 at equal FLOPs; smaller
    models are more compute-efficient at low budgets, while larger models reach
    lower loss with sufficient training.}
    \label{fig:mfu-vs-patch-size}
    \label{fig:sweep_results}
\end{figure}

Even for a fixed FLOP budget ($P\propto N^{-1}$) the patch size is an important lever. Reducing $N$ improves model FLOP utilization (MFU)---by reducing the size of the activations which scales as $O(Nd)=O(N\sqrt{P})=O(\sqrt{N})$ and in so doing increasing the overall arithmetic intensity of the architecture.
Another benefit of larger patches is that they expand the effective window size of the neighborhood attention at no extra cost, which improves the accuracy. Figure~\ref{fig:sweep_results}b shows that for constant FLOPs the loss decreases with patch size.

\textbf{Patch instability and pixel-space de-aliasing decoder.}
Patch tokenization introduces checkerboard artifacts in rollouts
\citep{zhangMakingConvolutionalNetworks2019,biAccurateMediumrangeGlobal2023},
most visibly in smooth fields like near-surface temperature.
We argue this is caused by a spectral bias of the patched architecture upon
initialization: the block-diagonal structure of the linearized update operator
gives $\rho(A)>1$, causing patch-scale eigenvectors to dominate the rollout.
A formal stability analysis is provided in
Appendix~\ref{sec:stability-analsis-details}.
We mitigate this with a lightweight decoder of PixelDiT pixel
blocks~\citep{yuPixelDiTPixelDiffusion2025}; within each block, we replace
the original linear-projection upsampling with bilinear upsampling followed by
depthwise convolutional refinement to enforce continuity across patch boundaries
(Figure~\ref{fig:dit-schematic}b,c).
Figure~\ref{fig:patch-artifact} in Appendix~\ref{sec:app_patch_dealiasing}
shows this substantially reduces patch-scale artifacts in 12-hour rollouts.
A small residual remains at some locations. We expect that architectural changes, such as anti-aliased resampling~\citep{zhangMakingConvolutionalNetworks2019} or decoders in which pixel tokens query the semantic context through cross-attention~\citep{he2026hyperdit}, possibly combined with the inference-time techniques in Appendix~\ref{sec:app_inference_tricks}, may help reduce it further.

\subsection{Grid-Invariant Position Encoding via Stereographic Projection}
\label{sec:stereorope}

Our architecture requires locally rectangular grids for efficiency, but tile-based training requires the same local physical patterns to be represented consistently across the sphere.
Standard embedding strategies for vision transformers, such as axial Rotary Position Embedding (RoPE) ~\cite{suRoFormerEnhancedTransformer2023,heoRotaryPositionEmbedding2024}, will fail at this since the same 2D index offset will have a different physical orientation and length depending on the location.

We address this with Stereographic RoPE (StereoRoPE; Figure~\ref{fig:method-overview}). This approach decouples the coordinate used for RoPE from the discretization of the input data, allowing us to choose the most convenient option for each.
 For each tile, we project grid points to a local stereographic tangent plane centered on the tile (Figure~\ref{fig:method-overview}d) and use the resulting coordinates as RoPE inputs. Because these coordinates approximate physical displacement rather than grid indices, attention patterns can transfer across tile locations and grid topologies. For a tile centered at $(\phi_0,\lambda_0)$, each token at $(\phi,\lambda)$ is mapped to tile-centered stereographic coordinates:
\begin{align}
    \cos c &=
    \sin\phi_0\sin\phi +
    \cos\phi_0\cos\phi\cos(\lambda-\lambda_0), \quad
    k = \frac{2}{1+\cos c}, \nonumber\\
    x &=
    \frac{k}{\ell_\mathrm{scale}}\cos\phi\sin(\lambda-\lambda_0), \quad
    y =
    \frac{k}{\ell_\mathrm{scale}}
    \bigl(\cos\phi_0\sin\phi -
    \sin\phi_0\cos\phi\cos(\lambda-\lambda_0)\bigr),
\end{align}
where $c$ is the angular separation from the tile center, $(x,y)$ are local East--North coordinates in the tile-centered tangent frame, and $\ell_\mathrm{scale}=\sqrt{4\pi/N_\mathrm{total}}$ normalizes coordinates across resolutions.
These coordinates replace row/column indices as inputs to axial 2D RoPE~\cite{heoRotaryPositionEmbedding2024}. The stereographic projection is chosen because it is conformal, which preserves local angles and facilitates learning vector operators such as advection $\mathbf{u}\cdot\nabla{q}$ where $\mathbf{u}$ is the velocity and $q$ a tracer quantity.

Because the local East--North frame rotates across the sphere (Figure~\ref{fig:method-overview}e), raw wind components $(U, V)$ encode the same physical wind differently at different tile positions---a frame inconsistency that would corrupt the grid-invariant encoding of StereoRoPE. We resolve this by projecting all wind vectors into the single East--North frame defined at the tile center.

Concretely, we form the 3D vector $\mathbf{v} = U\hat{e} + V\hat{n}$ using the per-point unit vectors $(\hat{e}, \hat{n})$ at $(\phi,\lambda)$, then project onto the tile-center tangent plane spanned by $(\hat{e}_0, \hat{n}_0)$---removing any radial leakage introduced by the change of frame---to obtain tile-center components:
\begin{equation}
    \mathbf{v}_\perp = \mathbf{v} - (\mathbf{v}\cdot\hat{r}_0)\hat{r}_0,
    \qquad
    U_\mathrm{loc} = \mathbf{v}_\perp\cdot\hat{e}_0,
    \qquad
    V_\mathrm{loc} = \mathbf{v}_\perp\cdot\hat{n}_0,
\end{equation}
where $\hat{r}_0$ is the radial unit vector at the tile center.

Section~\ref{sec:grid_invariance} tests StereoRoPE and wind rotation empirically with zero-shot rollouts on an unseen grid, ablating wind rotation and replacing StereoRoPE with standard axial RoPE, which indexes position by grid row and column.

\subsection{Training Objective and Multi-Step Fine-Tuning}
\label{sec:training_objective}

STRATA is first trained for one-step prediction with a channel-normalized Smoothed $\ell_1$ loss (Huber loss with $\delta=1$). One-step training minimizes single-step prediction error but allows artifacts to accumulate over autoregressive steps. We therefore fine-tune the model autoregressively for up to four steps, computing the loss on the final prediction. This multi-step fine-tuning substantially suppresses rollout instabilities and is key to stable 24-hour rollouts. Optimization details are in Appendix~\ref{sec:app_training_details}.

\section{Experiments}
\label{sec:results}

We evaluate STRATA along two axes: an isoFLOP scaling study to
characterize the compute--accuracy tradeoff at km-scale (\S\ref{sec:entropy}),
and a 24-hour autoregressive rollout to assess emulation fidelity
(\S\ref{sec:rollout}).
For the rollout, we adopt metrics appropriate for a chaotic dynamical system:
individual convective cells lose pointwise predictability within a few hours
\citep{surcelStudyScaleDependence2015}, so structural and statistical
fidelity---precipitation organization, error growth rate, and distributions---
are the meaningful tests, not pointwise agreement.

\subsection{Dataset}
\label{sec:data_training_rollout}
\label{sec:datasets}
\label{sec:app_training}
\label{sec:app_rollout}

STRATA emulates native SCREAM physics-grid output without regridding.  While the effective dynamics grid spacing of SCREAM is 3.3 km, its physics output data are saved every 10 minutes on a 4.9-km cubed-sphere grid ($6\times1024^2$ spectral elements, each with $2\times2$ physics cells, ${\approx}25$ million horizontal grid points) and subsampled from 128 raw SCREAM levels to 24 tropospheric vertical levels (surface to ${\sim}100$\,hPa; Figure~\ref{fig:vertical_levels}). We collect 28 days from three non-contiguous SCREAM simulations spanning diverse seasons (October, August, and January) and train on 17 days; rollout evaluation uses 6 held-out initializations at 12-hourly intervals from 2020-10-11 to 2020-10-14 (see Appendix~\ref{sec:app_training_data}). The model predicts the 10-minute evolution of 3D potential temperature, zonal and meridional wind, vertical velocity, geopotential height, and specific humidity fields, along with near-surface temperature; it also predicts diagnostic surface pressure and precipitation. Forcing inputs include solar zenith angle, surface geopotential, topography, latitude, and land fraction. A complete list of variable information is given in Appendix~\ref{sec:app_variables}.

\subsection{Scaling transformers for high-entropy data\label{sec:entropy}}

\begin{table}[t]
  \centering
  \caption{%
  \textbf{Energy efficiency and compute context.}
  \textbf{(a)} Throughput and energy efficiency of STRATA at two hardware scales
  compared to the SCREAM physics model it emulates.
  \textbf{(b)} Compute requirements of representative coarse AI weather models
  and STRATA; coarse models are compute context, not SCREAM-emulation baselines.
  The PFLOPs/SD gap includes
  $(6\,\text{h}/10\,\text{min})\times(25\,\text{km}/4.9\,\text{km})^2\approx900\times$
  from geometry alone.
  }
  \label{tab:hero-compute}
  \setlength{\tabcolsep}{3pt}
  
  \begin{minipage}[t]{0.47\linewidth}
  \centering
  {\small\textbf{(a) SCREAM emulation and energy}}\\[0.35em]
  \footnotesize
  \begin{tabular}{@{}lrr@{}}
  \toprule
  System & SDPD$^*$ & SD/MWh \\
  \midrule
  SCREAM               & 460  & 0.78 \\
  STRATA (16 H100s)    & 23.6 & 48.1 \\
  STRATA (512 H100s)   & 741  & 47.3 \\
  \bottomrule
  \end{tabular}

  \vspace{0.35em}
  \raggedright
  \footnotesize
  SD: simulated day; SDPD: simulated days per wall-clock day.
  $^*$SDPD depends critically on available hardware; see
  Table~\ref{tab:throughput_energy_accounting} for full accounting.
  \end{minipage}
  \hfill
  \begin{minipage}[t]{0.49\linewidth}
  \centering
  {\small\textbf{(b) Coarse AI compute context}}\\[0.35em]
  \footnotesize
  \begin{tabular}{@{}lrrr@{}}
  \toprule
  \textbf{Model} &
  \textbf{dt} &
  \textbf{TFLOPs/step} &
  \textbf{PFLOPs/SD} \\
  \midrule
  ACE2       & 6 h  & 0.48  & 0.002 \\
  Stormer    & 6 h  & 4.95  & 0.020 \\
  Atlas-CRPS & 12 h & 72.07 & 0.144 \\
  Aurora     & 6 h  & 47.00 & 0.188 \\
  \midrule
  DiT-S & 10 min & 1.2 & 1062 \\
  DiT-M & 10 min & 4.9 & 4335 \\
  \bottomrule
  \end{tabular}
  
  \vspace{0.35em}
  \raggedright
  \footnotesize
  TFLOPs/step is per global step for coarse models but per tile for
  DiT-S/M; PFLOPs/SD normalizes to a full simulated day globally.
  \end{minipage}
  \end{table}

A key design question for any emulator is how many FLOPs to allocate per forward pass: more FLOPs increase inference cost, but too few yield poor prediction quality. To understand this effect, we sweep over three iso-FLOP model tiers (DiT-S/M/L) and three horizontal patch sizes (1, 2, 4); see Appendix~\ref{sec:app_patch_sweep} and Table~\ref{tab:sweep} for configurations. Figure~\ref{fig:sweep_results}b shows that the loss decreases predictably with model FLOPs, with smaller models more compute-efficient at low training budgets and larger models reaching lower loss given sufficient training compute---consistent with \citet{hoffmannTrainingComputeOptimalLarge2022}. Within each tier, the larger patch sized models are more computationally efficient (Figure~\ref{fig:mfu-vs-patch-size}a). Models with a horizontal patch size $p_h=4$ run $1.5$--$2.9\times$ faster than $p_h=1$ models at equal FLOPs, with greater gains at smaller FLOP budgets.
Reasonable 24-hour rollout stability begins to emerge at and above the DiT-M tier
(${\sim}5$\,TFLOP), though biases remain present even at this scale;
models below this threshold exhibit more visible large-scale artifacts
(Figure~\ref{fig:model-size-ablation},
Appendix~\ref{sec:model-capacity-ablation}).

Consistent with a fundamental increase in underlying entropy, the computational scale required to succeed at simulating realistic convective-scale systems is 10$\times$ higher than naive extrapolations from established large-scale autoregressive AI weather models would predict based on spatial and temporal resolution differences alone (Table~\ref{tab:hero-compute}b). We postulate this increased computational load is required since storm-resolving data (in space and time) has higher information content due to the 3D nature of atmospheric convection. By contrast, the large-scale atmosphere behaves like a two-dimensional fluid, due to the strong impact of the Earth's rotation at those scales and the large aspect ratio.
A statistical analysis of the compressibility and covariance structure of the km-scale data supports this hypothesis (cf.\ Appendix~\ref{si:information-entropy}) and shows that the differential entropy per dimension ($-E_X[\log_2 p(X)]/d$) of 10-minute SCREAM increments is higher than that of 6-hour 0.25$^\circ$ ERA5 increments at every patch size from $1\times1$ to $8\times8$ grid cells---for example, 0.73 versus $-1.14$ bits/dim for $2\times2$ patches. Across patch sizes the gap is 0.4--1.9 bits/dim, a substantial fraction of the ${\approx}2$ bits/dim of unit-variance white noise, the maximum-entropy case. 

\subsection{Physical Realism, Predictability and Statistical Fidelity}
\label{sec:rollout}

Visualizations of the rollout quality provide a useful qualitative test, if done long after initial condition km-scale memory has vanished. Figure~\ref{fig:big-qv} shows that STRATA's self-generated dynamics resemble the ground truth across multiple phenomena -- tropical cyclone,
shallow clouds, midlatitude cold front, Intertropical Convergence Zone,
diurnal island convection, and orographic precipitation. Global videos provide a holistic view (\url{https://research.nvidia.com/labs/climate/strata/}).

\begin{figure}[t]
    \centering
    \includegraphics[width=\linewidth]{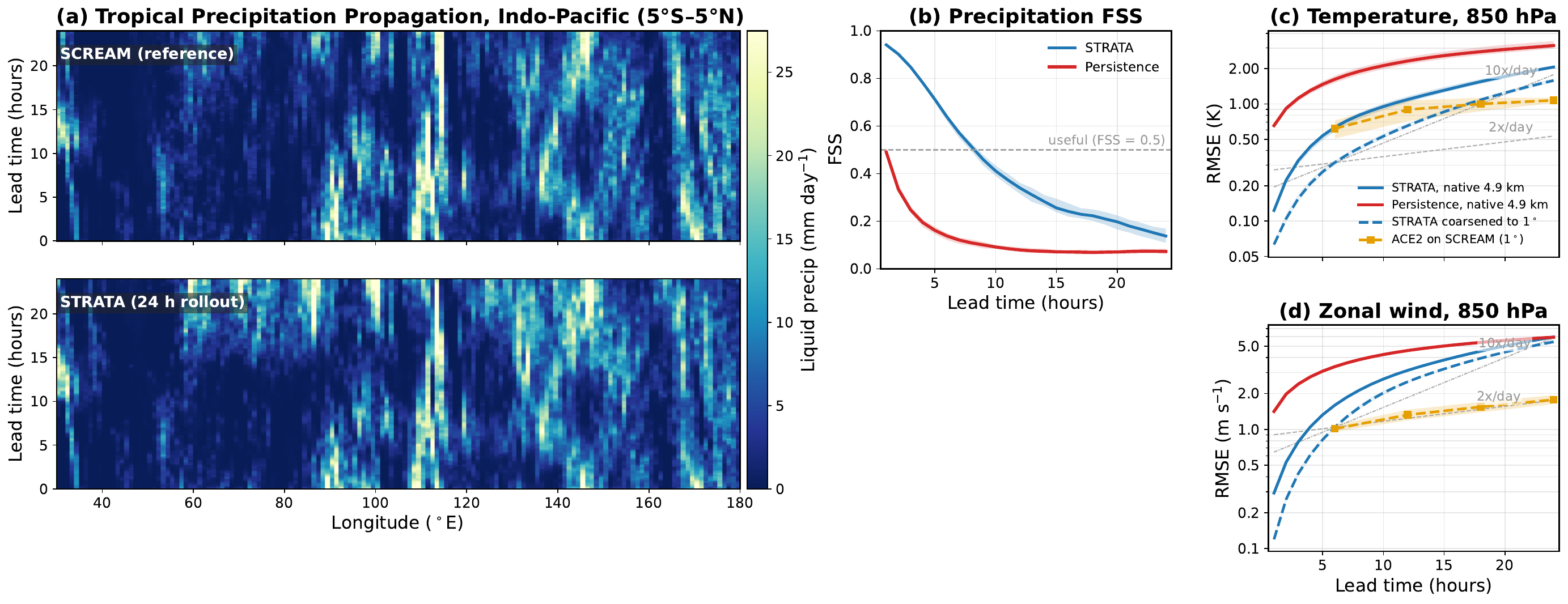}%
    \caption{%
    \textbf{Predictability diagnostics for 24-hour rollout.}
    \textbf{(a)} Tropical precipitation over the Indo-Pacific, averaged over
    5$^\circ$S--5$^\circ$N, shows coherent eastward and westward propagation of
    large-scale precipitation structures during STRATA free rollout.
    \textbf{(b)} Fractions Skill Score measures neighborhood-scale precipitation
    agreement with SCREAM; persistence is shown as an initial-condition memory
    reference rather than a competitive baseline.
    \textbf{(c,d)} Root-mean-square error for 850-hPa temperature and 850-hPa
    zonal wind on the native 4.9-km grid and after coarsening to
    the 1$^\circ$ ACE2 grid. ACE2, a coarse-resolution AI emulator, is shown as a reference for large-scale error growth rather than as a competing baseline. Lines show the mean over the 6 initialization times; shading shows their min--max range.
    }
    \label{fig:hovmoller}
    \label{fig:rmse}
\end{figure}

Quantitative error growth rates affirm good performance for the km-scale. The rainfall fractions skill score \citep{robertsScaleSelectiveVerificationRainfall2008} (FSS; Appendix~\ref{sec:app_rollout_metrics}) exceeds persistence (Figure~\ref{fig:hovmoller}b) indicating nontrivial maintenance of initialized organized rainfall structure. Meanwhile, tropical weather systems traverse the equatorial Indian and West Pacific at the correct speeds (phase tilt of rain streaks in Figure~\ref{fig:hovmoller}a). 

Limitations are also apparent in simulating the largest scale circulations. Here persistence and the coarse-resolution emulator ACE2~\citep{watt-meyerACE2AccuratelyLearning2024} serve as reference curves for realistic error growth rather than baselines to beat, since the skill scores test whether STRATA is as predictable as the km-scale simulation it emulates, not its value as a forecast. On time scales of 12-24 hours, where macroscale forecast errors dominate, total error growth is too fast (Figure~\ref{fig:hovmoller}c,d; slope of blue and blue-dashed) relative to expectations of an RMSE doubling-per-day (grey, yellow).
ACE2, trained on six decades of reanalysis to capture exactly these scales, follows the expected slower growth; STRATA, trained on 17 days of km-scale data, does not match it at synoptic scales; for slowly varying large-scale fields such as surface pressure and 500-hPa height, even persistence is more accurate (Figure~\ref{fig:scorecard} in Appendix~\ref{sec:app_rollout_metrics}).
We see two likely reasons. First, STRATA is trained only on 10-minute, km-scale targets, in which the slow evolution of the balanced large-scale flow is a weak signal: for geostrophically balanced motion, the wind tendency the model learns is not a leading-order term in the momentum budget. Second, 17 days of training data may be too short to sample the balanced large-scale states. Fine-tuning on larger tiles with an additional rollout loss at hourly and ${\geq}25$-km scales could potentially target the first reason directly, letting the model learn the large-scale balance from data.

\begin{figure}[t!]
    \centering
    \includegraphics[width=0.8\linewidth]{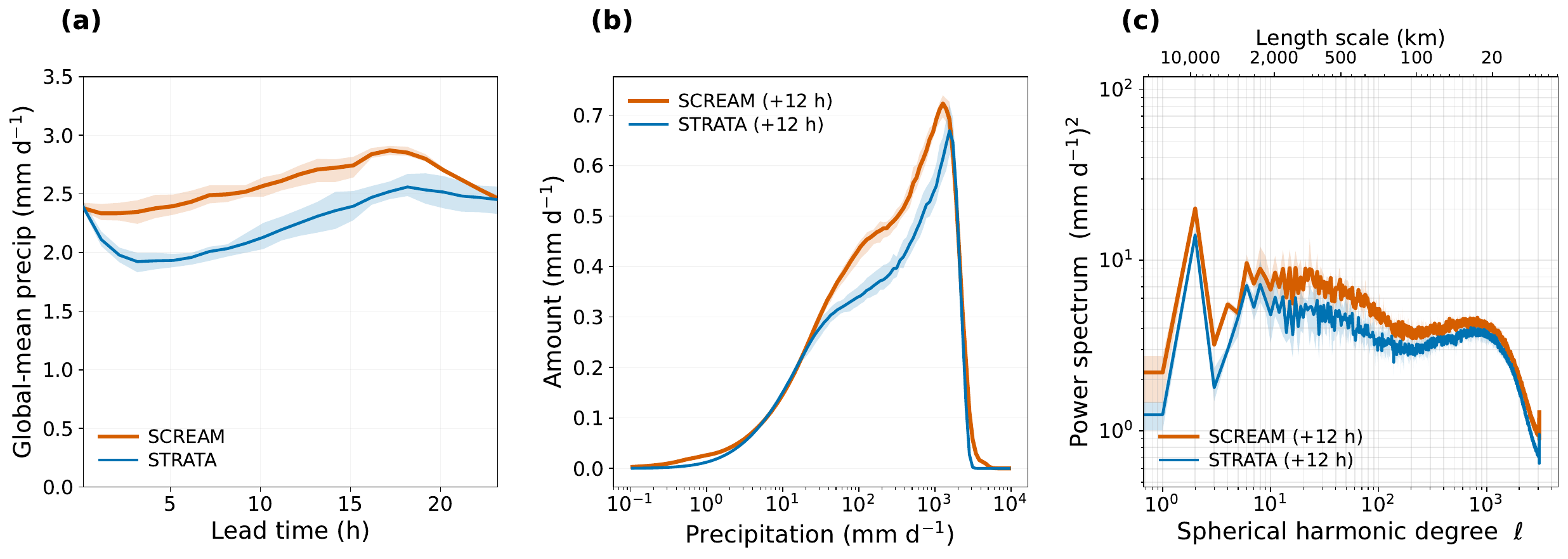}
    \caption{%
    \textbf{Precipitation statistics during rollout.}
    Solid lines show the mean over three rollouts initialized on different
    dates at the same time of day; shading shows the min--max range.
    \textbf{(a)} Global-mean precipitation over 24-hour rollouts.
    \textbf{(b)} Precipitation amount distribution at 12-hour lead time,
    showing each rain-rate bin's contribution to the global mean.
    \textbf{(c)} Spherical-harmonic power spectrum of precipitation at
    12-hour lead time.
    }
    \label{fig:precip-stats}
\end{figure}

Global mean rainfall is biased low by as much as 20\% (Figure~\ref{fig:precip-stats}a); an inference-time filter on the largest scales of vertical velocity limits this dry drift. Its global-mean part enforces mass continuity, while the rest is empirical (Appendix~\ref{sec:app_omega_filter}, Figure~\ref{fig:omega-filter-ablation}). The precipitation amount distribution
(Figure~\ref{fig:precip-stats}b) broadly matches SCREAM though the
moderate-to-intense rain rates contribute slightly less than in SCREAM,
consistent with the dry bias. Suppressed 12-h spectral power concentrated in the 100-2,000 km length scale regime again implicates imperfect synoptics (Figure~\ref{fig:precip-stats}c). 

Skill is similar on held-out initializations from the August and January simulations (Appendix~\ref{sec:app_seasons}). Despite the large-scale drift, the storm-scale dynamics STRATA learns also largely transfer to a season absent from training. We train a second model without the January simulation data, excluding it from training and checkpoint selection, and compare it with STRATA on three held-out January initializations (Figures~\ref{fig:loso-maps} and~\ref{fig:loso-lines} in Appendix~\ref{sec:app_seasons}). Relative to the same comparison on August, which both models saw, 850-hPa wind and potential temperature errors grow by up to about 20\%, while moisture and precipitation skill are unchanged. The learned convective dynamics thus carry over to the unseen season, likely because weeks of global km-scale output already contain a vast number of nearly independent convective events.

\subsection{Grid invariance}
\label{sec:grid_invariance}
\begin{table}[t]
\centering
\caption{Zero-shot transfer to a cubed sphere rotated by 45$^\circ$. $q_v$ RMSE at the model level nearest 850\,hPa on the rotated grid divided by that on the native grid, and the change in tropical-mean ($|\mathrm{lat}|<20^\circ$) rainfall at +3\,h. Initialized 2020-10-13 00\,UTC.}
\label{tab:grid-transfer}
\small
\begin{tabular}{lcccc}
\toprule
 & \multicolumn{3}{c}{$q_v$ RMSE, rotated / native} & Tropical rain \\
Model & +3\,h & +6\,h & +12\,h & +3\,h \\
\midrule
StereoRoPE + wind rotation & 1.54 & 1.24 & 1.07 & $-6\%$ \\
StereoRoPE, no wind rotation & 1.55 & 1.30 & 1.15 & $+9\%$ \\
Axial RoPE, no wind rotation & 2.90 & 2.02 & 1.49 & $+75\%$ \\
\bottomrule
\end{tabular}
\end{table}

StereoRoPE and wind vector rotation together should make learned operators transferable across grid topologies.
To test this, we train three models, separately from the production model  (Appendix~\ref{sec:app_ablation_protocols}), that differ only in position encoding and wind rotation:
StereoRoPE with wind rotation, StereoRoPE without wind rotation, and standard axial 2-D RoPE without wind rotation.
All three are trained on the native cubed-sphere grid only, and on that grid they are closely matched (Figure~\ref{fig:grid-invariance} in Appendix~\ref{sec:app_grid_types}).
We then run each zero-shot on a cubed sphere rotated by 45$^\circ$, which moves the centres of the two polar faces from the poles to 45$^\circ$N, 90$^\circ$W and 45$^\circ$S, 90$^\circ$E and places every face, edge and corner over new geography, and compare it with its own native-grid forecast at +3\,h (Table~\ref{tab:grid-transfer}).
Axial RoPE nearly triples its $q_v$ error and develops a spurious tropical rain band, raising tropical rainfall by 75\%.
Both StereoRoPE models keep tropical rainfall within 10\% of its native value and raise their $q_v$ error by about half, but without wind rotation thin diagonal rain streaks appear near the equator.
With both components the rotated forecast shows the fewest artifacts (only a weak residual over the eastern tropical South Atlantic) and most closely resembles the native-grid forecast.

\section{Conclusion}
\label{sec:conclusion}

STRATA is the first AI model to simulate storm-scale atmospheric dynamics globally, expanding the prior state of the art of 900\,km domain and 2-hour rollouts \citep{floraWoFSCastMachineLearning2025} to the full planet and 24-hour rollouts, at 4.9\,km and 10-minute resolution. Rollout evaluation confirms realistic simulation across the planet, spanning diverse convective regimes---from shallow marine stratocumulus and isolated continental storm lifecycles to organized deep convection, topographic precipitation, diurnal island convection, and midlatitude frontal bands.

With more than $50\times$ the energy efficiency of SCREAM (the physics model it emulates) and a throughput of 741 simulated days per wall-clock day at 512 H100 GPUs, STRATA makes global convection-resolving simulation far cheaper than the physics model itself. Our analysis has revealed emulating convective dynamics requires ${\sim}10\times$ more FLOPs per grid point than synoptic-scale forecasting, due to the higher information entropy of 3D convective dynamics.

The current local-tile model learns convective-scale physics well but not large-scale circulation. 
Coupling with a complementary low-resolution global emulator is a natural way to explore the next frontier, enabling century-scale simulations that quantify how the \emph{frequency} of extreme weather events changes with environmental conditions. 
Eventually, we hope the efficiencies of AI simulation might unlock the global ${\sim}$10-100-m scales needed to directly resolve low-cloud feedbacks---a major driver of climate sensitivity uncertainty \citep{schneiderClimateGoalsComputing2017} that remains computationally intractable
\citep{parishaniInsensitivityCloudResponse2018, teraiImpactResolvingSubkilometer2020, pengImprovingStratocumulusCloud2024}. 

\section{Data and Code Availability}
\label{sec:availability}
The STRATA code, including the model architecture and training and inference pipelines, is available at \url{https://github.com/NVlabs/STRATA}; code release \texttt{v2026.11.6} is the version at the time of publication, and since the repository will continue to be developed, we recommend its latest version for new work. The trained model weights are available at \url{https://huggingface.co/nvidia/strata}, and videos of global rollouts are on the project page, \url{https://research.nvidia.com/labs/climate/strata/}. The raw SCREAM output used for training and evaluation is available at \url{https://portal.nersc.gov/archive/home/n/ndk/www/STRATA2026}. As E3SM model output, it is distributed under the Creative Commons Attribution 4.0 International (CC BY 4.0) license (see \url{https://docs.e3sm.org/e3sm_data_docs/_build/html/index.html}). A training-ready version in Zarr format is available on Hugging Face, one repository per simulation: \url{https://huggingface.co/datasets/nvidia/STRATA-SCREAM-sdecadal} for sdecadal, and the companion repositories \texttt{nvidia/STRATA-SCREAM-sdy1} and \texttt{nvidia/STRATA-SCREAM-sdy2} for DYAMOND1 and DYAMOND2.

\section*{Acknowledgements}
We thank Dale Durran, Jindong Jiang, Peter Harrington, Qiyu Song, Zhiming Kuang, Jean Kossaifi, and Julius Berner for their advice and feedback on this work. This research used resources of the National Energy Research Scientific Computing Center (NERSC), a Department of Energy User Facility using NERSC award BER-ERCAP-0035294. 
Part of this work was performed under the auspices of the U.S. Department of Energy by Lawrence Livermore National Laboratory under Contract DE-AC52-07NA27344 and at Pacific Northwest National Laboratory under Contract DE-AC05-76RL01830.

\bibliographystyle{unsrtnat}
\bibliography{bibliography}

\newpage
\tableofcontents

\appendix
\setcounter{table}{0}
\renewcommand{\thetable}{S\arabic{table}}
\setcounter{figure}{0}
\renewcommand{\thefigure}{S\arabic{figure}}
\renewcommand{\theHtable}{S\arabic{table}}
\renewcommand{\theHfigure}{S\arabic{figure}}

\section{Background: Global Storm-Resolving Models}
\label{sec:app_gsrm_background}

Traditional climate models use approximate representations
of small-scale processes like thunderstorms and cloud physics \citep{arakawa2004cumulus},
which are a major source of uncertainty in climate projections.
Physics-based Global Storm-Resolving Models (GSRMs)
operate at kilometer-scale resolutions ($<5$\,km)
to explicitly resolve deep convection and complex orography,
demonstrating superior fidelity in capturing
precipitation extremes, diurnal cycles,
and mesoscale organization compared to coarser models
\citep{satoh2014non,caldwellConvectionPermittingSimulationsE3SM2021,maSuperiorDailySubDaily2022,hohenegger2023icon}.
However, their staggering computational and energy costs
fundamentally limit their utility
for large ensembles needed to optimally calibrate uncertain physics parameterizations and to sample internal variability alongside long-term ocean-coupled climate projections.
Recent supercomputing initiatives represent the state of the art:
the nextGEMS project demonstrated the feasibility
of 30-year coupled simulations
at 5--10\,km resolutions \citep{seguraNextGEMSEnteringEra2025},
while SCREAMv1 achieved 419.5 simulated days per day
at 3.3\,km resolution,
requiring 8{,}192 nodes of the Frontier supercomputer
\citep{taylorSimpleCloudResolvingE3SM2023,donahueExascaleSimpleCloudResolving2024}.
The EU Destination Earth initiative
has operationalized coupled multi-decadal digital twins
on EuroHPC systems,
achieving 120--230 simulated days per day
at 5--10\,km resolutions
\citep{hadadeDestinationEarthClimate2025}.
Despite these milestones,
the computational demands remain immense,
requiring dedicated access to the world's largest supercomputers.

\section{Architecture details}
\label{sec:app_architecture_details}

\subsection{STRATA Architecture and Production Hyperparameters}
\label{sec:app_arch_spec}

\begin{figure}[t]
    \centering
    \includegraphics[width=0.8\textwidth]{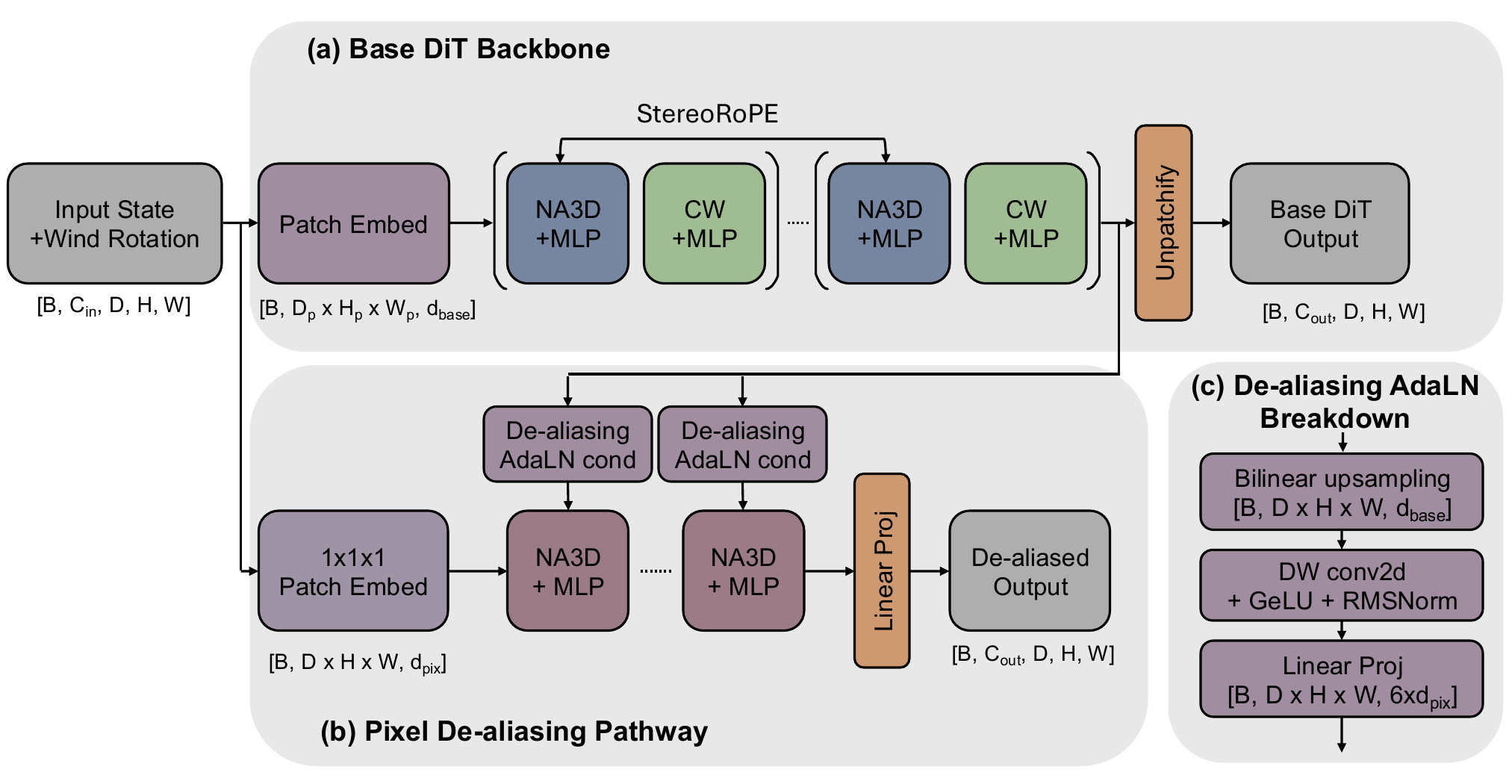}
    \caption{Architecture of STRATA with DiT backbone and PixelDiT-style pixel-space de-aliasing decoder.
    \textbf{(a)} The DiT backbone applies alternating 3D neighborhood attention
    (NA3D) and column-wise (CW) attention for transformer blocks with StereoRoPE position
    embeddings, followed by an unpatchify projection.
    \textbf{(b)} The pixel de-aliasing pathway re-embeds at full resolution and
    applies NA3D blocks conditioned via adaptive layer normalization (AdaLN)
    on the backbone output, then projects to the output fields.
    \textbf{(c)} Each AdaLN block uses bilinear upsampling, depthwise (DW)
    Conv2d with GeLU activation and RMSNorm, and a linear projection.
    $d_\mathrm{base}$: backbone embedding dimension; $d_\mathrm{pix}$:
    decoder embedding dimension.
    Tensor shapes are written as [batch $B$, channels $C_\mathrm{in}$ or $C_\mathrm{out}$,
    vertical levels $D$, tile height $H$, tile width $W$];
    $D_p\times H_p\times W_p$ is the grid of patches.
    MLP is the feed-forward sublayer of each block.
    The AdaLN projection outputs $6\times d_\mathrm{pix}$ channels: a scale, shift and gate
    for the attention and MLP sublayers.}
    \label{fig:dit-schematic}
\end{figure}

Figure~\ref{fig:dit-schematic} summarizes the production architecture used for
the rollout results, and Table~\ref{tab:arch} lists its main hyperparameters
and inference cost. The base DiT backbone produces patch-resolution semantic tokens,
while the de-aliasing decoder refines the output at full horizontal resolution.

\begin{table}[H]
\centering
\caption{Architecture hyperparameters and inference cost of STRATA.
Inference measured on a single H100 80\,GB GPU with \texttt{torch.compile} and
BF16 mixed precision, batch size~1, tile size~64x64.}
\label{tab:arch}
\small
\begin{tabular}{lll}
\toprule
\textbf{Component} & \textbf{Hyperparameter} & \textbf{Value} \\
\midrule
\multirow{6}{*}{DiT Backbone}
  & Embedding dimension       & 1024 \\
  & Transformer layers        & 24   \\
  & Attention heads           & 16   \\
  & NA3D kernel size          & 9    \\
  & Horizontal patch size     & 4    \\
  & Vertical patch size       & 1    \\
\midrule
\multirow{3}{*}{De-aliasing Decoder}
  & Embedding dimension       & 128  \\
  & Transformer layers        & 4    \\
  & NA3D kernel size          & 9    \\
\midrule
\multirow{3}{*}{Cost}
  & Total parameters          & 331.7\,M  \\
  & GFLOPs (per tile)         & 4{,}824   \\
\bottomrule
\end{tabular}
\end{table}

\subsection{Patch-Size and Model-Size Sweep}
\label{sec:app_patch_sweep}

Table~\ref{tab:sweep} defines the nine configurations used in the patch-size
and model-size scaling study in Figure~\ref{fig:sweep_results}. Model names
encode the FLOP tier (S/M/L) and horizontal patch size. Within each tier, depth
and width are adjusted so that models have approximately matched forward-pass
FLOPs while varying tokenization. Training protocols of all sweeps are in Table~\ref{tab:sweep-protocols}.

PixelDiT-S (same FLOP tier as DiT-S, ${\sim}1.2$\,TFLOP) and PixelDiT-M (
same FLOP tier as DiT-M, ${\sim}5$\,TFLOP) add a standard PixelDiT decoder with
linear-projection upsampling to the corresponding DiT tier.
PixelDiT-M shares the same backbone hyperparameters as Table~\ref{tab:arch}
but uses the original PixelDiT decoder, whose adaLN conditioning is a single linear projection of each patch token
(i.e., Figure~\ref{fig:dit-schematic}c without the upsampling and depthwise convolution).
PixelDiT-S uses a smaller backbone: 768 embedding dimension, 12 transformer
layers, 12 attention heads, $4\times4\times1$ patch size; pixel decoder:
64 embedding dimension, 2 transformer layers.

\begin{table}[H]
\centering
\caption{Model configurations for the scaling sweep.
Model names encode FLOP tier (S/M/L) and horizontal patch embedding size
(e.g., DiT-S-ps4 is the smallest FLOP tier with $4\times4$ horizontal patch embedding).
Within each FLOP tier, the number of layers and hidden dimension are adjusted to match
the target FLOP budget; training hyperparameters are held fixed across all configurations.
All these  nine models use an NA3D kernel size of 3; the PixelDiT and production models later adopted
kernel size 9 following further exploration (Table~\ref{tab:arch}).
}
\label{tab:sweep}
\small
\begin{tabular}{lrrr}
\toprule
\textbf{Model} & \textbf{Layers} & \textbf{Hidden dim $d$} & \textbf{TFLOPs} \\
\midrule
DiT-S-ps1 & 14 & 192  & 1.2 \\
DiT-S-ps2 & 14 & 384  & 1.2 \\
DiT-S-ps4 & 14 & 768  & 1.2 \\
\midrule
DiT-M-ps1 & 14 & 384  & 4.9 \\
DiT-M-ps2 & 14 & 768  & 4.9 \\
DiT-M-ps4 & 32 & 1024 & 5.0 \\
\midrule
DiT-L-ps1 & 14 & 768  & 19.5 \\
DiT-L-ps2 & 32 & 1024 & 19.8 \\
DiT-L-ps4 & 32 & 2048 & 19.8 \\
\bottomrule
\end{tabular}
\end{table}

\subsection{Patch De-aliasing}
\label{sec:app_patch_dealiasing}

\begin{figure}[p]
    \centering
    \includegraphics[width=0.5\textwidth]{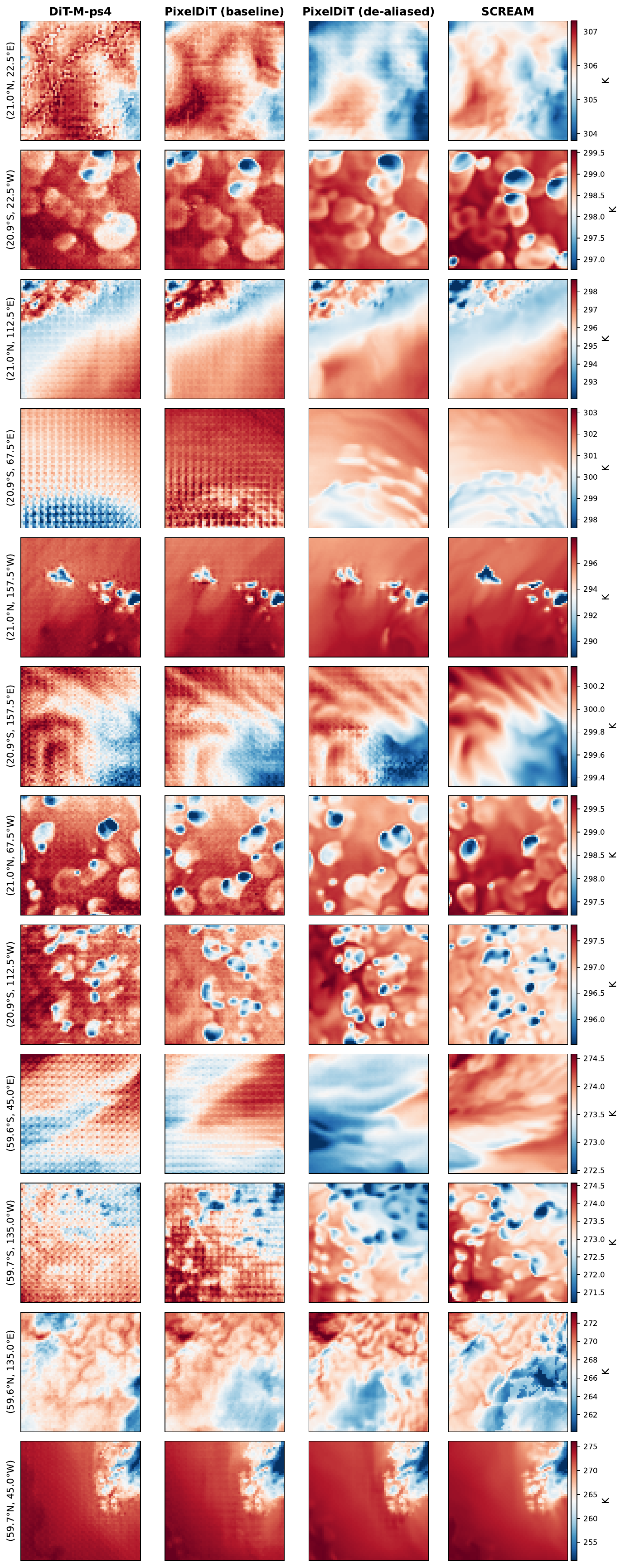}
    \caption{Near-surface temperature (K) at 12\,h lead time for 12 different locations
    comparing four model variants.
    \textbf{DiT-M-ps4}: patch-size-4 DiT without de-aliasing,
    showing visible checkerboard artifacts at the $4\times4$ patch scale.
    \textbf{PixelDiT-M-ps4 (baseline)}: PixelDiT without the de-aliasing conditioning,
    where patch-boundary artifacts persist.
    \textbf{PixelDiT-M-ps4 (dealiased)}: full STRATA de-aliasing decoder,
    producing spatially smooth fields consistent with the reference, although on a small portion of locations, the patch-scale artifacts are not fully fixed.
    \textbf{SCREAM}: high-resolution reference simulation.
    The dealiased column is the STRATA checkpoint used throughout the paper; the two models in the controlled ablation (Table~\ref{tab:dealias-ablation}) are trained separately and are not shown here.}
    \label{fig:patch-artifact}
\end{figure}

Patch-size-4 tokenization makes transformer inference substantially cheaper by
operating on fewer horizontal tokens, but direct decoding can introduce a
periodic patch-scale mode during rollout. This artifact is easiest to see in
smooth near-surface fields such as $T_{2\mathrm{m}}$. Figure~\ref{fig:patch-artifact}
compares model variants at 12\,h lead time: direct unpatchifying produces a
visible $4\times4$ checkerboard, while the PixelDiT-style two-stage decoder
reduces but does not fully remove patch-boundary structure. Our de-aliased
decoder replaces the original PixelDiT linear upsampling with bilinear
upsampling followed by a depthwise convolution, substantially reducing the
artifact while preserving the large-scale temperature structure.

\paragraph{Design of the conditioning path.}
We chose the de-aliasing decoder's conditioning path with an ablation during the development of STRATA. Each pixel block is conditioned on the semantic tokens through adaLN, which maps the 1024-dimensional semantic map at patch resolution to 768 modulation parameters (shift, scale and gate for the two sublayers of a 128-dimensional pixel block) at every pixel; the original PixelDiT decoder predicts each pixel's parameters directly from its own patch token. Starting from the same intermediate PixelDiT checkpoint with the semantic blocks frozen, we trained three variants on the same data: (a) project from 1024 to 768 dimensions at patch resolution, then upsample bilinearly to pixel resolution; (b) upsample the 1024-dimensional map bilinearly, then project at pixel resolution; (c) as (b), with an identity-initialized $5\times5$ depthwise convolution, a parameter-free RMSNorm and a GeLU between the upsampling and the projection, the design adopted in Figure~\ref{fig:dit-schematic}c. We compared them only on one-step metrics logged during training, not in rollouts or on the patch artifact. The three separate cleanly in both one-step validation loss and one-step skill (Table~\ref{tab:decoder-variants}), and the ordering holds at every logged training step. Variants (a) and (b) contain the same two operations and have identical parameter counts; they differ only in the width at which the upsampling is done. Projecting to 768 dimensions at patch resolution discards conditioning directions that interpolation cannot restore, while upsampling in the wider space keeps them. Moving the projection after the upsampling also moves it to pixel resolution, so (b) costs more FLOPs than (a). The depthwise convolution adds a further gain. Because its kernel starts as the identity, (c) begins as (b), and the gain is learned in training.

\begin{table}[h]
\centering
\caption{Variants of the de-aliasing decoder's conditioning path, compared during the development of STRATA. Each is trained from the same intermediate PixelDiT checkpoint with the semantic blocks frozen, and all are compared on one-step metrics at a matched training step. Skill is $1-\mathrm{MAE}/\mathrm{MAE}_{\mathrm{persistence}}$, median over channels. (c) is the design adopted in STRATA.}
\label{tab:decoder-variants}
\small
\begin{tabular}{llcc}
\toprule
 & Conditioning path & Val.\ loss $\downarrow$ & Skill $\uparrow$ \\
\midrule
(a) & project $\rightarrow$ upsample & 0.047 & 0.66 \\
(b) & upsample $\rightarrow$ project & 0.024 & 0.75 \\
(c) & upsample $\rightarrow$ DW conv, RMSNorm, GeLU $\rightarrow$ project & 0.015 & 0.81 \\
\bottomrule
\end{tabular}
\end{table}

\paragraph{Controlled decoder ablation.}
To compare the decoders fairly on one-step (offline) skill, rollout (online) skill and the patch artifact, we train two models from scratch that differ only in the decoder: the original PixelDiT decoder (``plain''), in which each pixel's adaLN modulation is projected directly from its own patch token, and our de-aliasing decoder, which upsamples the semantic map bilinearly and refines it with a depthwise convolution before the projection (Figure~\ref{fig:dit-schematic}c). We do not use the STRATA checkpoint for this comparison because it has a complicated multi-stage training history. All models use StereoRoPE in the semantic blocks. One difference to STRATA is that the two ablation models also apply it in the pixel blocks, which we enabled after training STRATA. Their training protocols are in Table~\ref{tab:ablation-models}. Offline, the de-aliasing decoder is 15\% worse in one-step validation loss, compared during pretraining at a matched budget of about 6M training images. Online, rollout skill is comparable: in 24-hour rollouts from 2020-10-13 00:00\,UTC, skill against persistence ($1-\mathrm{RMSE}/\mathrm{RMSE}_{\mathrm{persistence}}$) at +12\,h is 0.34 / 0.33 for $U$, 0.41 / 0.42 for $V$ and 0.50 / 0.52 for $\theta$ at model level 24, and 0.57 / 0.60 for $T_{2\mathrm{m}}$ (plain / de-aliasing). The one-step cost therefore does not carry into the rollout.

Table~\ref{tab:dealias-ablation} quantifies the patch artifact in these rollouts. We compute 1D power spectra along the rows and columns of each 128-pixel tile, trimming 8 pixels at each tile edge so that tile seams do not contribute, take the power at the 4-pixel patch wavelength relative to the mean power at wavelengths of 4.3--5.3 pixels, and divide by SCREAM's value at the same valid time, so 1.0 means the field is indistinguishable from SCREAM at the patch scale. The plain decoder's artifact grows with lead time, reaching 2.6 times SCREAM's level for $T_{2\mathrm{m}}$ at +24\,h, while the de-aliasing decoder stays within 5\% of SCREAM throughout. STRATA, rolled out from the same initialization, keeps $\theta$ and $q_v$ at SCREAM's level but retains a smaller $T_{2\mathrm{m}}$ excess that grows to 1.40 at +24\,h, consistent with the residual artifacts at some locations in Figure~\ref{fig:patch-artifact}.

\begin{table}[h]
\centering
\caption{Patch-scale artifact: power at the 4-pixel wavelength relative to slightly longer wavelengths, divided by SCREAM's value at the same valid time (1.0 = indistinguishable from SCREAM), in 24-hour rollouts from 2020-10-13 00:00\,UTC. Ablation-plain and ablation-de-aliasing are the controlled decoder pair, trained from scratch; STRATA is the model used throughout the paper. $\theta$ and $q_v$ at model level 24 (${\approx}848$\,hPa).}
\label{tab:dealias-ablation}
\small
\begin{tabular}{llcccc}
\toprule
 & Model & +6\,h & +12\,h & +18\,h & +24\,h \\
\midrule
\multirow{3}{*}{$T_{2\mathrm{m}}$} & ablation-plain       & 1.08 & 1.28 & 1.78 & 2.56 \\
                                   & ablation-de-aliasing & 1.01 & 0.99 & 0.99 & 0.99 \\
                                   & STRATA               & 1.02 & 1.04 & 1.16 & 1.40 \\
\midrule
\multirow{3}{*}{$\theta$}          & ablation-plain       & 1.04 & 1.19 & 1.57 & 2.25 \\
                                   & ablation-de-aliasing & 1.00 & 1.01 & 1.01 & 1.01 \\
                                   & STRATA               & 1.01 & 1.02 & 1.01 & 1.02 \\
\midrule
\multirow{3}{*}{$q_v$}             & ablation-plain       & 1.03 & 1.10 & 1.29 & 1.58 \\
                                   & ablation-de-aliasing & 1.01 & 1.02 & 1.04 & 1.05 \\
                                   & STRATA               & 1.01 & 1.02 & 1.02 & 1.02 \\
\bottomrule
\end{tabular}
\end{table}

\subsection{Inference-Time Mitigation of Patch Artifacts}
\label{sec:app_inference_tricks}

The patch artifact can also be reduced at inference without changing the model. With a patch size of 4, the artifact is a pattern that repeats every 4 pixels in each direction, locked to the patch grid of each tile. Shifting the tiling by each of the 16 possible offsets within a patch and averaging the 16 forecasts at every step would cancel this pattern in principle, but would make inference 16 times more expensive. A cheaper alternative is to add a random offset to the tiling at each rollout step, so that the patch grid does not stay in the same place from step to step. We find that this helps, but it does not prevent the patch artifact in DiT or PixelDiT. The pattern can also be estimated from the forecast itself, by averaging each of the 16 positions within a patch over a local block, and subtracted as a template. This removes part of the excess patch-scale power but not all of it, because the artifact's amplitude varies from place to place while the template within each block is fixed. None of these addresses the cause, so STRATA instead removes the artifact in the decoder (Appendix~\ref{sec:app_patch_dealiasing}).

\subsection{Patch Instability Analysis}
\label{sec:stability-analsis-details}

For simplicity we can focus on the one dimensional case of a length $n$ sequence. Let the flattened state vector be given by $x\in\mathbb{R}^{nc_{in}}$ where $c_{in}$ is the channel index. For simplicity, we focus on a simple network consisting of $Lx=DEx$ representing a linearized transformer model with a residual connection. Let $\ell$ be the patch-size. The embed operator $E$ represents patch embedding and can be expressed in matrix form as $E=diag([W_E, \ldots, W_E])$ where $W_E\in\mathbb{R}^{c \ell \times c_{hid}}$. $c_{hid}$ is the embedding dimension of the architecture. The decoding operation can also be expressed as a block-diagonal matrix $D=diag([W_D,\ldots,W_D])$ with $W_D\in\mathbb{R}^{c_{hid} \times c_{out} \ell}$. Putting this together we see that $L$ also has a block diagonal form $L=diag([W_L, ..., W_L])$ where  $W_L=W_D W_E$. This is a linear setting for ease of analysis. Moreover, it does represent an actual mode when there is a skip connection directly around the transformer blocks as in our architecture.

For any time marching procedure (e.g. weather rollout or diffusion sampling) the typical setup is to roll out the network recursively with dynamics given by
\begin{equation}
x_{j+1} = A x_j := \alpha I + \beta L x_j.    \label{eq:power}
\end{equation}
where $\alpha$ and $\beta$ are scalars or more generally diagonal matrices which repeat every $c_{out} l$ elements. Setting $\alpha=0,1$ correspond to state and tendency prediction, respectively. We note that $A$ is also block diagonal with all the blocks being $W_A=\alpha I + \beta W_D W_E$.
It is an elementary fact that when initialized with non-degenerate initial data  $x_0$ such an iteration converges as $x_j \rightarrow \lambda^j v_1 \langle x_0, v\rangle$, where $v$ is the eigenvector of $L$ whose eigenvalue $|\lambda|=\rho(A)$ has the largest norm, also known as the spectral radius. This is known as the power-iteration algorithm. The \emph{stability region} of the iteration is defined as the subset of $(W_E,\alpha,\beta,W_D)$ where $\rho(A)\leq1$.

For any block diagonal matrix like $A$, we can show that $\rho(A)=\rho(W_A)$ and that corresponding eigenvectors $v$ have patch-scale structure. To see this, let $v$ be an eigenvector of $W_L$ with eigenvalue $\lambda$, $i$ be one of the patches, and  $x[i\ell :(i + 1) \ell]=v$ (in Python notation) is the vector whose value is $v$ in patch $i$ and zero elsewhere. Then, we see that $x$ is an eigenvector of $A$. i.e.  $Ax=[0, \ldots,0, W_Av,0\ldots,0]=[0, \ldots,0, \lambda v,0\ldots,0] = \lambda x$. Therefore, the eigenvalues of $A$ are also eigenvalues of $W_A$, and in particular the largest one $\rho(A)=\rho(W_A)$. Because the same block $W_A$ is repeated in every patch, the leading eigenspace of $A$ consists of patch-local copies of the dominant eigenvector of $W_A$. Thus power iteration amplifies the same within-patch pattern independently in each patch, producing a visible patch-scale artifact; the amplitude (and, for complex eigenvalues, phase) in each patch is determined by the projection of the initial state onto that patch's eigenvector. 

For typical neural network initializations, we see that $\rho(A) > 1$ which means that a periodic patch-scale artifact will dominate the rollout. By the considerations above the spectral radius of A is 
\begin{equation}
\rho(A)=\max_{m} \sqrt{(\alpha + \beta \Re \lambda_m)^2 + (\beta \Im \lambda_m)^2}    
\end{equation}
where $\lambda_m$ are the eigenvalues of $W_D W_E$.
This explains why tendency prediction ($\alpha=1$) is observed to be less stable than state prediction \citep{bonevFourCastNet3Geometric2025a}.
In this case, $\rho(A)>1$ if $W_D W_E$ has any eigenvectors with a positive real component, something that is almost sure to happen.
Even state prediction $\alpha=0$ can be unstable for typical network initializations. With $\alpha=0$ and $\beta=1$ the spectral radius $\rho(A)=\rho(W_D W_E)\approx 2$ when using Kaiming initialization of the weights.

Like any initial bias this can be trained away given enough data, but this is more efficient for high frequency fields that project strongly onto sub-patch scale variations. To see why, we note that the projection of the input data $x_0$ onto $v$ can be bounded by $|\langle x_0, v \rangle|\leq || x-AvgPool_\ell x ||$, assuming $x_0$ is 0-centered. For smooth fields $||x - AvgPool_l (x)||\ll ||x||$ since the sub-patch variation only accounts for a small portion of the overall variance of the field. So, the component of single-step network prediction explained by patch artifact $\rho(A) \langle x_0, v \rangle v$ is small compared to the target magnitude $Std[x]$ and so is not suppressed quickly during training.

\subsection{Additional Stabilization Details}
\label{sec:app_arch_stabilization}

\paragraph{QK normalization and gated attention.}
Training in BF16 precision introduces attention instability: unconstrained query--key dot products can produce
pre-softmax logits on the order of $\pm100$ to $\pm1000$, causing gradient overflow.
We address this with QK RMSNorm~\cite{henryQueryKeyNormalizationTransformers2020}, which applies RMSNorm
(without learnable scale) to the query and key vectors before their dot product,
bounding the logit magnitude and enabling stable BF16 training. All attention blocks additionally use gated attention~\cite{qiuGatedAttentionLarge2025},
which applies a head-specific sigmoid gate to the attention output,
improving expressiveness and further stabilizing training.

\section{Dataset details}

STRATA is trained to emulate the Simple Cloud-Resolving E3SM Atmosphere
Model (SCREAM)~\cite{caldwellConvectionPermittingSimulationsE3SM2021}. We use
native SCREAM physics-grid output without regridding. Prognostic 3D fields have
global shape $(6,24,2048,2048)$ on the cubed-sphere grid; prognostic and
diagnostic 2D fields have shape $(6,2048,2048)$; boundary forcing 2D fields share the
same horizontal shape. Output is saved every 10\,minutes. SCREAM's main time step is 75\,s, with 9.375-s dynamical-core
substeps \citep{caldwellConvectionPermittingSimulationsE3SM2021}, so each 10-minute emulator step
spans 8 model time steps and 64 dynamical-core substeps.

\subsection{STRATA Vertical Grid}
\label{sec:app_vertical_grid}

SCREAM's 128-level vertical grid is aggressively subsampled for the emulator.
We retain 24 levels spanning the troposphere and tropopausal layer
(approximately ${\sim}100$\,hPa down to the surface in hybrid sigma-pressure coordinates),
discarding the stratosphere and above.

\begin{figure}
\centering
\includegraphics[width=0.45\textwidth]{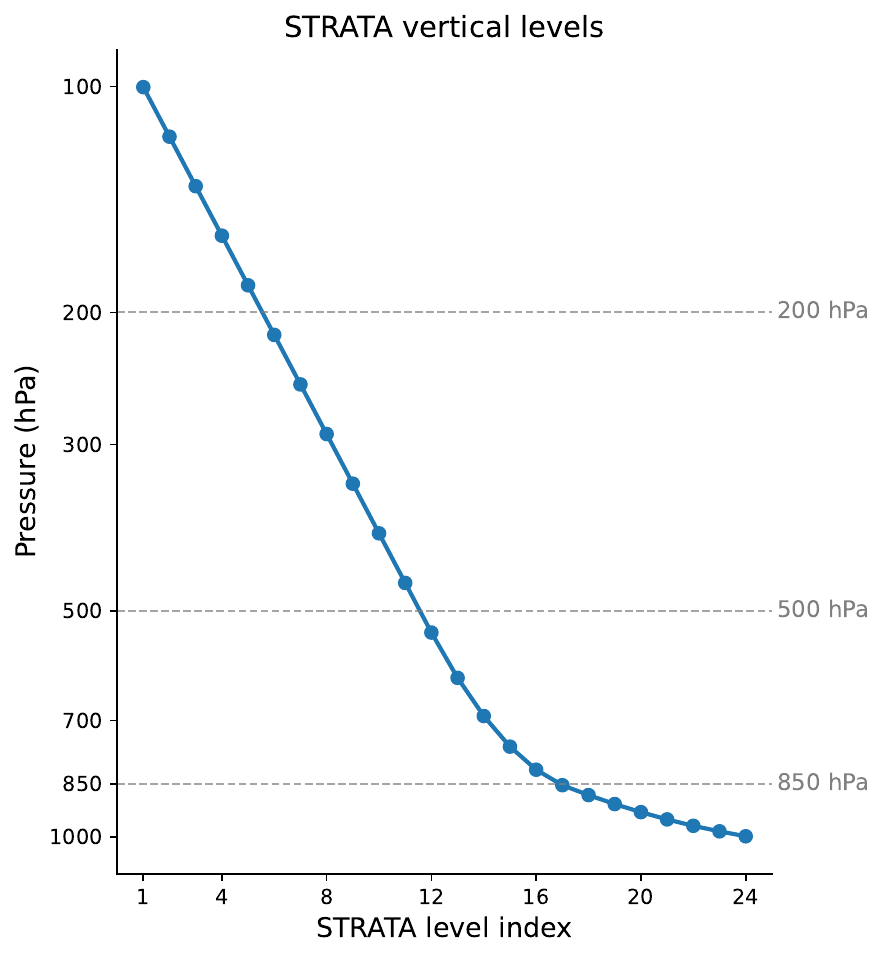}
\caption{Pressure (hPa, in SCREAM's hybrid sigma-pressure coordinates) of the STRATA vertical levels.
The training dataset stores 32 levels (every fourth of SCREAM's 128 native levels), of which STRATA
uses the lowest 24, numbered here from the top (${\sim}100$\,hPa) to the surface (${\sim}1000$\,hPa).
Dashed lines mark standard reference pressure levels.
Elsewhere in the paper, ``model level~24'' indexes the 32 stored levels; it is the level nearest
850\,hPa (${\approx}848$\,hPa).}
\label{fig:vertical_levels}
\end{figure}

\subsection{Training Data}
\label{sec:app_training_data}

We use three SCREAM simulations summarized in Table~\ref{tab:sims}.
The \emph{sdecadal} simulation is initialized on 1994-10-01 and runs for
14\,days.
The \emph{DYAMOND1} and \emph{DYAMOND2} simulations are initialized on
2016-08-01 and 2020-01-20, respectively, each running for 7\,days.
For all three simulations, the first 24\,h are excluded due to model spin-up.
The subsequent 9\,days of sdecadal (1994-10-02 to 1994-10-10) and 4\,days each
of DYAMOND1 (2016-08-02 to 2016-08-05) and DYAMOND2 (2020-01-21 to
2020-01-24) are used for training.
The only exception is the last 2.9\,M of the 37.3\,M pre-training samples of STRATA
(Appendix~\ref{sec:app_strata_protocol}), which were also drawn from the first 24\,h of sdecadal
and DYAMOND2; the decoder was replaced after pre-training, and all subsequent training
(25.8\,M samples) excludes spin-up.

The sdecadal simulation has an inconsistency: while the atmospheric initial
condition is from October 1994, the SSTs were erroneously prescribed from
October 2020 conditions.
We acknowledge this as a dataset error.
Nevertheless, we retain this simulation for training because SCREAM's
underlying physics solver is unchanged and operates consistently regardless
of the IC--SST combination; each time step is physically self-consistent and
represents a valid atmospheric state evolution produced by the same numerical
model used for all other simulations.

\begin{table}[H]
\centering
\caption{SCREAM simulations used in this study.}
\label{tab:sims}
\small
\begin{tabular}{lll}
\toprule
\textbf{Simulation} & \textbf{Period} & \textbf{Training period} \\
\midrule
sdecadal$^\dagger$  & 1994-10-01 -- 1994-10-14 & 1994-10-02 -- 1994-10-10 \\
DYAMOND1            & 2016-08-01 -- 2016-08-07 & 2016-08-02 -- 2016-08-05 \\
DYAMOND2            & 2020-01-20 -- 2020-01-26 & 2020-01-21 -- 2020-01-24 \\
\bottomrule
\multicolumn{3}{l}{$^\dagger$SSTs erroneously prescribed from October 2020 conditions.}
\end{tabular}
\end{table}

\paragraph{Evaluation initializations.}
All rollout evaluations use 6 initialization times from the
held-out portion of the sdecadal simulation (after the training period
ends on 2020-10-10), at 12-hourly intervals:
2020-10-11 12:00, 2020-10-12 00:00, 2020-10-12 12:00,
2020-10-13 00:00, 2020-10-13 12:00, and 2020-10-14 00:00 UTC.
Note that dates are reported as 2020 to reflect the prescribed SST
conditions, though the atmospheric initial conditions are from October 1994.
Unless otherwise noted, single-run snapshot visualizations use the
2020-10-13 00:00 UTC initialization.

\begin{figure}[t]
    \centering
    \includegraphics[width=1\linewidth]{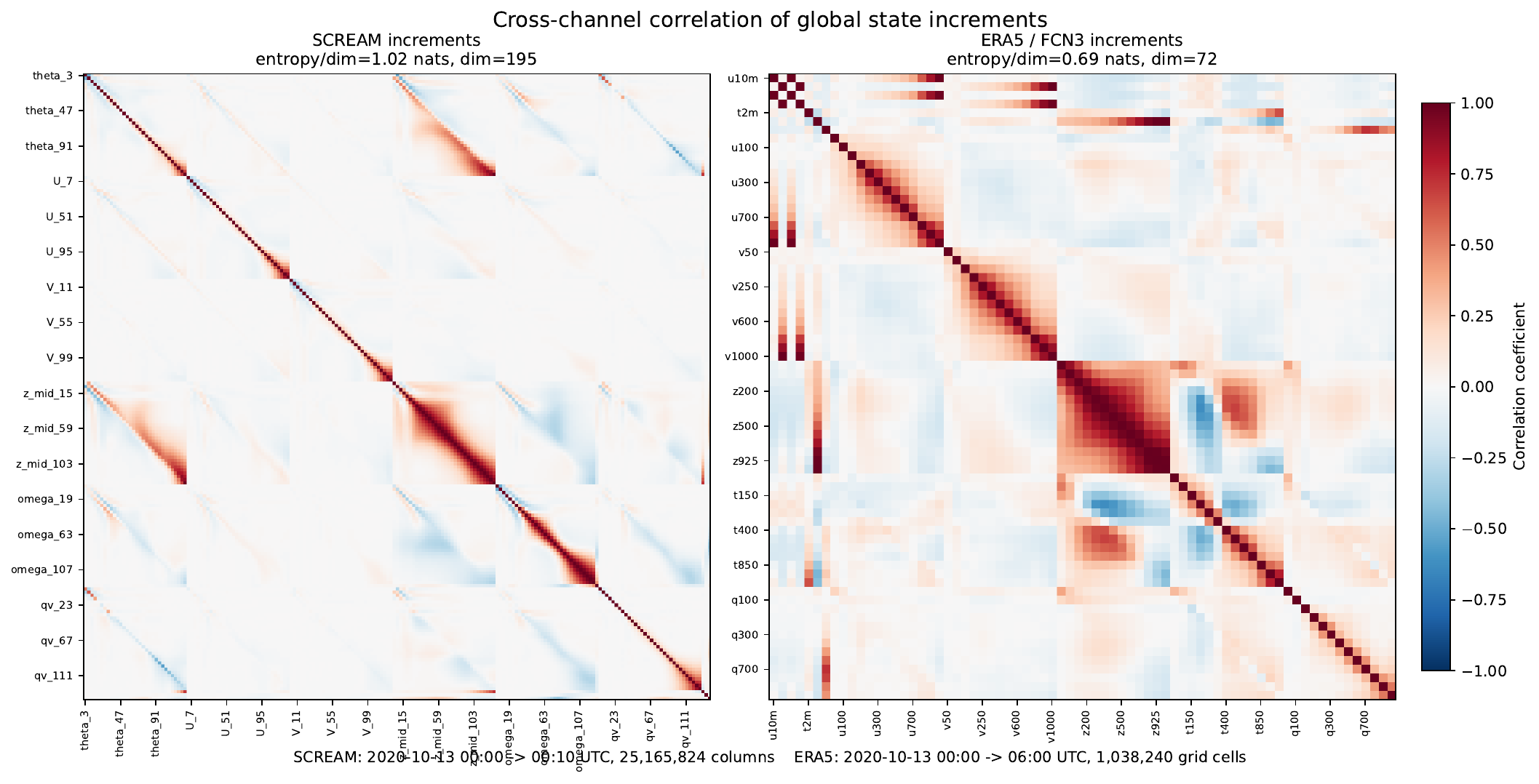}
    \caption{Correlation coefficient matrices of $x(t+\Delta t) - x(t)$ for SCREAM (left) and ERA5 (right).}

        \label{fig:cov}
\end{figure}

\section{Input and Output Variables}
\label{sec:app_variables}

The model ingests six 3D prognostic fields (potential temperature, zonal and meridional wind, geopotential height, vertical pressure velocity, and specific humidity) on all 24 vertical levels, plus 2-m temperature and five forcing fields (solar zenith angle, latitude, surface geopotential, topography, and land fraction).

Most prognostic variables are predicted as tendencies $\Delta x = x_{t+\Delta t} - x_t$; vertical pressure velocity is predicted as absolute state due to lower variance. Diagnostic outputs include precipitation fields and surface pressure.

\begin{table}[H]
\centering
\caption{STRATA input and output variables. ``3D'' variables have 24 vertical
levels; ``2D'' variables are single-level. Prediction mode applies only to
prognostic variables: \emph{tendency} means the model predicts
$\Delta x = x_{t+\Delta t}-x_t$; \emph{state} means the model predicts
$x_{t+\Delta t}$ directly.}
\label{tab:variables}
\small
\begin{tabular}{llllll}
\toprule
\textbf{Variable} & \textbf{Symbol} & \textbf{Units} & \textbf{Dims} & \textbf{Role} & \textbf{Pred.\ mode} \\
\midrule
\multicolumn{6}{l}{\textit{Prognostic (input \& output)}} \\
Potential temperature        & $\theta$             & K                        & 3D & Prognostic & Tendency \\
Zonal wind                   & $U$                  & m\,s$^{-1}$              & 3D & Prognostic & Tendency \\
Meridional wind              & $V$                  & m\,s$^{-1}$              & 3D & Prognostic & Tendency \\
Mid-level geopotential height & $z_\mathrm{mid}$    & m                        & 3D & Prognostic & Tendency \\
Vertical pressure velocity   & $\omega$             & Pa\,s$^{-1}$             & 3D & Prognostic & State    \\
Specific humidity            & $q_v$                & kg\,kg$^{-1}$            & 3D & Prognostic & Tendency \\
2-m temperature              & $T_{2\mathrm{m}}$    & K                        & 2D & Prognostic & Tendency \\
\midrule
\multicolumn{6}{l}{\textit{Forcing (input only)}} \\
Cosine solar zenith angle    & \texttt{coszr}       & --                       & 2D & Forcing    & -- \\
Cosine latitude              & $\cos\phi$           & --                       & 2D & Forcing    & -- \\
Sine latitude                & $\sin\phi$           & --                       & 2D & Forcing    & -- \\
Surface geopotential         & $\phi_s$             & m$^2$\,s$^{-2}$          & 2D & Forcing    & -- \\
Sub-grid std.\ dev.\ of topography & \texttt{sgh30} & m                        & 2D & Forcing    & -- \\
Land fraction                & \texttt{landfrac}    & --                       & 2D & Forcing    & -- \\
\midrule
\multicolumn{6}{l}{\textit{Diagnostic (output only)}} \\
Liquid precip.\ surface mass flux & \texttt{prec\_liq} & kg\,m$^{-2}$\,s$^{-1}$  & 2D & Diagnostic & State \\
Ice precip.\ surface mass flux    & \texttt{prec\_ice} & kg\,m$^{-2}$\,s$^{-1}$  & 2D & Diagnostic & State \\
Surface pressure             & $p_s$                & Pa                       & 2D & Diagnostic & State \\
\bottomrule
\end{tabular}
\end{table}

\section{Training details}
\label{sec:app_training_details}

This appendix gives the training protocol of STRATA (Appendix~\ref{sec:app_strata_protocol}); the protocols of the scaling sweeps and the ablation models are in Appendix~\ref{sec:app_ablation_protocols}.

\subsection{Training protocol of STRATA}
\label{sec:app_strata_protocol}

\paragraph{Tile size.}
All models are pre-trained on $64\times64$ tiles sampled from all
six cubed-sphere faces. The production STRATA model is further fine-tuned on
$128\times128$ tiles (see below).

\paragraph{Normalization.}
Each input and output channel is normalized independently
by subtracting its global mean and dividing by its global standard deviation,
computed over subsampled training data.

\paragraph{Optimization.}
We use AdamW with $\beta_1=0.9$, $\beta_2=0.999$, weight decay $10^{-5}$,
and the Smoothed $\ell_1$ loss (Huber loss with $\delta=1$). 
Training uses a global batch size of 8 or 16
(1 sample per GPU across 8 or 16 H100\,80\,GB GPUs)
with BF16 mixed precision and \texttt{torch.compile}.

\paragraph{Training stages.}
The production STRATA model is first pre-trained with the original PixelDiT decoder on $64\times64$ tiles
for 37.3\,M samples in three stages:
(i)~training of the PixelDiT-M of Table~\ref{tab:ablation-models} from scratch on sdecadal only
(damped cosine schedule from $5\times10^{-4}$, 18.0\,M samples);
(ii)~continued training on all three simulations (damped cosine schedule from $5\times10^{-5}$, 16.4\,M samples),
both with 3 cycles and a decay factor of $0.3162$ per cycle; and
(iii)~continued training at a constant learning rate of $10^{-5}$ for 2.9\,M samples, the only stage
that includes the first 24\,h (spin-up) of sdecadal and DYAMOND2 (Appendix~\ref{sec:app_training_data}).
The linear-projection upsampling of the decoder is then replaced with bilinear/convolutional upsampling
(Figure~\ref{fig:dit-schematic}), and the model is fine-tuned in three stages:
(1)~pixel-block warm-up with the DiT layers frozen, on sdecadal only (damped cosine schedule from
$5\times10^{-4}$, 5.2\,M samples);
(2)~joint fine-tuning of all layers (damped cosine schedule from $5\times10^{-5}$, 2 cycles,
decay factor $0.3162$, 13.3\,M samples), followed by 5.7\,M samples at a constant learning rate of $10^{-5}$;
(3)~large-tile fine-tuning on $128\times128$ tiles at a constant learning rate of $10^{-5}$, 1.6\,M samples.
This recipe reflects how the project developed: the de-aliasing decoder was introduced after
pre-training, and we kept the resulting model as the reference for all analyses in this paper.
In hindsight, it is not necessary: the ablation models of Table~\ref{tab:ablation-models},
including one with the same backbone and decoder as STRATA, are trained from scratch, without
pre-training or a change of decoder (Appendix~\ref{sec:app_ablation_protocols}), and we expect
that the production model can be trained in the same way.

\paragraph{Multi-step fine-tuning.}
After convergence of the one-step model,
we fine-tune autoregressively to improve multi-step rollout stability.
For an $N$-step fine-tuning pass,
the model is rolled out for $N-1$ steps under \texttt{torch.no\_grad()}
and a Smoothed $\ell_1$ loss is computed only on the final prediction;
forcing fields are provided at every rollout step.
We first fine-tune for 1{,}800 gradient steps with $N=2$,
then for a further 2{,}000 steps with $N=4$,
both at a constant learning rate of $10^{-5}$.
Multi-step fine-tuning suppresses ripple-style artifacts
that accumulate in one-step-trained rollouts, and it improves both precipitation and RMSE.
Table~\ref{tab:abl-multistep} compares the model before multi-step fine-tuning (1-step), after the $N=2$ stage (2-step) and after the $N=4$ stage (4-step, STRATA), in 24-hour rollouts from 2020-10-13 00:00\,UTC.
With 1-step training, precipitation runs away by +24\,h; the $N=2$ stage more than halves the excess and the $N=4$ stage removes it.
The 2-step model has the smallest precipitation bias at +6 and +12\,h, but this does not carry into RMSE: the 4-step model has the lowest RMSE in five of the six cells, the exception being $U$ at +6\,h, and its margin over 1-step grows with lead time to 15\% for $U$ and 17\% for $\theta$ at +24\,h.

\begin{table}[h]
\centering
\caption{Multi-step fine-tuning: global-mean liquid precipitation bias relative to SCREAM, and RMSE of $U$ and $\theta$ at model level 24 (${\approx}848$\,hPa), in 24-hour rollouts from 2020-10-13 00:00\,UTC. The 4-step model is STRATA.}
\label{tab:abl-multistep}
\small
\begin{tabular}{llccc}
\toprule
 & Fine-tuning & +6\,h & +12\,h & +24\,h \\
\midrule
\multirow{3}{*}{Precip.\ bias (\%)} & 1-step & $-12.7$ & $-7.3$ & $+85.2$ \\
                                    & 2-step & $-2.6$  & $-3.1$ & $+35.6$ \\
                                    & 4-step & $-19.8$ & $-17.1$ & $+2.0$ \\
\midrule
\multirow{3}{*}{$U$ RMSE (m\,s$^{-1}$)} & 1-step & 1.63 & 3.38 & 7.27 \\
                                        & 2-step & 1.58 & 3.20 & 6.77 \\
                                        & 4-step & 1.59 & 3.13 & 6.15 \\
\midrule
\multirow{3}{*}{$\theta$ RMSE (K)} & 1-step & 0.666 & 1.176 & 2.447 \\
                                   & 2-step & 0.632 & 1.116 & 2.200 \\
                                   & 4-step & 0.626 & 1.057 & 2.022 \\
\bottomrule
\end{tabular}
\end{table}

\subsection{Training protocols of the ablation models}
\label{sec:app_ablation_protocols}

Tables~\ref{tab:sweep-protocols} and~\ref{tab:ablation-models} list how every model other than STRATA was trained.
All use AdamW ($\beta_1=0.9$, $\beta_2=0.999$, weight decay $10^{-5}$), the Smoothed $\ell_1$ loss, BF16 mixed precision and, except in the batch-size sweep, one sample per GPU, so the global batch equals the number of GPUs; all are trained from scratch.
The scaling sweeps (Table~\ref{tab:sweep-protocols}) are compared only on one-step test loss, on the final day of the sdecadal simulation, which is held out from training.
The models in Table~\ref{tab:ablation-models} are evaluated in 24-hour rollouts, so the table also lists their multi-step fine-tuning and inference setup.

\begin{table}[!htbp]
\centering
\caption{Training protocols of the scaling sweeps. All models are DiT backbones without a pixel decoder (NA3D kernel 3, StereoRoPE with wind rotation, gated attention), trained from scratch on sdecadal with $64\times64$ tiles and one-step training only, and compared on test loss.
\emph{Batch:} global batch; 8 GPUs throughout.
\emph{Learning rate:} (a) constant $10^{-4}$, no warm-up; (b) as (a), then, for DiT-S-ps4, DiT-M-ps4 and DiT-L-ps2 only, a damped cosine schedule; (c) $6.25\times10^{-5}\times$ batch, 20{,}000-step warm-up, then constant.}
\label{tab:sweep-protocols}
\small
\begin{tabular}{@{}lrrrcl@{}}
\toprule
Sweep & Dim & Layers & Patch & Batch & Learning rate \\
\midrule
isoFLOP (\S\ref{sec:entropy}, B.2) & \multicolumn{3}{c}{9 models, Table~\ref{tab:sweep}} & 8 & (a), (b) \\
\midrule
\multirow{6}{*}{Width vs depth (K.1)} & 2048 & 8 & 4 & 8 & (a) \\
 & 1536 & 14 & 4 & 8 & (a) \\
 & 1024 & 32 & 4 & 8 & (a) \\
 & 256 & 8 & 8 & 8 & (a) \\
 & 192 & 14 & 8 & 8 & (a) \\
 & 128 & 32 & 8 & 8 & (a) \\
\midrule
Batch size (K.3) & 768 & 14 & 4 & 8--256 & (c) \\
\bottomrule
\end{tabular}
\end{table}

\begin{table}[!htbp]
\centering
\caption{Training protocols of the models used in the ablations, with STRATA for reference (its full recipe is in Appendix~\ref{sec:app_strata_protocol}). All models use NA3D kernel 9 and patch $4\times4\times1$, and all are trained from scratch and then multi-step fine-tuned with $N=2$ and $N=4$. All rotate winds into the tile frame, except the no-wind-rotation and axial-RoPE models. All are rolled out with the $\omega$ filter and the negative-$q_v$ fixer, on $128\times128$ tiles with a 16-cell halo if trained on $128\times128$ tiles, and on $32\times32$ tiles with a 16-cell halo otherwise.
Backbone gives the embedding dimension and number of layers; PixelDiT-S also has a smaller decoder (64 dimensions, 2 layers).
Batch is the global batch of the base training stage; later stages may differ.
Schedules: (a) damped cosine from $5\times10^{-4}$; (b) constant $10^{-4}$, then damped cosine, then constant $10^{-5}$ on $128\times128$ tiles; (c) as (b) without the $128\times128$ stage.
All ablation models except LOSO omit the precipitation ReLU; PixelDiT-S and -M also omit the $q_v$ softplus and gradient clipping; and LOSO uses its final weights rather than the best-validation checkpoint.}
\label{tab:ablation-models}
\scriptsize
\setlength{\tabcolsep}{3.5pt}
\begin{tabular}{@{}lclllcccr@{}}
\toprule
Model & Backbone & Decoder & Position enc. & Training data & Tile & Batch & Sched. & Samples \\
\midrule
STRATA (reference) & 1024 / 24 & de-aliasing & StereoRoPE & all three & 64$\to$128 & 8--16 & App.~\ref{sec:app_strata_protocol} & --- \\
\midrule
\multicolumn{9}{@{}l}{\emph{Model size (Appendix~\ref{sec:model-capacity-ablation})}} \\
PixelDiT-S & 768 / 12 & PixelDiT & StereoRoPE & sdecadal & 64 & 16 & (a) & 32.9\,M \\
PixelDiT-M & 1024 / 24 & PixelDiT & StereoRoPE & sdecadal & 64 & 16 & (a) & 18.0\,M \\
\midrule
\multicolumn{9}{@{}l}{\emph{Decoder (Appendix~\ref{sec:app_patch_dealiasing}) and grid transfer (Section~\ref{sec:grid_invariance})}} \\
Plain decoder & 1024 / 24 & PixelDiT & StereoRoPE$^\dagger$ & all three & 64$\to$128 & 8 & (b) & 13.5\,M \\
De-aliasing decoder$^\ddagger$ & 1024 / 24 & de-aliasing & StereoRoPE$^\dagger$ & all three & 64$\to$128 & 8 & (b) & 17.8\,M \\
No wind rotation & 1024 / 24 & de-aliasing & StereoRoPE$^\dagger$ & all three & 64$\to$128 & 8 & (b) & 17.9\,M \\
Axial RoPE & 1024 / 24 & de-aliasing & axial RoPE$^\dagger$ & all three & 64$\to$128 & 8 & (b) & 17.8\,M \\
\midrule
\multicolumn{9}{@{}l}{\emph{Unseen season (Appendix~\ref{sec:app_seasons})}} \\
LOSO & 1024 / 24 & de-aliasing & StereoRoPE$^\dagger$ & sdecadal, DYAMOND1 & 64 & 16 & (c) & 15.5\,M \\
\bottomrule
\multicolumn{9}{@{}l}{$^\dagger$In both the semantic and the pixel blocks; STRATA and PixelDiT-S/M use it in the semantic blocks only.} \\
\multicolumn{9}{@{}l}{$^\ddagger$The same model serves as the StereoRoPE + wind rotation model of Section~\ref{sec:grid_invariance}.}
\end{tabular}
\end{table}

\section{Inference}
\label{sec:app_inference_details}

\subsection{Tile extraction and DUO-style boundary padding}

SCREAM's cubed-sphere grid is defined by equiangular meshes
on each cube face.
To construct halo cells beyond a face boundary,
we extend the face's equiangular mesh on its own plane
past the boundary—preserving the same angular spacing
as the interior—and then project these extended planar coordinates
back onto the sphere.
The state values at these extended grid points are interpolated
from the 4 nearest neighbors on the adjacent face,
weighted by inverse distance.
The resulting halo is a seamless continuation
of the face's own grid pattern.

\paragraph{Tile stitching}
Unless stated otherwise, each face is divided into $128\times128$ tiles, each padded with a 16-pixel halo, so the model sees $160\times160$ inputs and adjacent padded tiles share a 32-pixel border region.
Each tile's output is weighted by a 2D Kaiser--Bessel-derived window, the outer product of two 1D windows of length 160 with shape parameter $\beta = 10\pi$,
which tapers smoothly to zero at the edges,
and the final value at each grid cell is the weighted average
of all overlapping tile outputs.
If the window is instead a 2D top-hat that is 1 in the interior and 0 in the halo region, the stitching operation reduces to simply a crop.

To see how much the window matters, we compare it with three simpler ways of combining neighboring tiles: no overlap at all (halo 0), the top-hat window, and uniform weights over the overlap (Table~\ref{tab:abl-blending}). A seam can show up as a step, where the field jumps at a tile boundary, or as a kink, where its slope changes abruptly, so we measure the mean first and second derivatives across tile boundaries, over every tile period on all six faces, relative to SCREAM's own values. Without overlap, the seams are two orders of magnitude stronger than in SCREAM. The top-hat window and uniform weights reduce them but still leave visible seams, because in both the weight drops abruptly at the edge of the overlap instead of tapering to zero. The Kaiser--Bessel-derived window brings both close to SCREAM's level, with a small residual kink.

\begin{table}[h]
\centering
\caption{Tile-seam step (first derivative) and kink (second derivative) in $T_{2\mathrm{m}}$ at +24\,h for four ways of combining neighboring tiles, as multiples of SCREAM's own value (1.0 = no seam), in rollouts from 2020-10-13 00:00\,UTC.}
\label{tab:abl-blending}
\small
\begin{tabular}{lcc}
\toprule
Combination of neighboring tiles & Step & Kink \\
\midrule
No overlap (halo 0)                        & 138  & 185  \\
Top-hat window (halo as context, cropped)  & 2.7  & 4.6  \\
Uniform weights over the overlap           & 6.2  & 10.7 \\
Kaiser--Bessel-derived window (STRATA)     & 1.1  & 1.8  \\
\bottomrule
\end{tabular}
\end{table}

\paragraph{Linear operator interpretation}

For implementation, it is helpful to express padding and stitching as sparse linear operators $P$ and $S$, respectively.
Let $\mathbf{x}$ be the un-padded input and $\mathbf{y}=P\mathbf{x}$ be the padded output.
Concretely, the rows of $P$ expressed in matrix form have 4 non-zero elements corresponding to the weights of the 4 nearest neighbors.
The stitching $S$ is an approximate inverse of $P$ expressed as
\begin{equation}
   S=D^{-1} P^T W 
   \label{eq:stitch}
\end{equation}
where $W$ is diagonal matrix consisting of the weights of the window and $D=P^T W \mathbf{1}$ is the sum of the incoming weights; $\mathbf{1}=[1,\ldots,1]^T$.
Interpreted in terms of message passing,  $P$ sends a unique message from each unpadded grid point and $S$ takes a weighted average of the return communications.

Eq.~\ref{eq:stitch} is helpful for implementing the stitching operation because $P^T (W \mathbf{y})$ is just the vector-Jacobian product of the forward operator $P$ applied on a $W \mathbf{y}$, and can in principle be computed using the automatic differentiation feature of modern deep learning frameworks.

\paragraph{Distributed Implementation}
The tile based inference approach also enables scalability to a distributed environment. Each rank processes a subset of tiles and model compute is independent across tiles. However, the DUO-style boundary padding outlined above needs to be consistent across processes.
We can do this by performing a communication so every process has all the data needed for the DUO-style halo cell computation to populate the halos for all local tiles.
This communication needs to be performed once per time-step to keep tiles in sync and ensure seamless continuation of fields across tile boundaries.
In a simple implementation, the padding operator $P \mathbf{x}$ can be applied using an all-gather followed by interpolation to the process-owned tiles using a sliced version of $P$.
In block matrix form, this is given by 
\begin{equation}
    P=diag([P_1, \ldots, P_n]) G,
\end{equation}
where $P_j$ is the slice of $P$ owned by process $j$ and $G$ is an all-gather.
This is simple, but most of the data communicated is unused, so it is more efficient to communicate only the tile boundary data needed for the 4 nearest neighbor interpolation. This reduces the all-gather to a local peer-to-peer communication pattern leveraging locality of the halo cell neighbor interpolation.

To implement the stitching operator $S$, the all-gather framing is helpful in light of Eq. \ref{eq:stitch}.
The only expression in that equation requiring communication is $P^T$.
In this form, 
\[P^T=G^T diag([P_1^T, \ldots, P_n^T])\]
Since the second part of this expression is block-diagonal it can be applied locally on each process independently. Then, the transpose of the all-gather $G^T \mathbf{v}$ can be implemented as a distributed reduce-scatter operation.

\begin{figure}[H]
  \centering
  \includegraphics[width=0.95\linewidth]{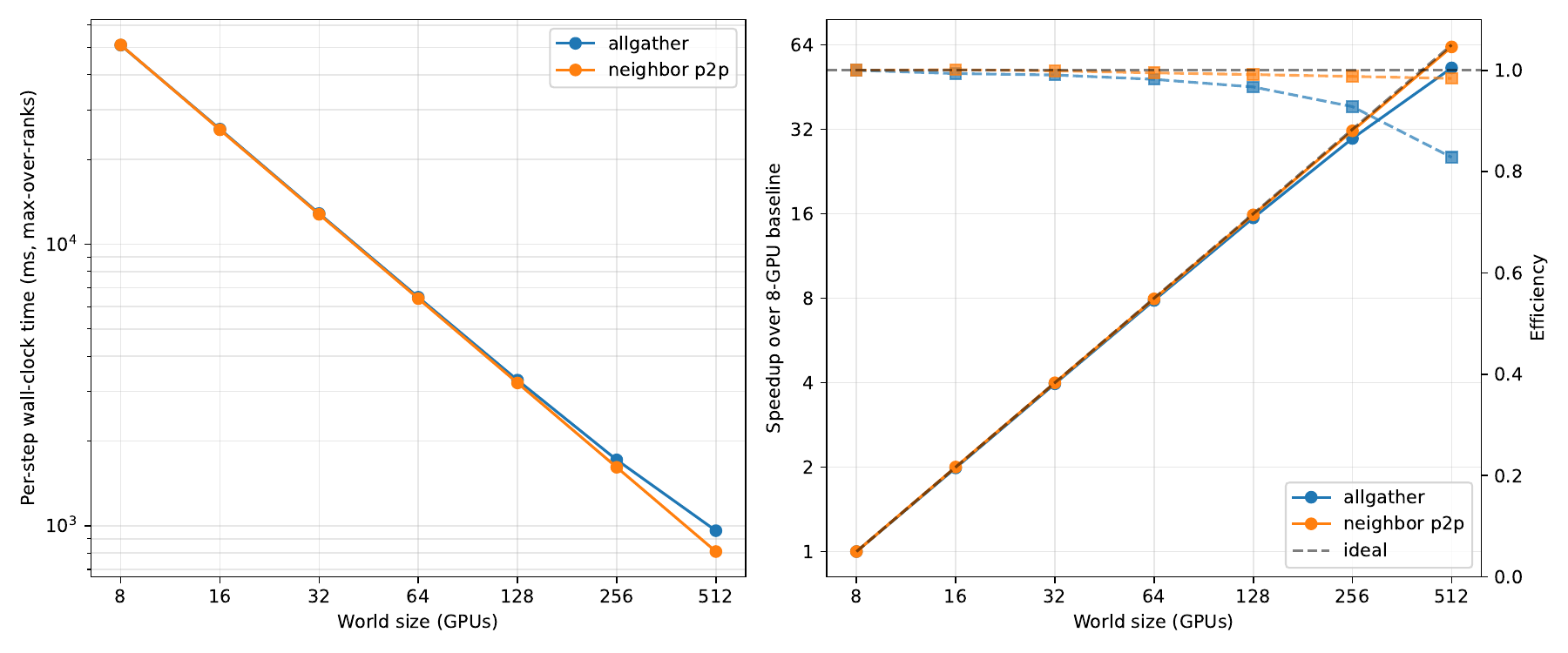}
  \caption{%
  \textbf{Scalability of tile based distributed inference.}
  Left: Per step wall-clock time. Right: Speedup over 8 GPU baseline in solid lines on the left y-axis and strong scaling efficiency in dashed lines on the right y-axis. Two DUO-style boundary padding implementations are shown: allgather (blue) and neighbor peer-to-peer (orange).}
  \label{fig:scaling}
\end{figure}

\paragraph{Scaling} Figure~\ref{fig:scaling} shows the scalability of this distributed inference through a strong scaling test from 8 to 512 GPUs using a 2D top-hat window function. Both the all-gather and neighbor peer-to-peer algorithms have very good scalability with the all-gather algorithm yielding 82.8\% strong scaling efficiency at 512 GPUs and the neighbor peer-to-peer algorithm yields 98.4\% efficiency at the same scale.

\subsection{Spectral filter to control large-scale vertical velocity}
\label{sec:app_omega_filter}
\begin{figure}[H]
  \centering
  \includegraphics[width=0.5\linewidth]{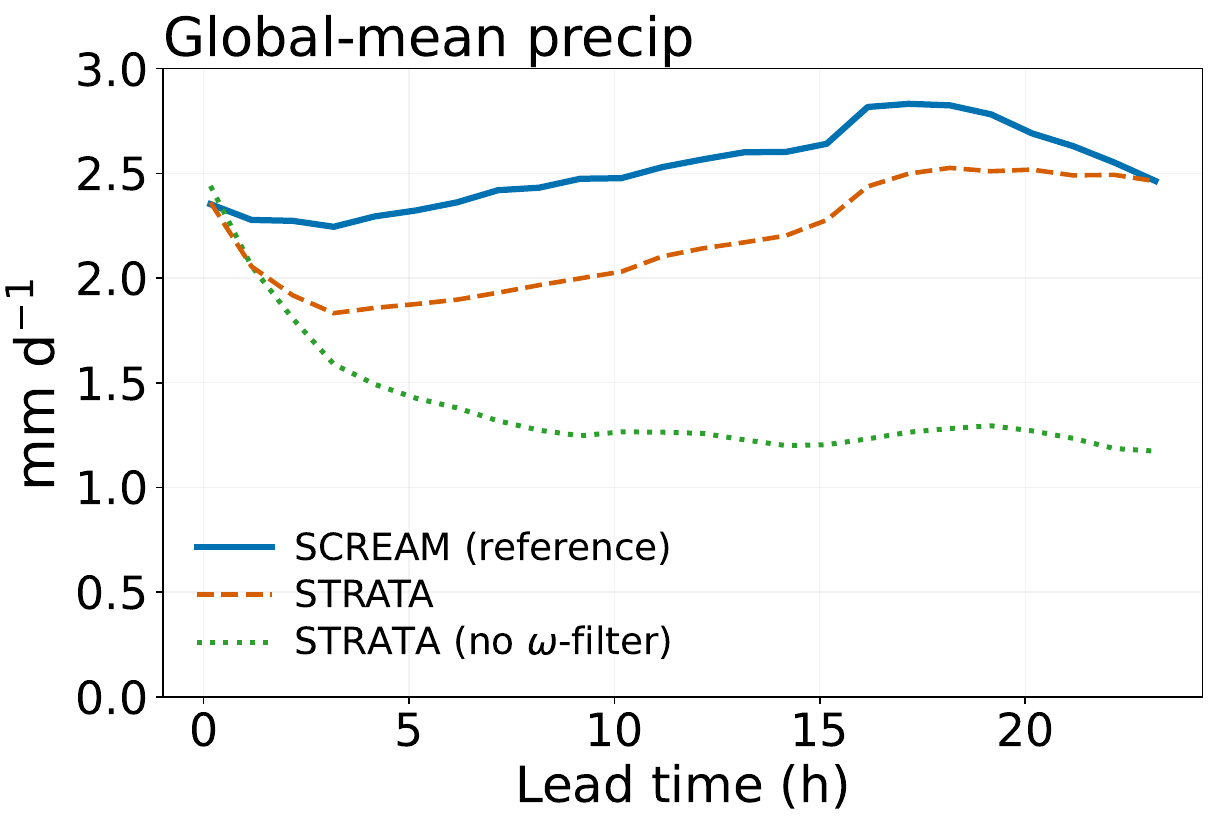}
  \caption{%
  \textbf{Effect of the large-scale $\omega$ filter on global-mean precipitation.}
  Global-mean liquid precipitation rate over a 24-hour rollout for
  SCREAM (reference, solid blue), STRATA with the $\omega$ filter
  (dashed orange), and STRATA without the filter (dotted green).
  Without the filter, unconstrained drift in the leading spherical-harmonic
  modes of $\omega$ causes global-mean precipitation to collapse to
  ${\sim}1.0$\,mm\,d$^{-1}$ by 24\,h---roughly half the reference value.
  The filter substantially reduces this dry drift while leaving the
  fine-scale precipitation structure unchanged.}
  \label{fig:omega-filter-ablation}
\end{figure}

Tile-based training is blind to global constraints: because tiles are processed independently, the model has no mechanism to keep the global-mean vertical pressure velocity $\omega$ near zero, as mass continuity requires, and the largest spherical-harmonic scales of $\omega$ drift during the rollout.
After each rollout step we compute the real spherical harmonic transform of $\omega$ at each vertical level, zero the degrees $\ell = 0, 1, 2$ (nine modes), and invert the transform. Without the filter, global-mean precipitation falls to about half of SCREAM's by 24\,h (Figure~\ref{fig:omega-filter-ablation}).

Removing $\ell = 0$ enforces a physical constraint at inference: it projects $\omega$ onto zero global mean, which SCREAM satisfies closely (at 850\,hPa its global mean is nearly 400 times smaller than its anomaly RMS). Removing $\ell = 1, 2$ is an empirical choice. At 850\,hPa these degrees hold only 0.7\% of $\omega$'s anomaly, compared with 77\% for $q_v$ and 91\% for $\theta$, so removing them there changes little; this is also why we filter only $\omega$. Closer to the surface, however, topography and land--sea contrast lock real planetary-scale vertical motion into these degrees. Their share of $\omega$'s anomaly is 2.6\% at 935\,hPa, 7.1\% at 970\,hPa and over 10\% at the lowest model level, so there the filter removes real structure. We therefore tested a restricted filter that removes only $\ell = 0$ on the six lowest model levels (below about 880\,hPa at global-mean surface pressure) and $\ell = 0, 1, 2$ above. In a 24-hour rollout from 2020-10-13 00:00\,UTC it runs stably and leaves the near-surface $\ell = 1, 2$ structure in place, and its global-mean precipitation stays within 1 percentage point of the full filter at 6-hourly leads and within 2.3 percentage points at every hourly lead. Weakening the filter near the surface thus does not bring back the dry drift. All other results in the paper use the full filter.

\subsection{Negative \texorpdfstring{$q_v$}{qv} and precipitation fixer.}
\label{sec:app_qvfix}
Without explicit constraints the model can predict physically inadmissible
negative specific humidity $q_v$ or precipitation.
We apply differentiable constraints in the forward pass,
active during both fine-tuning and rollout,
so the model learns outputs consistent with the constraints.
For $q_v$, let $r = \Delta q_v / \max(q_v^t,\,\varepsilon)$
with $\varepsilon=10^{-14}$\,kg\,kg$^{-1}$.
We constrain $r$ via $\tilde{r} = \mathrm{softplus}(r+1;\,\beta=10) - 1$
and set $q_v^{t+1} = (1+\tilde{r})\,\max(q_v^t,\,\varepsilon) \geq 0$.
During rollout only, an additional hard floor of $10^{-15}$\,kg\,kg$^{-1}$
eliminates any residual numerical negatives.
For surface precipitation, the predicted value is denormalized,
passed through ReLU, and renormalized.

\section{Throughput and Energy Accounting}
\label{sec:app_throughput_energy}

We report inference throughput in simulated days per wall-clock day (SDPD) and
energy efficiency in simulated days per megawatt-hour
(\(\mathrm{SD\,MWh^{-1}}\)). For STRATA, one simulated day contains
\(N_\mathrm{step}=24\,\mathrm{h}/10\,\mathrm{min}=144\) autoregressive steps.
If \(t_\mathrm{step}\) is the measured wall-clock latency of one global
autoregressive step, then
\begin{equation}
    \mathrm{SDPD}
    = \frac{86400}{N_\mathrm{step} t_\mathrm{step}} .
\end{equation}
Given an average power draw \(P\) in MW for the hardware included in the
accounting boundary, the corresponding energy-normalized throughput is
\begin{equation}
    \mathrm{SD\,MWh^{-1}}
    = \frac{\mathrm{SDPD}}{24P}.
\end{equation}
Equivalently, if \(t_\mathrm{SD}=N_\mathrm{step}t_\mathrm{step}\) is the
wall-clock seconds required for one simulated day, then
\(\mathrm{SD\,MWh^{-1}}=1/(P t_\mathrm{SD}/3600)\).
For SCREAM, we use the same conversion from its reported SDPD and average
system power, and compute speedups as the ratio of
\(\mathrm{SD\,MWh^{-1}}\) between STRATA and SCREAM. Under the assumptions
in Table~\ref{tab:throughput_energy_accounting}, the production STRATA configuration achieves
\(48.1\,\mathrm{SD\,MWh^{-1}}\), a ${>}50\times$ energy-efficiency
improvement over the SCREAM Frontier run
($0.78\,\mathrm{SD\,MWh^{-1}}$).
Using an emissions factor of \(367\,\mathrm{kg\,CO_2\,MWh^{-1}}\) (\url{https://www.eia.gov/tools/faqs/faq.php?id=74&t=11}), this
corresponds to \(131.1\) simulated days per ton of CO$_2$
vs.\ \(2.12\) for SCREAM.

\begin{table}[H]
\centering
\caption{Throughput and energy accounting assumptions for the SCREAM and
production STRATA comparison.
SCREAM SDPD from \citet{taylorSimpleCloudResolvingE3SM2023};
Frontier system power (24{,}607\,kW) from the TOP500 November 2025
list (\url{https://top500.org/lists/top500/2025/11/}).
Both systems are assumed to draw their full rated power; in practice
utilization is lower, so the reported SD\,MWh$^{-1}$ values are
conservative lower bounds.}
\label{tab:throughput_energy_accounting}
\small
\begin{tabular}{lrrrrr}
\toprule
System & Hardware & Power (kW) & SDPD & \(\mathrm{SD\,MWh^{-1}}\) & SD/tCO$_2$ \\
\midrule
STRATA & 2 DGX H100 & 20.4 & 23.6 & 48.1 & 131.1 \\
STRATA & 64 DGX H100 & 652.8 & 741 & 47.3 & 128.9 \\
SCREAM     & Frontier     & 24607 & 460    & 0.78  & 2.12 \\
\bottomrule
\end{tabular}
\end{table}

\paragraph{Including data generation and training.}
The figures above cover inference only. Building STRATA also had two one-off costs: generating the 28 simulated days of SCREAM output used for training and evaluation, at SCREAM's own $0.78\,\mathrm{SD\,MWh^{-1}}$ (35.9\,MWh), and training the model, 38 hours on 16 H100 GPUs at the 20.4\,kW of Table~\ref{tab:throughput_energy_accounting} (0.78\,MWh, about 2\% of the total). The training cost is that of one from-scratch run with the production configuration. Amortizing the resulting 36.7\,MWh over an application that needs $N$ simulated days gives an effective efficiency of $N/(36.7 + N/48.1)\,\mathrm{SD\,MWh^{-1}}$ (Table~\ref{tab:energy_amortization}). STRATA breaks even with SCREAM at about 29 simulated days, and its advantage grows with $N$ toward the inference-only value. A 100-member ensemble of 30-day forecasts (3{,}000 simulated days), for example, is about $39\times$ more efficient than SCREAM.

\begin{table}[H]
\centering
\caption{Effective energy efficiency of STRATA including the one-off cost of data generation and training (36.7\,MWh), as a function of the number of simulated days $N$ an application needs.}
\label{tab:energy_amortization}
\small
\begin{tabular}{lcccccc}
\toprule
$N$ (simulated days) & 29 & 100 & 365 & 1{,}000 & 3{,}650 & $\infty$ \\
\midrule
Effective $\mathrm{SD\,MWh^{-1}}$ & 0.78 & 2.6 & 8.2 & 17.4 & 32.4 & 48.1 \\
Relative to SCREAM & $1.0\times$ & $3.3\times$ & $10.6\times$ & $22.3\times$ & $41.6\times$ & $61.7\times$ \\
\bottomrule
\end{tabular}
\end{table}

\section{Rollout Metrics}
\label{sec:app_rollout_metrics}

\begin{figure}[t]
  \centering
  \includegraphics[width=\linewidth]{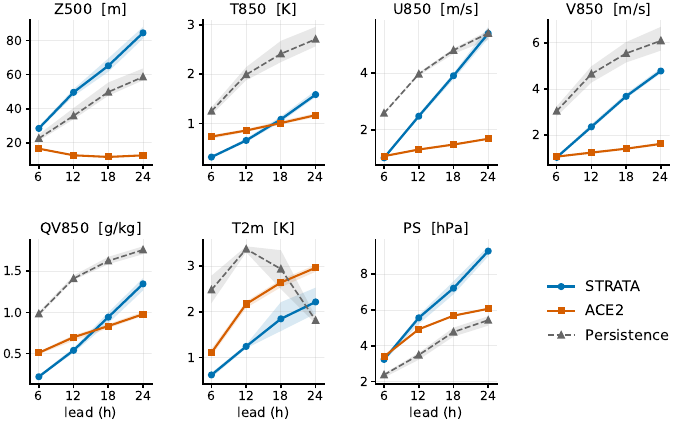}
  \caption{RMSE of STRATA, ACE2 and persistence against SCREAM coarsened to ACE2's 1$^\circ$ grid, at 6, 12, 18 and 24\,h lead, averaged over the six October initialization times (shading: min--max). Fields are on true isobaric surfaces or the surface; $T_{850}$ is absolute temperature, and grid points where the surface lies above 850\,hPa are masked. ACE2 has no vapour-only humidity, so its $q_{850}$ uses total water as a proxy. Persistence is more accurate than STRATA for surface pressure at every lead and for $Z_{500}$, which vary slowly at large scales where STRATA drifts (Section~\ref{sec:rollout}). Persistence error in $T_{2\mathrm{m}}$ falls from 12 to 24\,h because the initializations are at fixed local times, so the 24\,h forecast returns to the initial diurnal phase.}
  \label{fig:scorecard}
\end{figure}

\paragraph{Choice of metrics and reference curves.}
We score continuous fields by area-weighted RMSE, the standard for global data-driven emulators~\citep{raspWeatherBench2Benchmark2024a}, and precipitation by the FSS, the standard metric at convective scales. We do not report anomaly correlation: it is closely related to RMSE and requires a SCREAM climatology, which we do not have. ACE2 is not scored on precipitation because it outputs 6-hourly averages rather than 10-minute instantaneous rates, and we add no further baselines because no other global storm-resolving AI emulator exists. Figure~\ref{fig:scorecard} compares STRATA, ACE2 and persistence for seven fields at 6--24\,h lead.

\paragraph{Precipitation amount distribution.}
We follow the amount-distribution diagnostic of \citet{pendergrass2014two} as described by 
\citet{koopermanRobustEffectsCloud2016}, adapted from daily accumulated
rainfall to instantaneous precipitation rates.
For a sample set \(S=\{(k,t)\}\) over grid cells \(k\) and frames \(t\), let
\(a_k\) be cell area and \(r_{kt}\) be precipitation rate in mm\,day\(^{-1}\).
For logarithmic rain-rate bin
\(B_i=\{(k,t):R_i^\ell \le r_{kt} < R_i^r\}\) with
\(\Delta\ln R=\ln(R_i^r/R_i^\ell)\), we compute
\begin{equation}
  p_i =
  \frac{1}{N_T\,\Delta\ln R}
  \sum_{(k,t)\in B_i} a_k r_{kt},
  \qquad
  N_T=\sum_{(k,t)\in S} a_k .
\end{equation}
Dry samples are included in \(N_T\) but not assigned to any rainy bin.
Thus \(\sum_i p_i\Delta\ln R\) equals the area-weighted global-mean
precipitation rate contributed by the plotted rain-rate range.

\paragraph{Fractions Skill Score (FSS).}
We follow standard practice to reward realistic spatial structure of organized rainfall systems without 
penalizing misplacement of individual convective cells by using the Fractions Skill Score
\citep{robertsScaleSelectiveVerificationRainfall2008}.
Let $\hat{r}(\mathbf{x},t)$ and $r(\mathbf{x},t)$ denote the model and
SCREAM reference precipitation rates (mm\,day$^{-1}$) at grid cell
$\mathbf{x}$ and lead time $t$.
For a threshold $T$ (mm\,day$^{-1}$), we form binary exceedance
indicators
\begin{equation}
  I_{\hat{r}}(\mathbf{x}) = \mathbf{1}[\hat{r}(\mathbf{x}) > T],
  \qquad
  I_r(\mathbf{x}) = \mathbf{1}[r(\mathbf{x}) > T].
\end{equation}
Each indicator field is then averaged over a square $n\times n$ window
of neighboring grid cells to obtain local exceedance fractions:
\begin{equation}
  F_{\hat{r}}(\mathbf{x})
    = \frac{1}{n^2}\sum_{\mathbf{x}'\in\mathcal{N}_n(\mathbf{x})}
      I_{\hat{r}}(\mathbf{x}'),
  \qquad
  F_r(\mathbf{x})
    = \frac{1}{n^2}\sum_{\mathbf{x}'\in\mathcal{N}_n(\mathbf{x})}
      I_r(\mathbf{x}'),
\end{equation}
where $\mathcal{N}_n(\mathbf{x})$ is the $n\times n$ neighborhood
centered on $\mathbf{x}$.
The FSS is
\begin{equation}
\mathrm{FSS} = 1 - \frac{\displaystyle\sum_{\mathbf{x}}
  (F_{\hat{r}}(\mathbf{x}) - F_r(\mathbf{x}))^2}
  {\displaystyle\sum_{\mathbf{x}}
  (F_{\hat{r}}(\mathbf{x})^2 + F_r(\mathbf{x})^2)},
\end{equation}
ranging from 0 (no overlap) to 1 (perfect agreement).
The pooling is applied independently per cubed-sphere face with reflect
padding (edge values mirrored outward) and does not cross face boundaries.
To obtain a single global score, numerator and denominator are accumulated
across all six faces before forming the ratio.
We report FSS against a persistence baseline (SCREAM truth held constant
at the initial time $t_0$), with the same binarization, pooling, and
aggregation applied to both.

\begin{figure}[H]
  \centering
  \includegraphics[width=0.99\textwidth]{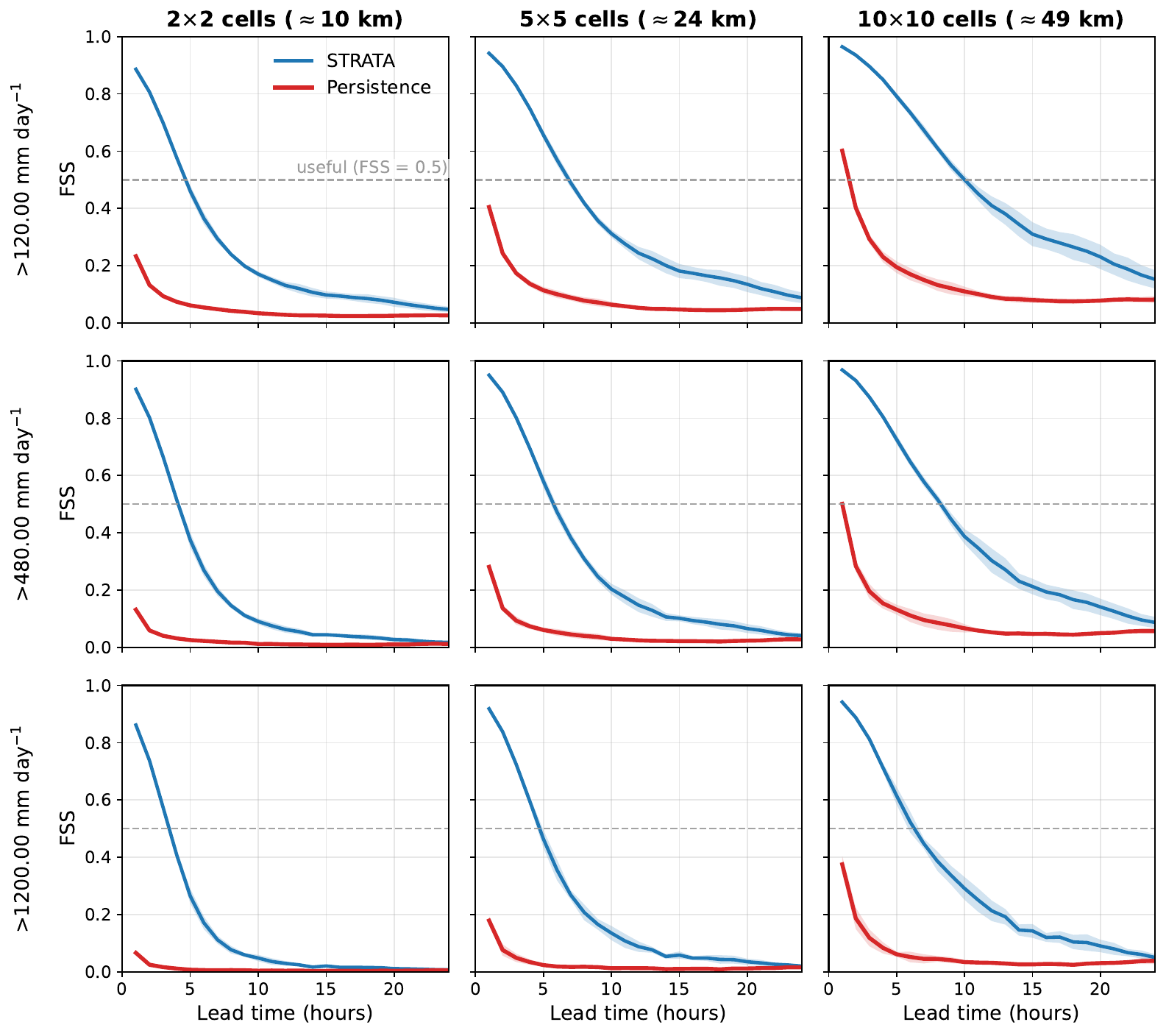}
  \caption{%
  \textbf{FSS sensitivity to precipitation threshold and neighborhood size.}
  Each panel shows Fractions Skill Score (FSS) versus lead time for STRATA
  (blue) and persistence (red), with shading indicating the spread across
  initializations. Rows use fixed precipitation thresholds of 120, 480, and
  1200\,mm\,day$^{-1}$, corresponding to increasingly intense rainfall.
  Columns use pooling neighborhoods of $2\times2$, $5\times5$, and $10\times10$
  native grid cells (${\approx}$10, 24, and 49\,km). The dashed line marks
  FSS = 0.5, a common scale-dependent reference for useful neighborhood
  agreement. As expected, FSS decreases faster for smaller neighborhoods and
  higher precipitation thresholds.
  }
  \label{fig:fss-grid}
\end{figure}

\section{Further rollout diagnostics}
\label{sec:app_rollout_results}

\begin{figure}[H]
  \centering
  \includegraphics[width=0.99\textwidth]{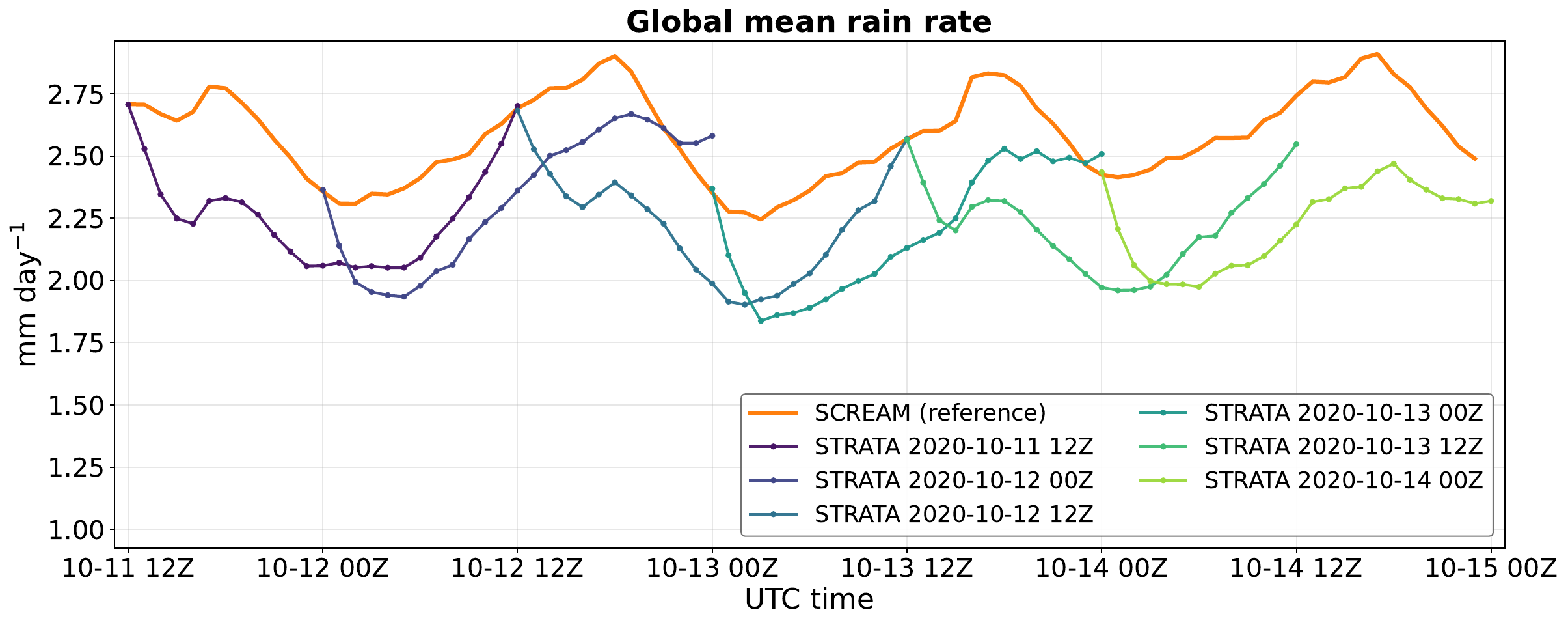}
  \caption{
  Time series of global-mean liquid precipitation rate for SCREAM (ground truth) and for independent STRATA simulations initialized at different times in the held-out test period. All STRATA simulations show the same pattern: global-mean precipitation first declines and then rebounds.
  }
  \label{fig:precip-ts-ensemble}
\end{figure}

\begin{figure}[H]
  \centering
  \includegraphics[width=0.99\textwidth]{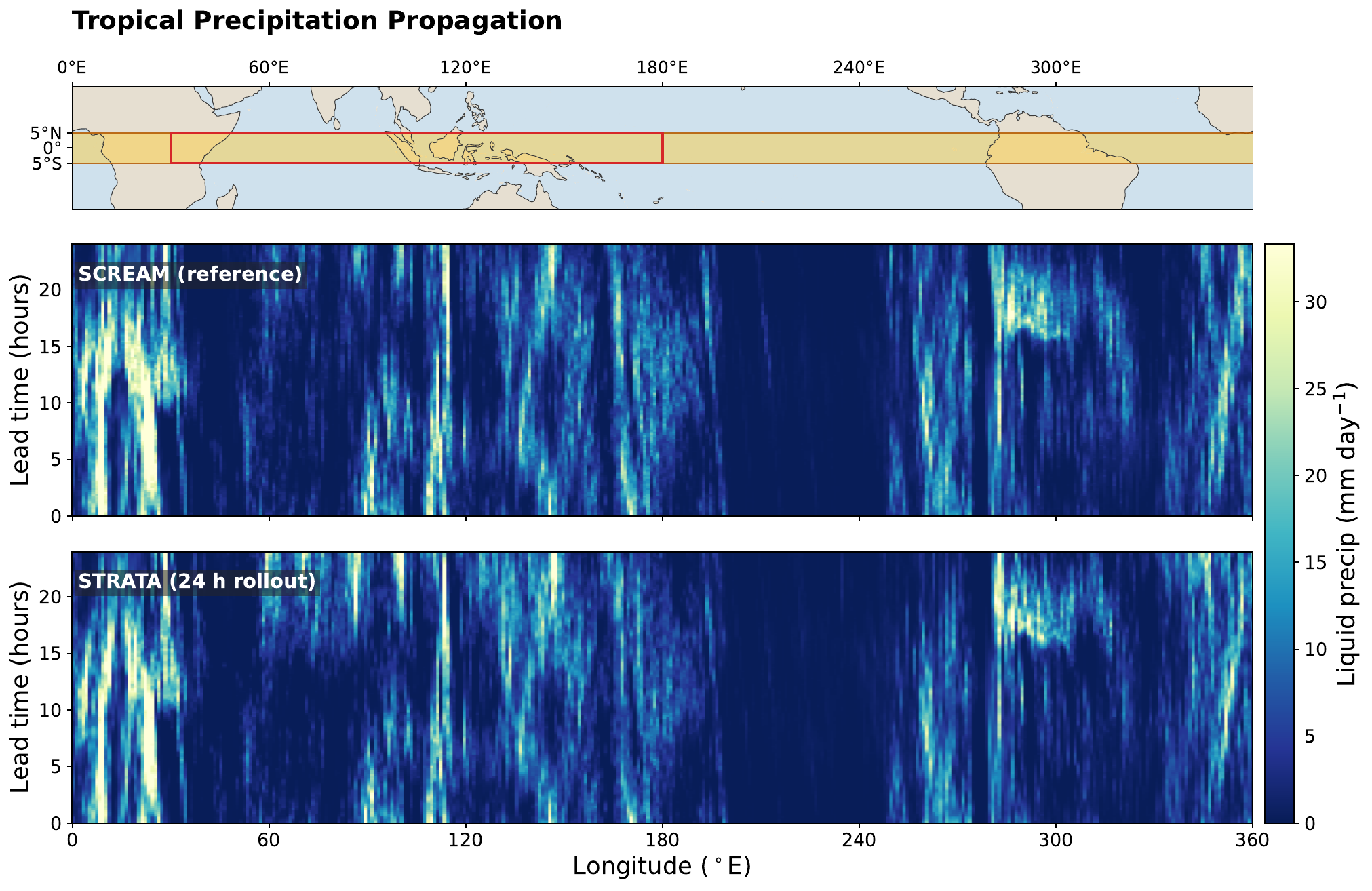}
  \caption{\textbf{Global tropical precipitation propagation over 24-hour rollout.}
  Top: geographic reference showing the 5$^\circ$S--5$^\circ$N averaging band
  (yellow); the red box marks the Indo-Pacific region shown in
  Figure~\ref{fig:hovmoller}a.
  Middle and bottom: Hovm\"{o}ller diagrams (longitude vs.\ lead time) of
  liquid precipitation averaged over 5$^\circ$S--5$^\circ$N for SCREAM
  (reference) and STRATA.
  Slanted features indicate propagating precipitation systems; their slope
  encodes propagation speed and direction.
  STRATA reproduces the propagation patterns of tropical convection
  throughout the 24-hour rollout.}
  \label{fig:hovmoller-global}
\end{figure}

  \begin{figure}[H]
  \centering
  \includegraphics[width=0.85\textwidth]{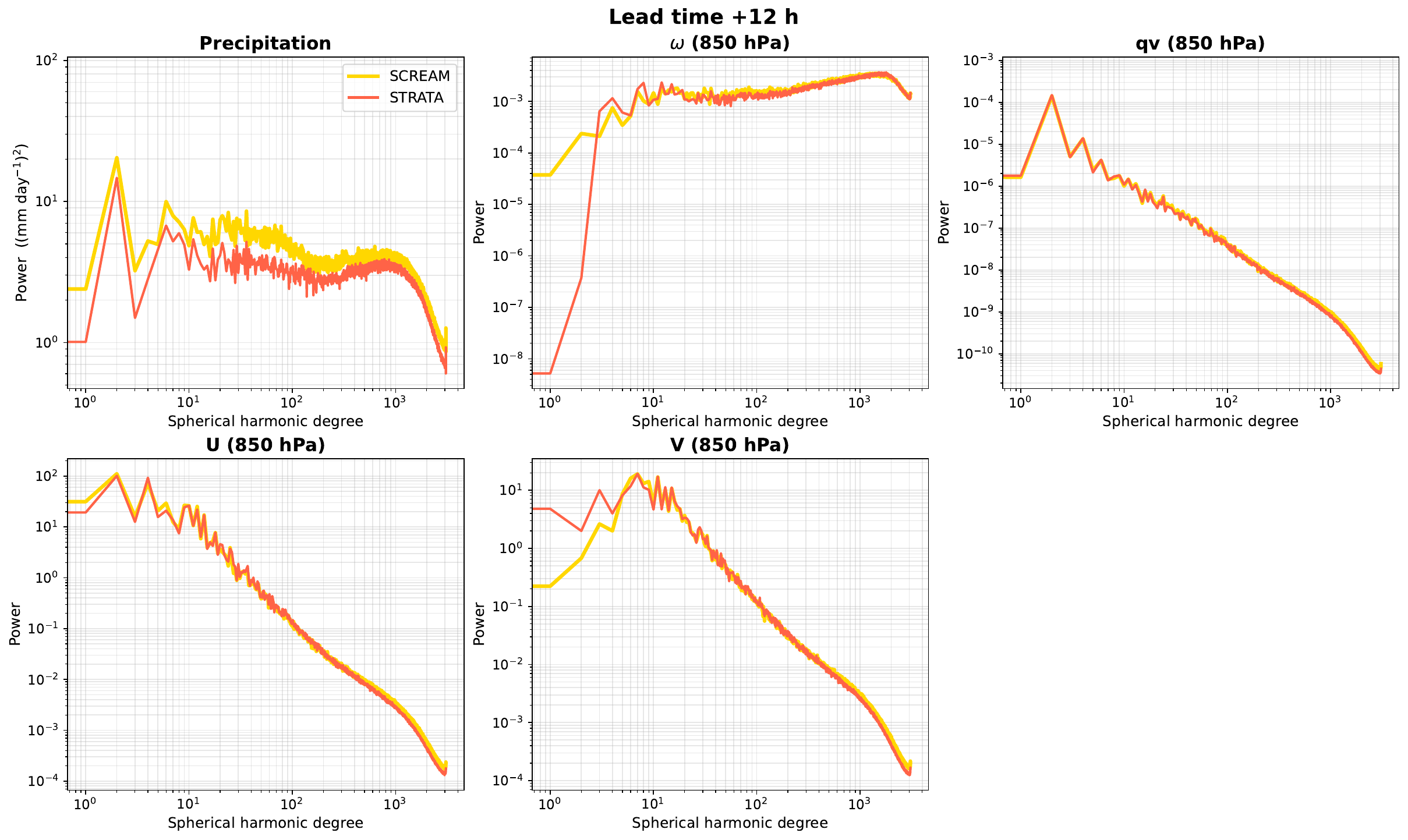}
  \caption{Spherical harmonic power spectra at lead time 12\,h comparing STRATA (orange) and SCREAM (yellow)
  for surface precipitation, vertical velocity $\omega$ at 850\,hPa, specific humidity $q_v$ at 850\,hPa, zonal and meridional wind at 850\,hPa.
  STRATA reproduces the SCREAM spectra closely at most scales; precipitation power is suppressed at 100--2,000\,km (Section~\ref{sec:rollout}).
  The suppressed power at the three lowest spherical harmonic degrees in $\omega$ reflects the
  large-scale spectrum filter applied after each rollout step (Appendix~\ref{sec:app_inference_details}).}
  \label{fig:spectra_frame71}
\end{figure}

\begin{figure}[H]
  \centering
  \includegraphics[width=0.99\textwidth]{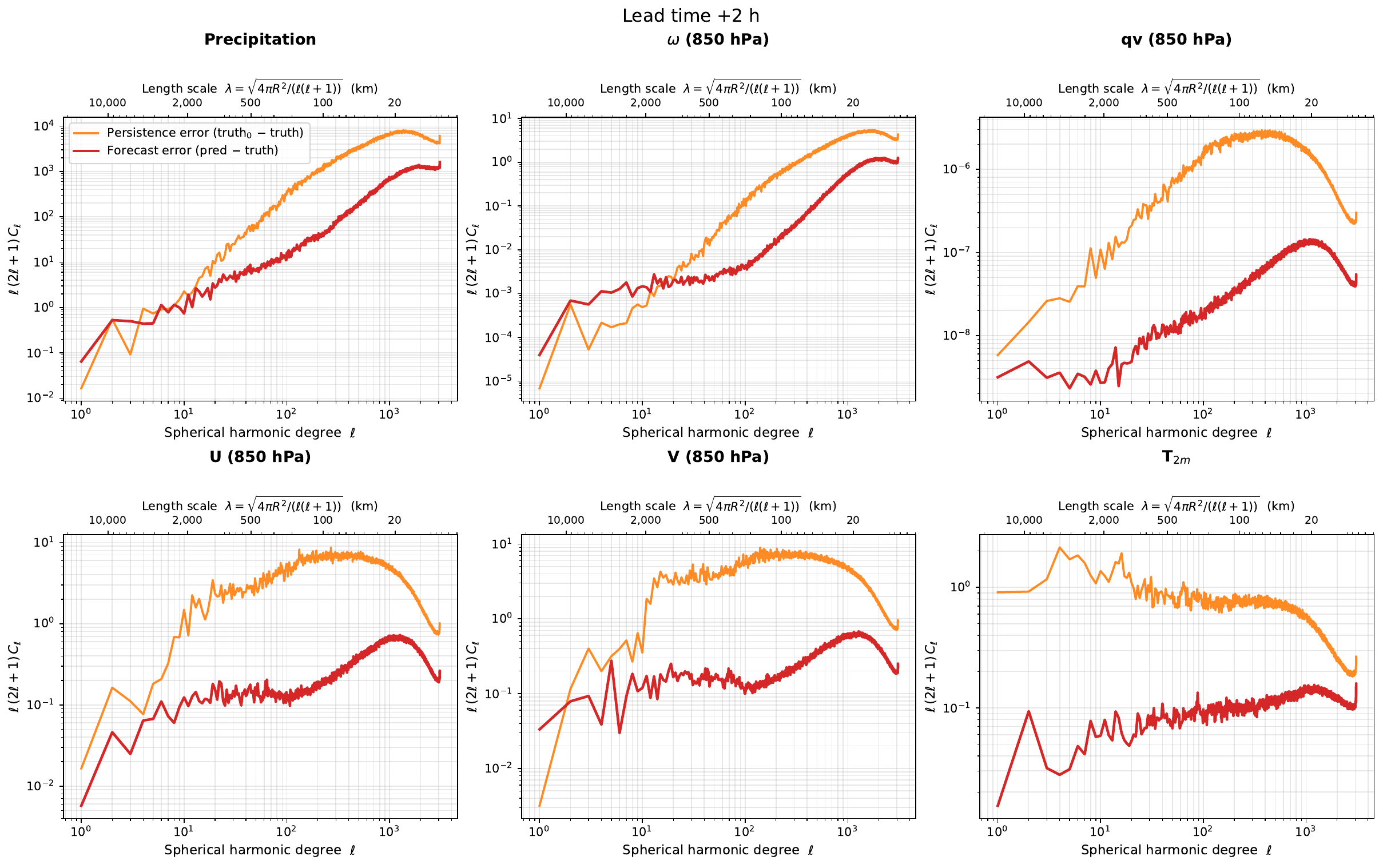}
  \includegraphics[width=0.99\textwidth]{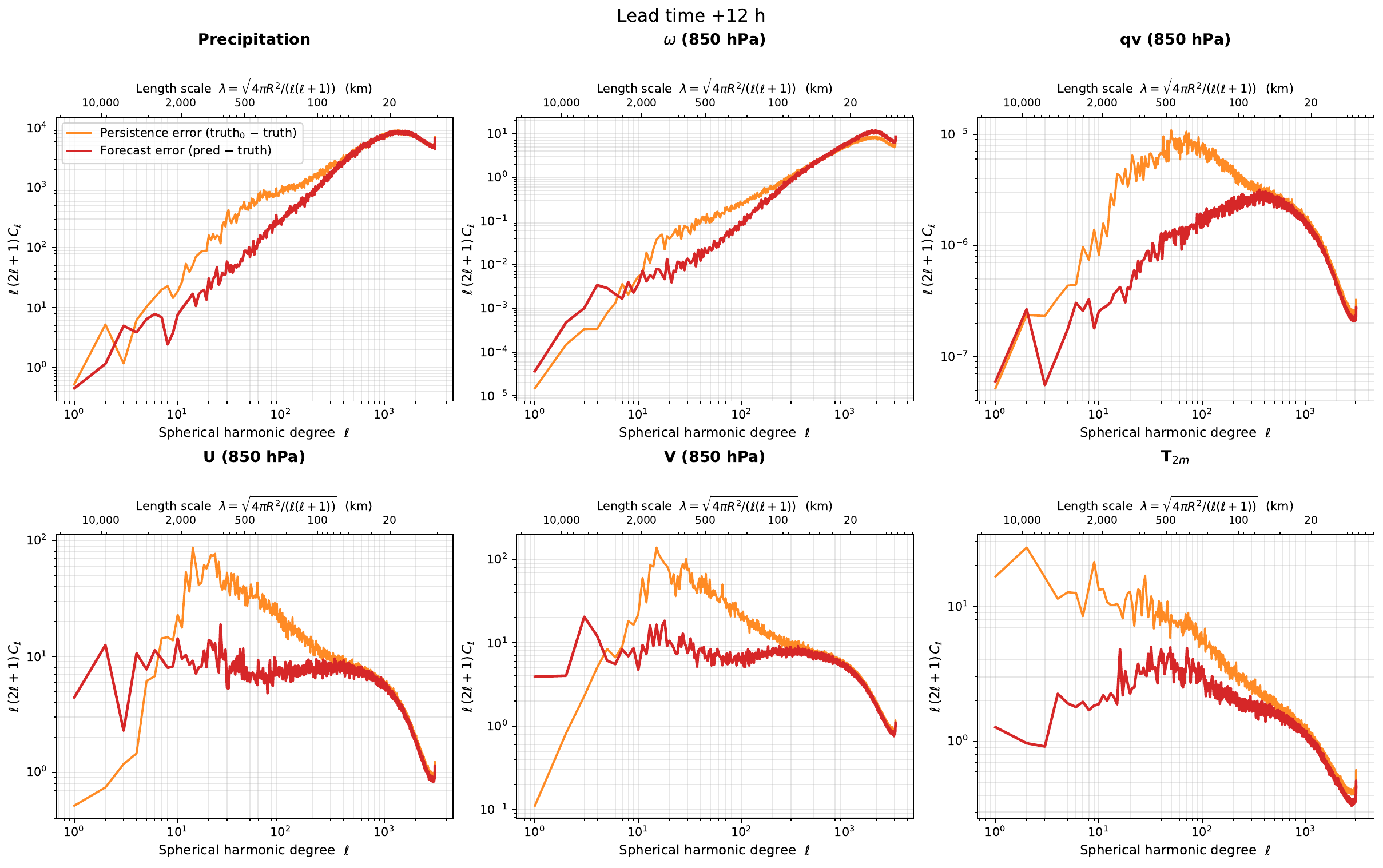}
  \caption{%
  \textbf{Error power spectra at 2-hour and 12-hour lead time.}
  Spherical-harmonic power spectra of STRATA forecast error
  (red; prediction minus SCREAM reference) and persistence error
  (orange; initial SCREAM state minus reference) for precipitation,
  850-hPa vertical pressure velocity $\omega$, 850-hPa specific humidity $q_v$,
  850-hPa zonal wind $U$, 850-hPa meridional wind $V$, and 2-m temperature
  $T_{2\mathrm{m}}$. The top figure shows 2-hour lead time and the bottom figure
  shows 12-hour lead time. At short lead time, STRATA errors are concentrated
  mainly at smaller scales; by 12 hours, error power increases at larger scales
  across several variables.}
  \label{fig:error-spectra}
\end{figure}

\begin{figure}[H]
  \centering
  \includegraphics[width=\textwidth]{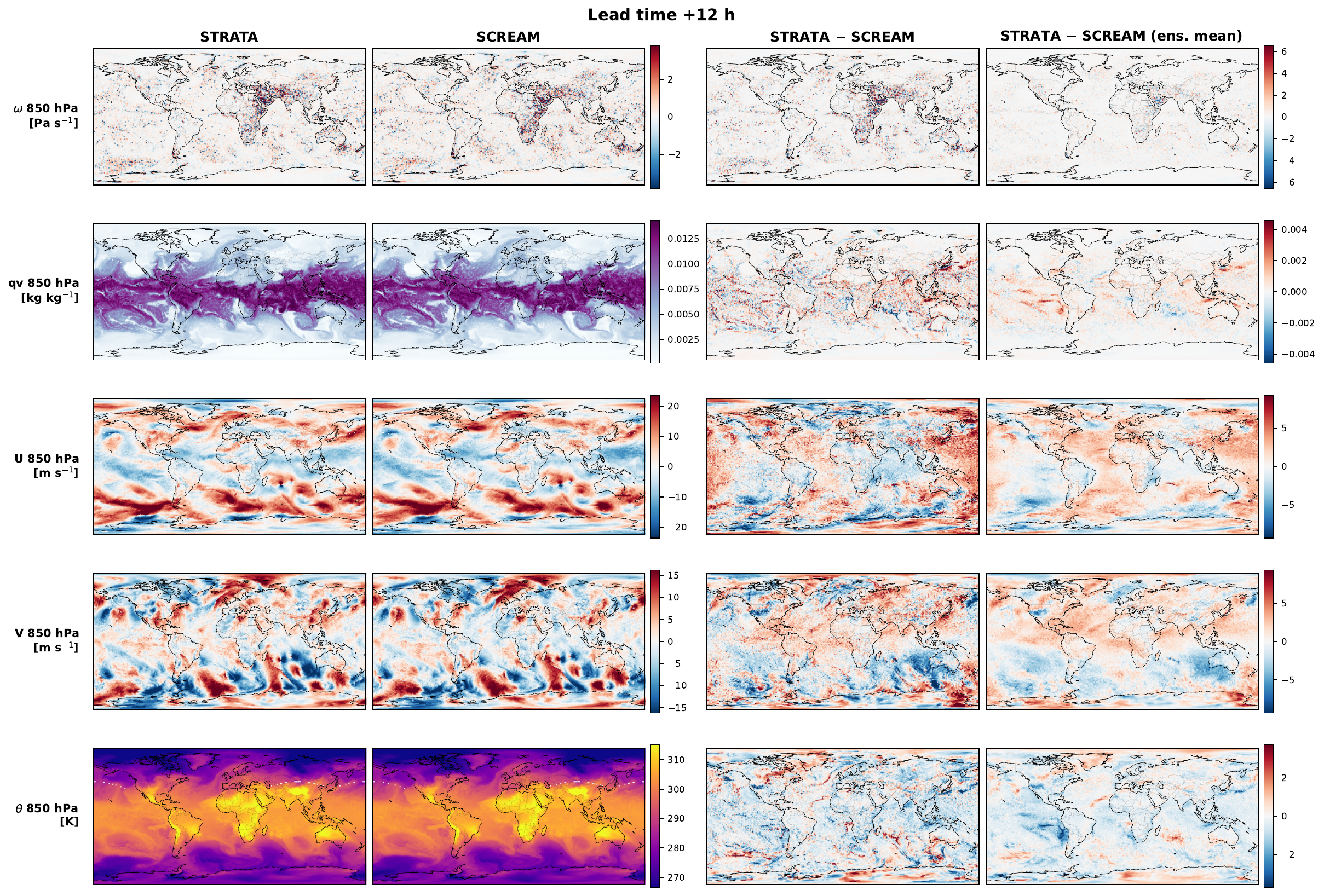}
  \caption{Global snapshot and error maps at 12-hour lead time. Rows show
  850-hPa vertical pressure velocity $\omega$, specific humidity $q_v$, zonal wind
  $U$, meridional wind $V$, and potential temperature $\theta$. Columns show the
  STRATA rollout, the corresponding SCREAM reference, the pointwise STRATA--SCREAM
  difference for the shown initialization, and the STRATA--SCREAM difference
  averaged across six rollout initializations.}
  \label{fig:app-global-snapshot-12h}
\end{figure}

\begin{figure}[H]
  \centering
  \includegraphics[width=\textwidth]{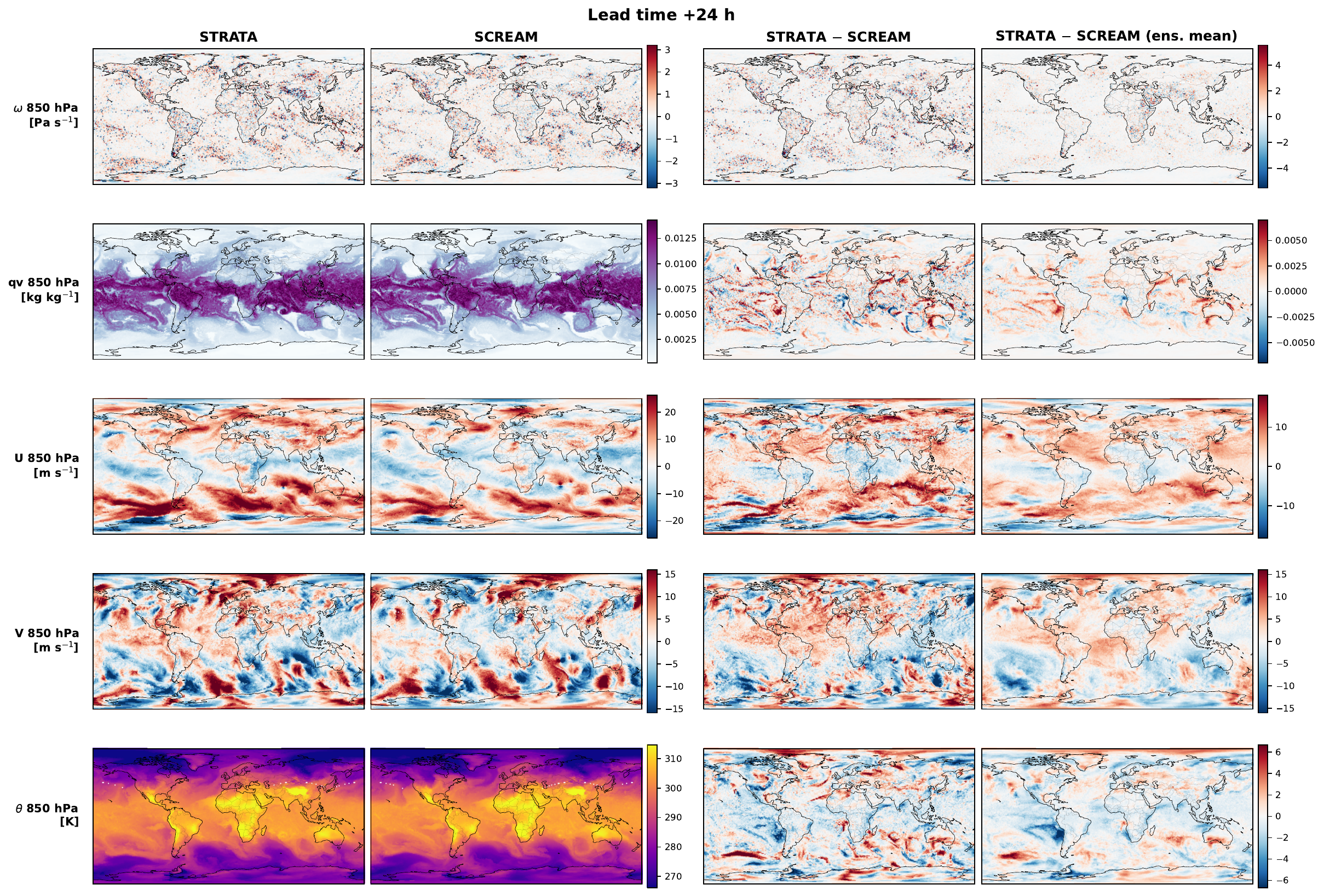}
  \caption{Same as Figure~\ref{fig:app-global-snapshot-12h}, but at 24-hour lead time.}
  \label{fig:app-global-snapshot-24h}
\end{figure}

\subsection{Conservation and large-scale balance}
\label{sec:app_conservation}

STRATA enforces no conservation law, either in the architecture or in the loss, so any conservation during a rollout is emergent. Table~\ref{tab:conservation} gives the drift of global dry-air mass, column water vapour and total energy relative to the run's own initial state, for the 24-hour rollout initialized at 2020-10-13T00, with the same diagnostic applied to SCREAM as a control. STRATA drifts much more than SCREAM in all three, most of all in water vapour.

\begin{table}[h]
\centering
\caption{Drift (\%) of area-weighted global column integrals relative to the initial state, for the rollout initialized at 2020-10-13T00 and for SCREAM over the same period.}
\label{tab:conservation}
\begin{tabular}{llccc}
\toprule
 & & +6\,h & +12\,h & +24\,h \\
\midrule
\multirow{2}{*}{Dry-air mass} & STRATA & $+0.09$ & $+0.24$ & $+0.51$ \\
 & SCREAM & $0.00$ & $0.00$ & $0.00$ \\
\multirow{2}{*}{Column water vapour} & STRATA & $+0.95$ & $+2.26$ & $+4.82$ \\
 & SCREAM & $+0.22$ & $+0.40$ & $+0.30$ \\
\multirow{2}{*}{Total energy} & STRATA & $+0.10$ & $+0.32$ & $+0.76$ \\
 & SCREAM & $-0.01$ & $0.00$ & $-0.01$ \\
\bottomrule
\end{tabular}
\end{table}

These are drifts, not closed budgets. We do not carry the surface and top-of-atmosphere fluxes, so the moisture and energy budgets cannot be closed. Dry-air mass is surface pressure minus the weight of the column water vapour, which removes vapour but not condensate. Total energy is the sum of internal, potential, kinetic and latent energy, with the latent term counting vapour only. All column integrals cover only the 24 retained levels (surface to ${\sim}100$\,hPa), and subselecting levels is not itself conservative.

The large-scale flow also moves away from balance. At 250\,hPa, after a 1500-km Gaussian low-pass filter, the ageostrophic fraction of the wind, $|\mathbf{v}-\mathbf{v}_g|/|\mathbf{v}|$, stays near 0.10 in SCREAM over the first 12 hours in both the northern (40--70$^\circ$N) and southern (32--60$^\circ$S) midlatitudes, whereas in STRATA it rises steadily to 0.17 and 0.16. This is consistent with the large-scale errors discussed in \S\ref{sec:rollout}. Enforcing conservation, or penalizing drift in the loss, is a direction for future work.

\subsection{Evaluation across seasons}
\label{sec:app_seasons}

\begin{figure}[h]
  \centering
  \IfFileExists{figures/loso_vs_strata.pdf}{%
    \includegraphics[width=\linewidth]{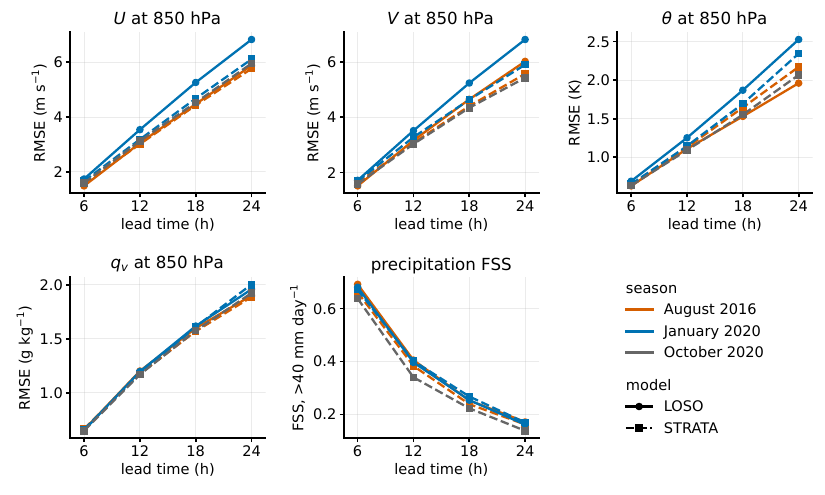}}{%
    \fbox{\parbox[c][5cm][c]{0.9\linewidth}{\centering Placeholder: figures/loso\_vs\_strata.pdf}}}
  \caption{RMSE at model level 24 (${\approx}848$\,hPa) and precipitation FSS ($>$40\,mm\,day$^{-1}$, higher is better) against lead time, pooled over three initializations per season (six for October). Orange: August (seen by both models); blue: January (unseen by the LOSO model); grey: October (STRATA only). Solid with circles: LOSO model; dashed with squares: STRATA.}
  \label{fig:loso-lines}
\end{figure}

Besides the six October initializations (Appendix~\ref{sec:app_training_data}), we evaluate on three held-out initializations in each of DYAMOND1 (2016-08-06 12:00, 2016-08-06 18:00 and 2016-08-07 00:00 UTC) and DYAMOND2 (2020-01-25 12:00, 2020-01-25 18:00 and 2020-01-26 00:00 UTC), 
all after the end of each simulation's training period (Table~\ref{tab:sims}). 
RMSE errors are computed on the native grid with area weighting and pooled over initializations. 
The rollout errors of STRATA (here it refers to the model that saw all three seasons) are similar in all three seasons (Figure~\ref{fig:loso-lines}, dashed).

\begin{figure}[h]
  \centering
  \IfFileExists{figures/loso_unseen_season.pdf}{%
    \includegraphics[width=\linewidth]{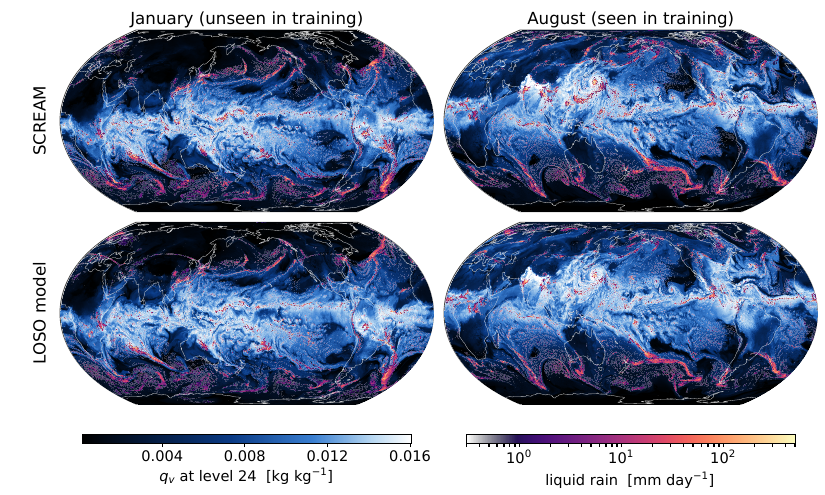}}{%
    \fbox{\parbox[c][4cm][c]{0.9\linewidth}{\centering Placeholder: figures/loso\_unseen\_season.pdf}}}
  \caption{$q_v$ at model level 24 (${\approx}848$\,hPa, blue) with liquid precipitation overlaid at +12\,h, for SCREAM (top) and the LOSO model (bottom), which was not trained on January. Left: January 2020 (DYAMOND2, unseen in training), initialized 2020-01-25 12:00\,UTC. Right: August 2016 (DYAMOND1, seen in training), initialized 2016-08-06 12:00\,UTC.}
  \label{fig:loso-maps}
\end{figure}

To test a season absent from training, we train a second model on sdecadal (October) and DYAMOND1 (August) only, excluding DYAMOND2 (January) from training. Its January rollouts remain realistic at the storm scale (Figure~\ref{fig:loso-maps}). Figure~\ref{fig:loso-lines} compares this leave-one-season-out (LOSO) model with STRATA, which saw all three seasons, on both August and January. For $q_v$ and precipitation the four curves nearly coincide. For wind and potential temperature the LOSO model is worse in January than in August by more than STRATA is: 4--11\% more for wind and 7--19\% more for temperature from +6 to +24\,h. January is somewhat harder for both models, which is why STRATA serves as the reference.

One caveat is that the LOSO model is a separate training run that also differs from STRATA in configuration: it is trained on two sources, applies StereoRoPE in the pixel decoder, and is trained and rolled out on 64$\times$64 tiles (at inference, 32$\times$32 tiles with a 16-cell halo). Comparing each model's January error with its own August error removes most of this difference. Its full training protocol is in Table~\ref{tab:ablation-models}.

\section{Information Entropy: Details\label{si:information-entropy}
}

Section \ref{sec:entropy} argued that km-scale data is higher entropy and therefore requires more compute for a given level of accuracy than coarser resolution modeling.

To support this interpretation, we compute the differential entropy of spatial patches of size $p$ of $dx\in\mathbb{R}^{c p^2}$ assuming a Gaussian distribution $dx\sim\mathcal{N}(0, \Sigma)$. The number of output dimensions is $d=cp^2$. To make this comparable across datasets with different units we unit normalize so that $\Sigma_{ii}=1$. We estimate $\Sigma$ on 90\% of the patch samples and regularize it by shrinkage toward a scaled identity, choosing the shrinkage parameter that maximizes the Gaussian log-likelihood of the held-out 10\%. For ERA5 we pool four consecutive 6-hour increments, which gives the largest ($8\times8$) covariance more than ten samples per dimension. For SCREAM we use a single 10-minute increment (2020-10-13 00:00 UTC), whose ${\sim}25$ million grid cells already supply far more patch samples.
The entropy in bits/dim is $H/d = \frac{1}{2d\ln 2}\left(d + d\ln(2\pi) + \ln |\Sigma|\right)$. Table~\ref{tab:block-entropy} shows this quantity for several patch sizes. SCREAM carries more information per dimension than ERA5 at every patch size, by 0.4--1.9 bits/dim for patches up to $4\times4$ and 0.7 bits/dim at $8\times8$.
We suspect that this additional entropy increases the computational demand (harmful) but at the same time the information density is helpful for training models on a limited number of samples.

Qualitatively, this can be visualized comparing the covariance structure of the state-vector difference $dx=x(t+\Delta t) - x(t)$ for SCREAM and ERA5 in a single atmospheric column. Figure \ref{fig:cov} shows that ERA5 targets have much stronger correlation across vertical levels and variables.

\begin{table}[t]
\centering
\begin{tabular}{lrrrr}
\toprule
Dataset & 1x1 & 2x2 & 4x4 & 8x8 \\
\midrule
SCREAM & \num{1.374} & \num{0.728} & \num{0.191} & \num{-0.222} \\
ERA5 & \num{0.984} & \num{-1.139} & \num{-1.371} & \num{-0.949} \\
\bottomrule
\end{tabular}\\
{\small Channels: SCREAM \num{148}, ERA5 \num{72}.}

\caption{Differential entropy (bits/dim) of spatial patches assuming a Gaussian distribution. The entropy of white noise $\Sigma=I$ is $\frac{1}{2}\log_2(2\pi e)\approx2$. For ERA5 this analysis uses the same 72 output channels as FCN3\citep{bonevFourCastNet3Geometric2025a}. }
\label{tab:block-entropy} %
\end{table}

\section{Additional sensitivity test}

\subsection{Width versus Depth Scaling}
\label{sec:app-width-depth}

Given a fixed compute budget (FLOPs), should one prefer a wider model (larger embedding dimension) or a deeper one (more layers)?
\citet{yuScalingLawsGlobal2026} find that global weather forecasting models consistently favor increased width over depth,
a trend that differs from behavior observed in language models.
We examine whether this holds across the full range of model sizes relevant to STRATA.

\begin{figure}
    \centering
    \includegraphics[width=\linewidth]{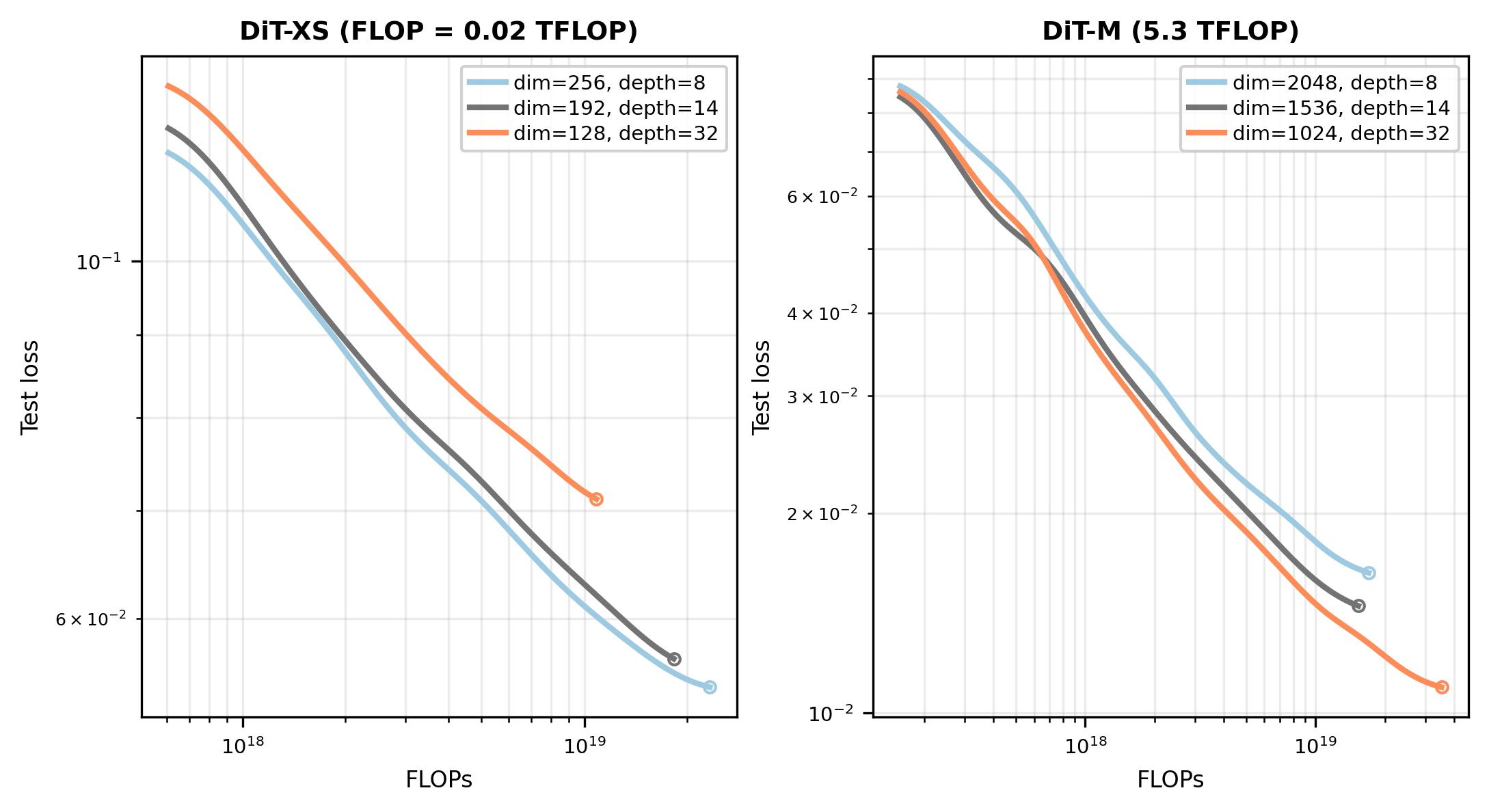}
    \caption{%
    \textbf{Width vs.\ depth scaling at fixed FLOPs.}
    Test loss as a function of model size for iso-FLOP sweeps that increase width (embedding dimension)
    or depth (number of layers).
    At small model sizes, wider models outperform deeper ones, consistent with \citet{yuScalingLawsGlobal2026}.
    At larger model sizes, deeper models show a marginal advantage, a regime not explored in prior work.}
    \label{fig:width-vs-depth-ablation}
\end{figure}

Figure~\ref{fig:width-vs-depth-ablation} shows test loss as a function of model scale under iso-FLOP width and depth sweeps (training protocol in Table~\ref{tab:sweep-protocols}).
At smaller model sizes, scaling width outperforms scaling depth at matched FLOPs, consistent with \citet{yuScalingLawsGlobal2026}.
We interpret this as the embedding dimension being a bottleneck at small scale:
a narrow model lacks sufficient channel capacity to represent the diverse atmospheric variables and their interactions,
so increasing width provides a larger per-token information budget that depth alone cannot compensate.
At larger model sizes, however, this advantage reverses: deeper models show a slight edge over wider ones at matched FLOPs.
This large-FLOP regime is not covered by \citet{yuScalingLawsGlobal2026}, and our results suggest that once the embedding
dimension is wide enough to carry the necessary information, additional depth provides more expressive power per FLOP.
These findings informed our choice of the STRATA architecture, which uses a relatively wide semantic encoder paired with a moderate number of layers.

\subsection{Inference Tile Size}

\begin{figure}
    \centering
    \includegraphics[width=0.75\linewidth]{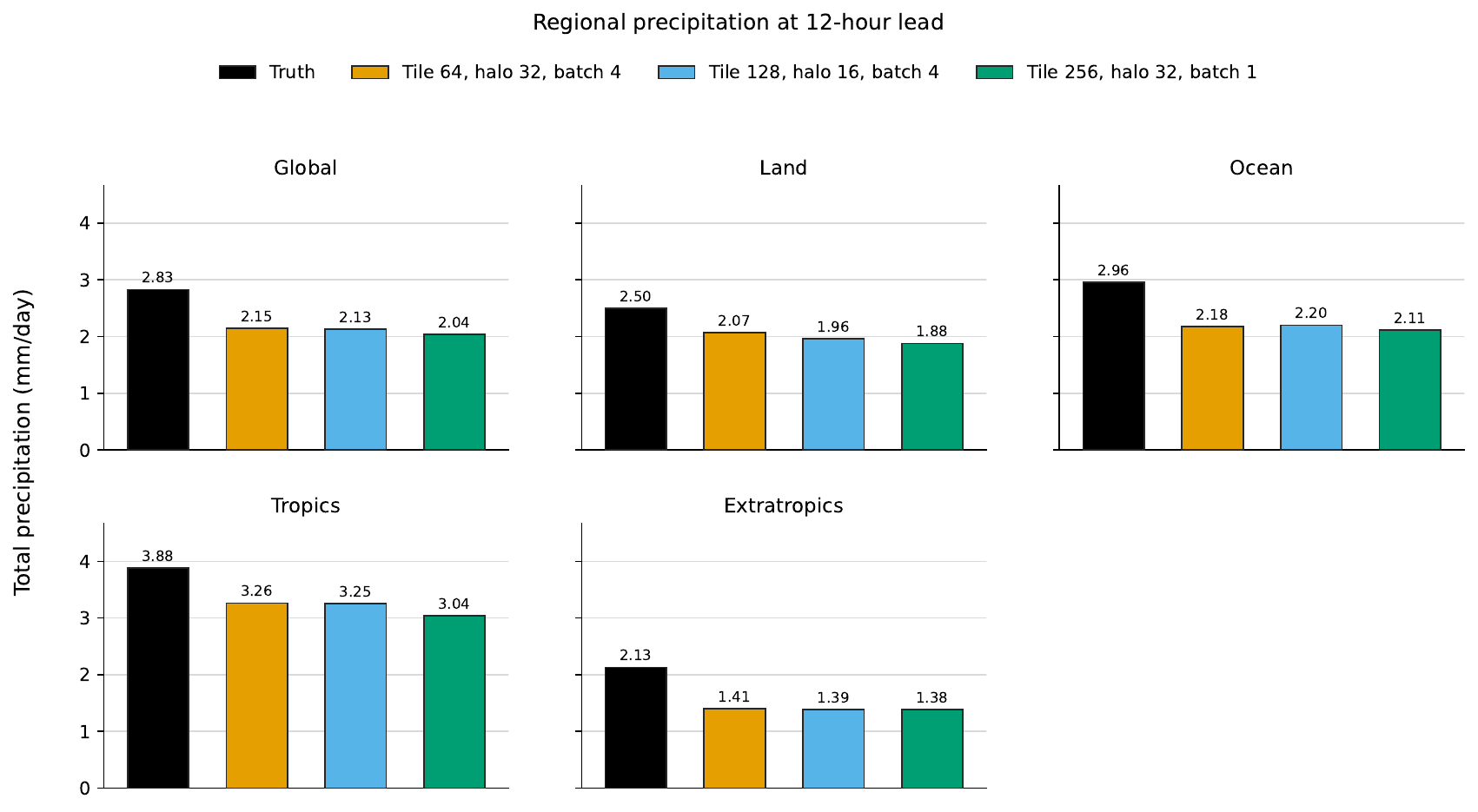}
    \caption{Precipitation bias at 12 hour lead time for various tile and halo sizes during inference.}
    \label{fig:tile-size-sweep}
\end{figure}

Using a large tile size at inference is advantageous because it means there is less computation wasted on the halo regions.
However, since the model was fine-tuned with a differing tile size of 128, there is some risk of exposure bias when running it with larger tile sizes only for inference. Figure \ref{fig:tile-size-sweep} shows that using a larger padded size ``tile size + 2 halo'' during inference results in a marginally worse dry bias especially in the tropics. However, the difference is small compared to the overall bias versus the ground truth.

\subsection{Effective Batch Size and Training Tile Size}

Larger training tiles increase the effective batch size, since each tile covers more samples of independent storms.
However, large-batch training is known to exhibit diminishing returns: beyond a critical batch size,
increasing the batch further yields no improvement in convergence per unit of data processed,
and may even degrade final loss~\citep{goyalAccurateLargeMinibatch2018,mccandlishEmpiricalModelLargeBatch2018}.
This raises the question of whether km-scale atmospheric emulation benefits from the very large batch sizes
that large tiles would provide.

\begin{figure}
  \centering
  \includegraphics[width=0.5\linewidth]{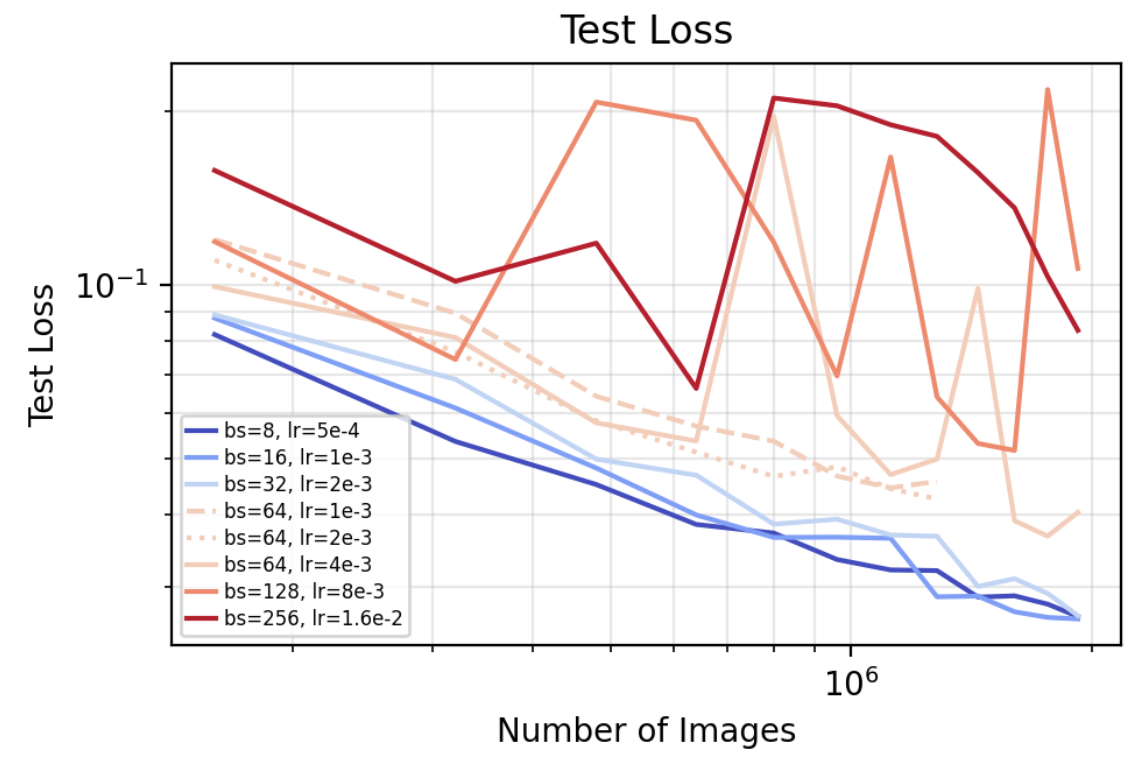}
  \caption{%
  \textbf{Batch size ablation.}
  Test loss as a function of batch size under linear learning rate scaling. Each
  sample is a $64\times64$ tile.
  Training is stable and achieves the lowest loss at batch sizes up to 32.
  Smaller batch sizes show a marginal benefit in per-sample efficiency.
  Beyond batch size 32, training becomes unstable as the linear scaling rule breaks down,
  indicating a critical batch size beyond which further increases are counterproductive.}
  \label{fig:batch-size-ablation}
\end{figure}

To answer this, we perform a batch size sweep on the SCREAM dataset, scaling the learning rate linearly
with batch size following the linear scaling rule~\citep{goyalAccurateLargeMinibatch2018}. All samples are $64\times64$ tiles; the model and training protocol are in Table~\ref{tab:sweep-protocols}.
As shown in Figure~\ref{fig:batch-size-ablation}, training with batch sizes up to 32 is stable, and smaller batch sizes achieve marginally lower test loss when evaluated
at equal total number of training samples---consistent with the large-batch generalization gap reported
in prior work.
Increasing the batch size beyond this regime leads to training instability: the linear learning rate
scaling rule breaks down, and loss diverges or fails to converge, indicating that the gradient noise
at smaller batch sizes plays a beneficial regularization role that cannot simply be compensated by a
proportionally larger learning rate.

Together, these results suggest that training on moderate tile sizes ($64\times64$ or $128\times128$)
is not only computationally efficient but also well-matched to the effective batch size regime where
optimization is most stable.
As mentioned in the main text, this is further supported by the physics of the problem: at a 10-minute timestep, the maximum range
of information propagation is bounded by the acoustic wave speed (${\sim}340$\,m/s), limiting the
physical context required within a single tile.
Large tiles therefore do not provide proportionally more useful physical context per training step,
reinforcing the case for compact, compute-efficient tiles.

\subsection{Model capacity and rollout stability}
\label{sec:model-capacity-ablation}

\begin{figure}[t]
  \centering
  \includegraphics[width=0.7\linewidth]{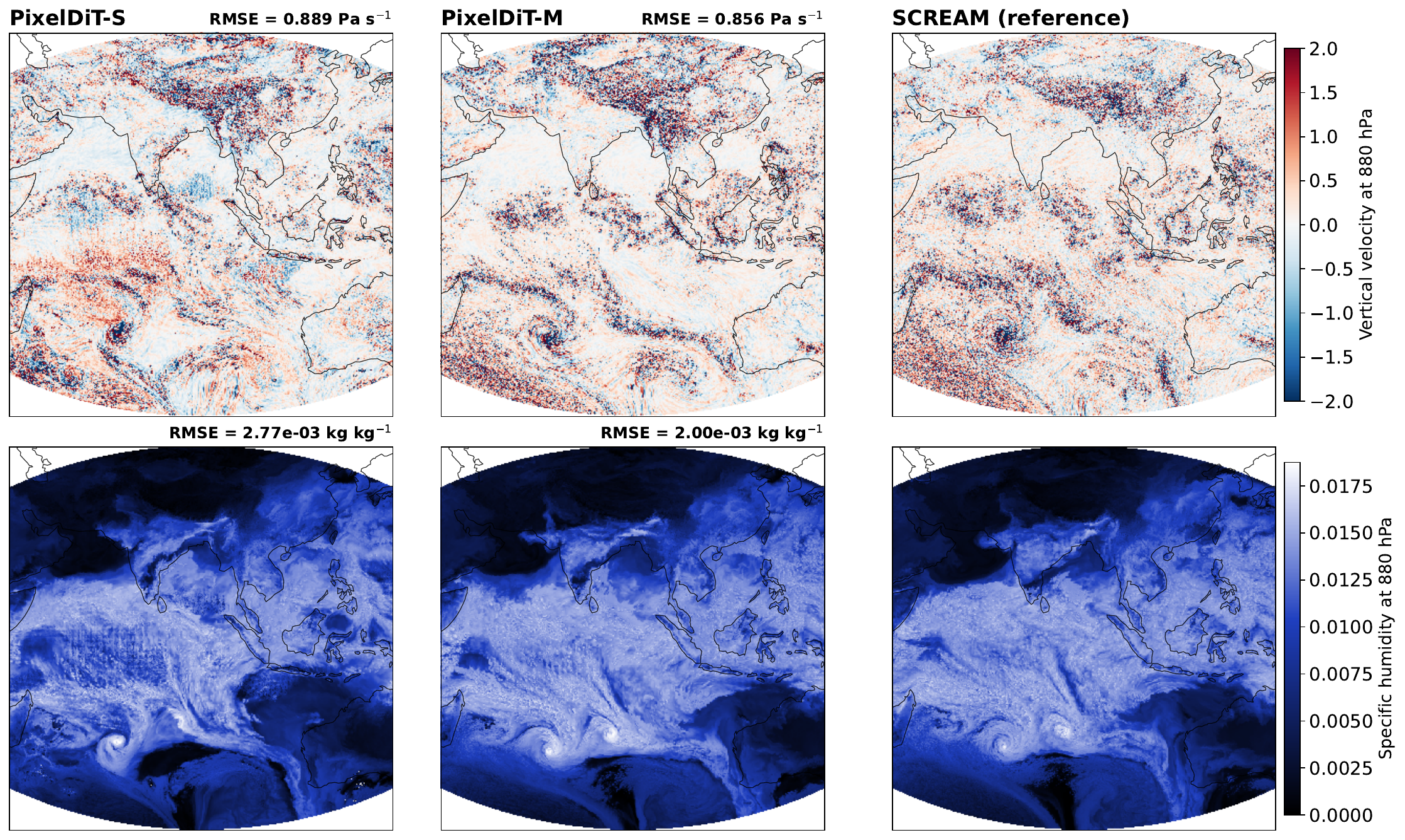}
  \caption{%
  \textbf{Rollout skill on smaller vs larger models.}
  Snapshots of several fields (vertical velocity, specific humidity) for small (1.2 TFLOP, left) and middle (4.9 TFLOP, middle) FLOP-size models and reference SCREAM ground truth (right) at 24-hour rollout lead time. The smaller model shows more pronounced artifacts with larger humidity RMSE. RMSE values of vertical velocity and specific humidity are reported on the forecast panels are evaluated on 4$\times$ coarsened fields for the shown cube face to focus more on large-scale error instead of pixel-wise agreement. 
  }
  \label{fig:model-size-ablation}
\end{figure}

Figure~\ref{fig:model-size-ablation} illustrates a characteristic failure mode
of the smaller FLOP-size model (${\sim}1.2$\,TFLOP) at 24-hour lead time. The two models are PixelDiT-S and PixelDiT-M; their training protocols are in Table~\ref{tab:ablation-models}.
Behind the dominant speckle of km-scale dynamics, the vertical velocity $\omega$ field shows large, coherent planetary-scale
anomalies---broad regions of anomalous ascent and
subsidence absent in both the SCREAM reference and the larger model rollouts.
These patterns are inconsistent with mass continuity, which requires
$\omega$ averaged over thousand-kilometer scales to remain near zero,
suggesting the smaller model lacks sufficient capacity to maintain
physically balanced large-scale circulation.
The specific humidity field shows a corresponding degradation: the
unrealistic large-scale descent drives spurious drying that further
degrades forecast quality.
Empirically, larger FLOP-size models are less prone to such artifacts at
24-hour lead times; however, a more principled constraint on large-scale
$\omega$ is likely needed to ensure stability at longer horizons.

\subsection{Zero-shot transfer: maps and other grid types}
\label{sec:app_grid_types}
\begin{figure}[h]
    \centering
    \includegraphics[width=\linewidth]{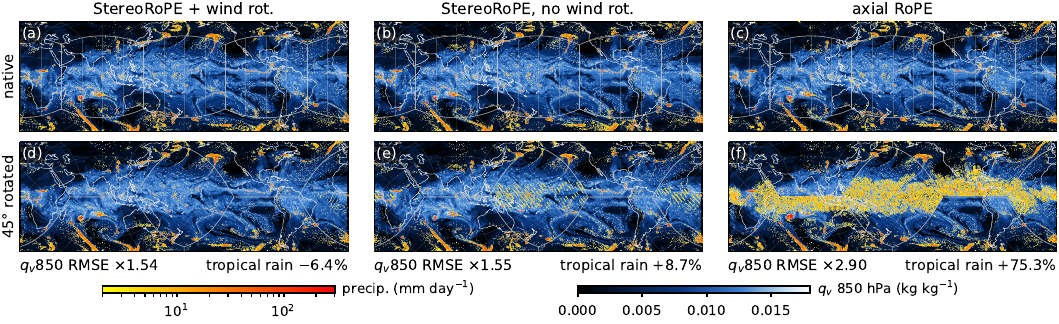}
    \caption{%
    \textbf{Zero-shot transfer to an unseen grid.} $q_v$ at the model level nearest 850\,hPa (blue) with liquid precipitation overlaid (yellow--red; blank below 2\,mm\,day$^{-1}$) at +3\,h, initialized 2020-10-13 00\,UTC. Top: native cubed sphere (training grid). Bottom: cubed sphere rotated by 45$^\circ$ (polar face centres at 45$^\circ$N, 90$^\circ$W and 45$^\circ$S, 90$^\circ$E), unseen in training. Columns: (a,d)~StereoRoPE with wind rotation; (b,e)~StereoRoPE without wind rotation; (c,f)~axial 2-D RoPE without wind rotation. Thin white lines mark the cube-face edges of each grid. Labels give the rotated/native $q_v$ RMSE ratio and the change in tropical-mean ($|\text{lat}|<20^\circ$) rainfall. Table~\ref{tab:grid-transfer} summarizes these statistics.
    }
    \label{fig:grid-invariance}
\end{figure}

\begin{figure}[h]
    \centering
    \includegraphics[width=0.85\linewidth]{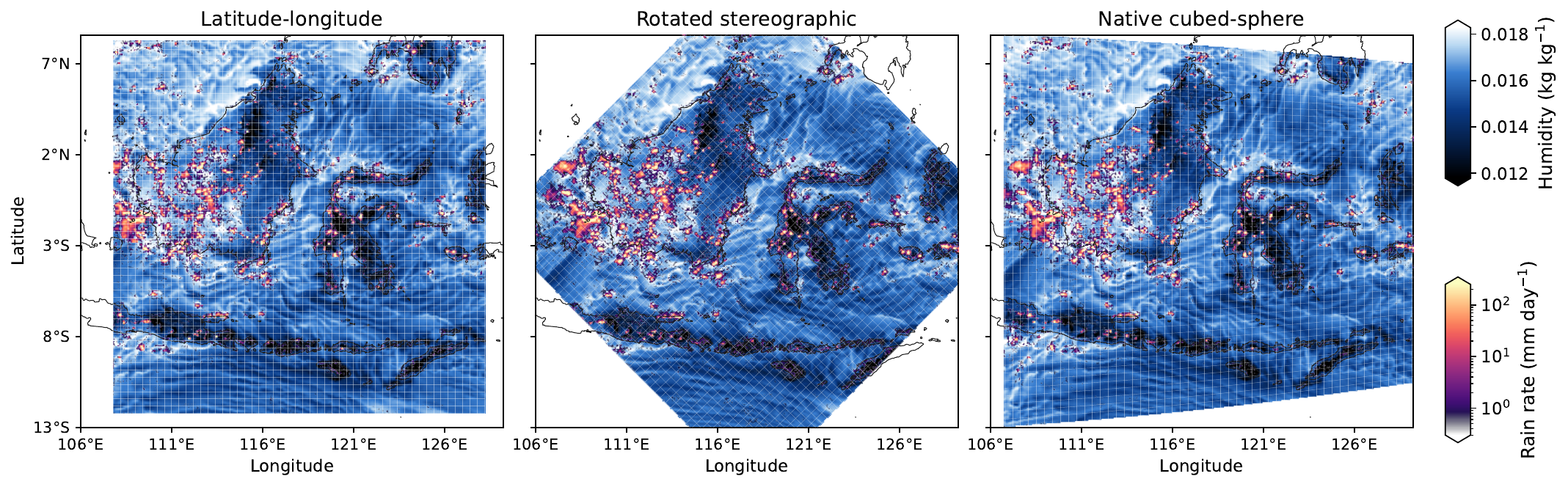}
    \caption{Near-surface humidity and rain rate from a 3-hour zero-shot regional STRATA rollout over Indonesia on three grids of similar nominal resolution: regular latitude--longitude (left), oblique stereographic rotated by 45$^\circ$ (middle), and the native SCREAM cubed sphere (right, training grid).}
    \label{fig:grid-types}
\end{figure}

Figure~\ref{fig:grid-invariance} shows maps from the global rollouts behind Table~\ref{tab:grid-transfer} for the three models of Section~\ref{sec:grid_invariance}, on the native and the rotated cubed sphere.
Figure~\ref{fig:grid-types} shows a separate, regional experiment at higher detail: STRATA is rolled out zero-shot for 3 hours only within a domain over Indonesia, on two discretizations it never saw in training, a regular latitude--longitude grid and an oblique stereographic grid rotated by 45$^\circ$, and compared with a rollout over the same domain on the native cubed-sphere grid. The precipitation and humidity fields remain qualitatively consistent across the three grids. The agreement is not exact, since interpolation, masking and grid resolution differ between the setups, but the spatial organization of precipitation and moisture is preserved.

\newpage
\input{checklist.tex}

\end{document}

%% file: checklist.tex
\section*{NeurIPS Paper Checklist}

\begin{enumerate}

\item {\bf Claims}
    \item[] Question: Do the main claims made in the abstract and introduction accurately reflect the paper's contributions and scope?
    \item[] Answer: \answerYes{} %
    \item[] Justification: Contributions are enumerated in the introduction bullet points; the abstract states scope as 24-hour global km-scale emulation with documented limitations of large-scale biases.

\item {\bf Limitations}
    \item[] Question: Does the paper discuss the limitations of the work performed by the authors?
    \item[] Answer: \answerYes{} %
    \item[] Justification: §\ref{sec:rollout} discusses large-scale error growth faster than SCREAM and its likely causes, the dry precipitation bias and the partly empirical vertical-velocity filter; §\ref{sec:app_arch} notes the residual patch artifact; Appendix~\ref{sec:app_conservation} reports conservation drift and loss of large-scale balance.

\item {\bf Theory assumptions and proofs}
    \item[] Question: For each theoretical result, does the paper provide the full set of assumptions and a complete (and correct) proof?
    \item[] Answer: \answerYes{} %
    \item[] Justification: The patch instability analysis in Appendix B.5 provides a formal stability analysis under a linearized model assumption, explicitly stated.

    \item {\bf Experimental result reproducibility}
    \item[] Question: Does the paper fully disclose all the information needed to reproduce the main experimental results of the paper to the extent that it affects the main claims and/or conclusions of the paper (regardless of whether the code and data are provided or not)?
    \item[] Answer: \answerYes{} %
    \item[] Justification: Architecture in Tables~\ref{tab:arch}--\ref{tab:sweep}; training hyperparameters in Appendix~\ref{sec:app_training_details}; data in Appendix~\ref{sec:app_training_data}; inference settings (tiling, halo, blending window, filters) in Appendix~\ref{sec:app_inference_details}; code and data are publicly available (Section 6).

\item {\bf Open access to data and code}
    \item[] Question: Does the paper provide open access to the data and code, with sufficient instructions to faithfully reproduce the main experimental results, as described in supplemental material?
    \item[] Answer: \answerYes{} %
    \item[] Justification: Code, trained model weights, the raw SCREAM output and a training-ready Zarr version of the data are publicly available (Section~\ref{sec:availability}).

\item {\bf Experimental setting/details}
    \item[] Question: Does the paper specify all the training and test details (e.g., data splits, hyperparameters, how they were chosen, type of optimizer) necessary to understand the results?
    \item[] Answer: \answerYes{} %
    \item[] Justification: Dataset and training setup in §\ref{sec:datasets}; full hyperparameters in Appendix~\ref{sec:app_training_details}; architecture in Appendix~\ref{sec:app_arch_spec}.

\item {\bf Experiment statistical significance}
    \item[] Question: Does the paper report error bars suitably and correctly defined or other appropriate information about the statistical significance of the experiments?
    \item[] Answer: \answerYes{} %
    \item[] Justification: Rollout metrics (Figure~\ref{fig:rmse}, Figure~\ref{fig:scorecard}) show the min--max range across six initialization times; Appendix~\ref{sec:app_seasons} evaluates across three seasons.

\item {\bf Experiments compute resources}
    \item[] Question: For each experiment, does the paper provide sufficient information on the computer resources (type of compute workers, memory, time of execution) needed to reproduce the experiments?
    \item[] Answer: \answerYes{} %
    \item[] Justification: GPU type (H100 80GB), count and batch size in Appendix~\ref{sec:app_training_details}; training time, inference throughput and energy in Appendix~\ref{sec:app_throughput_energy} (Table~\ref{tab:throughput_energy_accounting}), including the one-off cost of data generation and training (Table~\ref{tab:energy_amortization}).
    
\item {\bf Code of ethics}
    \item[] Question: Does the research conducted in the paper conform, in every respect, with the NeurIPS Code of Ethics \url{https://neurips.cc/public/EthicsGuidelines}?
    \item[] Answer: \answerYes{} %
    \item[] Justification: The authors have reviewed the NeurIPS Code of Ethics, and the research conforms to it.

\item {\bf Broader impacts}
    \item[] Question: Does the paper discuss both potential positive societal impacts and negative societal impacts of the work performed?
    \item[] Answer: \answerYes{} %
    \item[] Justification: STRATA aims to reduce the energy cost of km-scale climate simulation. Limitations (documented biases, 17-day training scope) should be understood before using outputs in high-stakes climate decisions.
    
\item {\bf Safeguards}
    \item[] Question: Does the paper describe safeguards that have been put in place for responsible release of data or models that have a high risk for misuse (e.g., pre-trained language models, image generators, or scraped datasets)?
    \item[] Answer: \answerNA{} %
    \item[] Justification: The model emulates atmospheric physics and poses no significant misuse risk.

\item {\bf Licenses for existing assets}
    \item[] Question: Are the creators or original owners of assets (e.g., code, data, models), used in the paper, properly credited and are the license and terms of use explicitly mentioned and properly respected?
    \item[] Answer: \answerYes{} %
    \item[] Justification: All third-party assets used are properly cited: SCREAM, NATTEN, DiT, PixelDiT, and Duo-Grid padding.

\item {\bf New assets}
    \item[] Question: Are new assets introduced in the paper well documented and is the documentation provided alongside the assets?
    \item[] Answer: \answerYes{} %
    \item[] Justification: The SCREAM training dataset and STRATA model are publicly available (Section 6); architecture and training details are fully documented in the appendix.

\item {\bf Crowdsourcing and research with human subjects}
    \item[] Question: For crowdsourcing experiments and research with human subjects, does the paper include the full text of instructions given to participants and screenshots, if applicable, as well as details about compensation (if any)? 
    \item[] Answer: \answerNA{} %
    \item[] Justification: Paper does not involve crowdsourcing or human subjects.

\item {\bf Institutional review board (IRB) approvals or equivalent for research with human subjects}
    \item[] Question: Does the paper describe potential risks incurred by study participants, whether such risks were disclosed to the subjects, and whether Institutional Review Board (IRB) approvals (or an equivalent approval/review based on the requirements of your country or institution) were obtained?
    \item[] Answer: \answerNA{} %
    \item[] Justification: Paper does not involve human subjects research.

\item {\bf Declaration of LLM usage}
    \item[] Question: Does the paper describe the usage of LLMs if it is an important, original, or non-standard component of the core methods in this research? Note that if the LLM is used only for writing, editing, or formatting purposes and does \emph{not} impact the core methodology, scientific rigor, or originality of the research, declaration is not required.
    \item[] Answer: \answerNA{} %
    \item[] Justification: LLMs were used only for writing assistance, not as a core methodological component.

\end{enumerate}

%% file: main.bbl
\begin{thebibliography}{60}
\providecommand{\natexlab}[1]{#1}
\providecommand{\url}[1]{\texttt{#1}}
\expandafter\ifx\csname urlstyle\endcsname\relax
  \providecommand{\doi}[1]{doi: #1}\else
  \providecommand{\doi}{doi: \begingroup \urlstyle{rm}\Url}\fi

\bibitem[Bony et~al.(2015)Bony, Stevens, Frierson, Jakob, Kageyama, Pincus, Shepherd, Sherwood, Siebesma, Sobel, Watanabe, and Webb]{bonyCloudsCirculationClimate2015}
Sandrine Bony, Bjorn Stevens, Dargan M.~W. Frierson, Christian Jakob, Masa Kageyama, Robert Pincus, Theodore~G. Shepherd, Steven~C. Sherwood, A.~Pier Siebesma, Adam~H. Sobel, Masahiro Watanabe, and Mark~J. Webb.
\newblock Clouds, circulation and climate sensitivity.
\newblock \emph{Nature Geoscience}, 8\penalty0 (4):\penalty0 261--268, April 2015.
\newblock ISSN 1752-0908.
\newblock \doi{10.1038/ngeo2398}.

\bibitem[Schneider et~al.(2017)Schneider, Teixeira, Bretherton, Brient, Pressel, Sch{\"a}r, and Siebesma]{schneiderClimateGoalsComputing2017}
Tapio Schneider, Jo{\~a}o Teixeira, Christopher~S. Bretherton, Florent Brient, Kyle~G. Pressel, Christoph Sch{\"a}r, and A.~Pier Siebesma.
\newblock Climate goals and computing the future of clouds.
\newblock \emph{Nature Climate Change}, 7\penalty0 (1):\penalty0 3--5, January 2017.
\newblock ISSN 1758-6798.
\newblock \doi{10.1038/nclimate3190}.

\bibitem[Caldwell et~al.(2021)Caldwell, Terai, Hillman, Keen, Bogenschutz, Lin, Beydoun, Taylor, Bertagna, Bradley, Clevenger, Donahue, Eldred, Foucar, Golaz, Guba, Jacob, Johnson, Krishna, Liu, Pressel, Salinger, Singh, Steyer, Ullrich, Wu, Yuan, Shpund, Ma, and Zender]{caldwellConvectionPermittingSimulationsE3SM2021}
P.~M. Caldwell, C.~R. Terai, B.~Hillman, N.~D. Keen, P.~Bogenschutz, W.~Lin, H.~Beydoun, M.~Taylor, L.~Bertagna, A.~M. Bradley, T.~C. Clevenger, A.~S. Donahue, C.~Eldred, J.~Foucar, J.-C. Golaz, O.~Guba, R.~Jacob, J.~Johnson, J.~Krishna, W.~Liu, K.~Pressel, A.~G. Salinger, B.~Singh, A.~Steyer, P.~Ullrich, D.~Wu, X.~Yuan, J.~Shpund, H.-Y. Ma, and C.~S. Zender.
\newblock Convection-{{Permitting Simulations With}} the {{E3SM Global Atmosphere Model}}.
\newblock \emph{Journal of Advances in Modeling Earth Systems}, 13\penalty0 (11):\penalty0 e2021MS002544, November 2021.
\newblock ISSN 1942-2466, 1942-2466.
\newblock \doi{10.1029/2021MS002544}.

\bibitem[Segura et~al.(2025)Segura, {Pedruzo-Bagazgoitia}, Weiss, M{\"u}ller, Rackow, Lee, {Dolores-Tesillos}, Benedict, Aengenheyster, Aguridan, Arduini, Baker, Bao, Bastin, Baulenas, Becker, Beyer, Bockelmann, Br{\"u}ggemann, Brunner, Cheedela, Das, Denissen, Dragaud, Dziekan, Ekblom, Engels, Esch, Forbes, Frauen, Freischem, {Garc{\'i}a-Maroto}, Geier, Gierz, {Gonz{\'a}lez-Cervera}, Grayson, Griffith, Gutjahr, Haak, Hadade, Haslehner, {ul Hasson}, Hegewald, Kluft, Koldunov, Koldunov, K{\"o}lling, Koseki, Kosukhin, Kousal, Kuma, Kumar, Li, Maury, Meindl, Milinski, Mogensen, Niraula, Nowak, Praturi, Proske, Putrasahan, Redler, Santuy, S{\'a}rm{\'a}ny, Schnur, Scholz, Sidorenko, Sp{\"a}t, S{\"u}tzl, Takasuka, Tompkins, Uribe, Valentini, Veerman, Voigt, Warnau, Wachsmann, Wac{\l}awczyk, Wedi, Wieners, Wille, Winkler, Wu, Ziemen, Zimmermann, Bender, Bojovic, Bony, Bordoni, Brehmer, Dengler, Dutra, Faye, Fischer, {van Heerwaarden}, Hohenegger, J{\"a}rvinen, Jochum, Jung, Jungclaus, Keenlyside, Klocke, Konow,
  Klose, Malinowski, Martius, Mauritsen, Mellado, Mieslinger, Mohino, Paw{\l}owska, {Peters-von Gehlen}, Sarr{\'e}, Sobhani, Stier, Tuppi, Vidale, Sandu, and Stevens]{seguraNextGEMSEnteringEra2025}
Hans Segura, Xabier {Pedruzo-Bagazgoitia}, Philipp Weiss, Sebastian~K. M{\"u}ller, Thomas Rackow, Junhong Lee, Edgar {Dolores-Tesillos}, Imme Benedict, Matthias Aengenheyster, Razvan Aguridan, Gabriele Arduini, Alexander~J. Baker, Jiawei Bao, Swantje Bastin, Eul{\`a}lia Baulenas, Tobias Becker, Sebastian Beyer, Hendryk Bockelmann, Nils Br{\"u}ggemann, Lukas Brunner, Suvarchal~K. Cheedela, Sushant Das, Jasper Denissen, Ian Dragaud, Piotr Dziekan, Madeleine Ekblom, Jan~Frederik Engels, Monika Esch, Richard Forbes, Claudia Frauen, Lilli Freischem, Diego {Garc{\'i}a-Maroto}, Philipp Geier, Paul Gierz, {\'A}lvaro {Gonz{\'a}lez-Cervera}, Katherine Grayson, Matthew Griffith, Oliver Gutjahr, Helmuth Haak, Ioan Hadade, Kerstin Haslehner, Shabeh {ul Hasson}, Jan Hegewald, Lukas Kluft, Aleksei Koldunov, Nikolay Koldunov, Tobias K{\"o}lling, Shunya Koseki, Sergey Kosukhin, Josh Kousal, Peter Kuma, Arjun~U. Kumar, Rumeng Li, Nicolas Maury, Maximilian Meindl, Sebastian Milinski, Kristian Mogensen, Bimochan Niraula, Jakub
  Nowak, Divya~Sri Praturi, Ulrike Proske, Dian Putrasahan, Ren{\'e} Redler, David Santuy, Domokos S{\'a}rm{\'a}ny, Reiner Schnur, Patrick Scholz, Dmitry Sidorenko, Dorian Sp{\"a}t, Birgit S{\"u}tzl, Daisuke Takasuka, Adrian Tompkins, Alejandro Uribe, Mirco Valentini, Menno Veerman, Aiko Voigt, Sarah Warnau, Fabian Wachsmann, Marta Wac{\l}awczyk, Nils Wedi, Karl-Hermann Wieners, Jonathan Wille, Marius Winkler, Yuting Wu, Florian Ziemen, Janos Zimmermann, Frida A.-M. Bender, Dragana Bojovic, Sandrine Bony, Simona Bordoni, Patrice Brehmer, Marcus Dengler, Emanuel Dutra, Saliou Faye, Erich Fischer, Chiel {van Heerwaarden}, Cathy Hohenegger, Heikki J{\"a}rvinen, Markus Jochum, Thomas Jung, Johann~H. Jungclaus, Noel~S. Keenlyside, Daniel Klocke, Heike Konow, Martina Klose, Szymon Malinowski, Olivia Martius, Thorsten Mauritsen, Juan~Pedro Mellado, Theresa Mieslinger, Elsa Mohino, Hanna Paw{\l}owska, Karsten {Peters-von Gehlen}, Abdoulaye Sarr{\'e}, Pajam Sobhani, Philip Stier, Lauri Tuppi, Pier~Luigi Vidale, Irina
  Sandu, and Bjorn Stevens.
\newblock {{nextGEMS}}: Entering the era of kilometer-scale {{Earth}} system modeling.
\newblock \emph{Geoscientific Model Development}, 18\penalty0 (20):\penalty0 7735--7761, October 2025.
\newblock ISSN 1991-959X.
\newblock \doi{10.5194/gmd-18-7735-2025}.

\bibitem[Taylor et~al.(2023)Taylor, Caldwell, Bertagna, Clevenger, Donahue, Foucar, Guba, Hillman, Keen, Krishna, Norman, Sreepathi, Terai, White, Salinger, McCoy, Leung, Bader, and Wu]{taylorSimpleCloudResolvingE3SM2023}
Mark Taylor, Peter~M. Caldwell, Luca Bertagna, Conrad Clevenger, Aaron Donahue, James Foucar, Oksana Guba, Benjamin Hillman, Noel Keen, Jayesh Krishna, Matthew Norman, Sarat Sreepathi, Christopher Terai, James~B. White, Andrew~G Salinger, Renata~B McCoy, Lai-yung~Ruby Leung, David~C. Bader, and Danqing Wu.
\newblock The {{Simple Cloud-Resolving E3SM Atmosphere Model Running}} on the {{Frontier Exascale System}}.
\newblock In \emph{Proceedings of the {{International Conference}} for {{High Performance Computing}}, {{Networking}}, {{Storage}} and {{Analysis}}}, {{SC}} '23, pages 1--11, New York, NY, USA, November 2023. Association for Computing Machinery.
\newblock ISBN 979-8-4007-0109-2.
\newblock \doi{10.1145/3581784.3627044}.

\bibitem[Donahue et~al.(2024)Donahue, Caldwell, Bertagna, Beydoun, Bogenschutz, Bradley, Clevenger, Foucar, Golaz, Guba, Hannah, Hillman, Johnson, Keen, Lin, Singh, Sreepathi, Taylor, Tian, Terai, Ullrich, Yuan, and Zhang]{donahueExascaleSimpleCloudResolving2024}
A.~S. Donahue, P.~M. Caldwell, L.~Bertagna, H.~Beydoun, P.~A. Bogenschutz, A.~M. Bradley, T.~C. Clevenger, J.~Foucar, C.~Golaz, O.~Guba, W.~Hannah, B.~R. Hillman, J.~N. Johnson, N.~Keen, W.~Lin, B.~Singh, S.~Sreepathi, M.~A. Taylor, J.~Tian, C.~R. Terai, P.~A. Ullrich, X.~Yuan, and Y.~Zhang.
\newblock To {{Exascale}} and {{Beyond}}---{{The Simple Cloud-Resolving E3SM Atmosphere Model}} ({{SCREAM}}), a {{Performance Portable Global Atmosphere Model}} for {{Cloud-Resolving Scales}}.
\newblock \emph{Journal of Advances in Modeling Earth Systems}, 16\penalty0 (7):\penalty0 e2024MS004314, 2024.
\newblock ISSN 1942-2466.
\newblock \doi{10.1029/2024MS004314}.

\bibitem[Vaswani et~al.(2023)Vaswani, Shazeer, Parmar, Uszkoreit, Jones, Gomez, Kaiser, and Polosukhin]{vaswaniAttentionAllYou2023}
Ashish Vaswani, Noam Shazeer, Niki Parmar, Jakob Uszkoreit, Llion Jones, Aidan~N. Gomez, Lukasz Kaiser, and Illia Polosukhin.
\newblock Attention {{Is All You Need}}, August 2023.

\bibitem[Pathak et~al.(2022)Pathak, Subramanian, Harrington, Raja, Chattopadhyay, Mardani, Kurth, Hall, Li, Azizzadenesheli, Hassanzadeh, Kashinath, and Anandkumar]{pathakFourCastNetGlobalDatadriven2022}
Jaideep Pathak, Shashank Subramanian, Peter Harrington, Sanjeev Raja, Ashesh Chattopadhyay, Morteza Mardani, Thorsten Kurth, David Hall, Zongyi Li, Kamyar Azizzadenesheli, Pedram Hassanzadeh, Karthik Kashinath, and Animashree Anandkumar.
\newblock {{FourCastNet}}: {{A Global Data-driven High-resolution Weather Model}} using {{Adaptive Fourier Neural Operators}}, February 2022.

\bibitem[Bi et~al.(2023)Bi, Xie, Zhang, Chen, Gu, and Tian]{biAccurateMediumrangeGlobal2023}
Kaifeng Bi, Lingxi Xie, Hengheng Zhang, Xin Chen, Xiaotao Gu, and Qi~Tian.
\newblock Accurate medium-range global weather forecasting with {{3D}} neural networks.
\newblock \emph{Nature}, 619\penalty0 (7970):\penalty0 533--538, July 2023.
\newblock ISSN 1476-4687.
\newblock \doi{10.1038/s41586-023-06185-3}.

\bibitem[Lam et~al.(2023)Lam, {Sanchez-Gonzalez}, Willson, Wirnsberger, Fortunato, Alet, Ravuri, Ewalds, {Eaton-Rosen}, Hu, Merose, Hoyer, Holland, Vinyals, Stott, Pritzel, Mohamed, and Battaglia]{lamGraphCastLearningSkillful2023}
Remi Lam, Alvaro {Sanchez-Gonzalez}, Matthew Willson, Peter Wirnsberger, Meire Fortunato, Ferran Alet, Suman Ravuri, Timo Ewalds, Zach {Eaton-Rosen}, Weihua Hu, Alexander Merose, Stephan Hoyer, George Holland, Oriol Vinyals, Jacklynn Stott, Alexander Pritzel, Shakir Mohamed, and Peter Battaglia.
\newblock {{GraphCast}}: {{Learning}} skillful medium-range global weather forecasting, August 2023.

\bibitem[Chen et~al.(2023)Chen, Zhong, Zhang, Cheng, Xu, Qi, and Li]{chenFuXiCascadeMachine2023}
Lei Chen, Xiaohui Zhong, Feng Zhang, Yuan Cheng, Yinghui Xu, Yuan Qi, and Hao Li.
\newblock {{FuXi}}: A cascade machine learning forecasting system for 15-day global weather forecast.
\newblock \emph{npj Climate and Atmospheric Science}, 6\penalty0 (1):\penalty0 190, November 2023.
\newblock ISSN 2397-3722.
\newblock \doi{10.1038/s41612-023-00512-1}.

\bibitem[Bodnar et~al.(2024)Bodnar, Bruinsma, Lucic, Stanley, Vaughan, Brandstetter, Garvan, Riechert, Weyn, Dong, Gupta, Thambiratnam, Archibald, Wu, Heider, Welling, Turner, and Perdikaris]{bodnarFoundationModelEarth2024}
Cristian Bodnar, Wessel~P. Bruinsma, Ana Lucic, Megan Stanley, Anna Vaughan, Johannes Brandstetter, Patrick Garvan, Maik Riechert, Jonathan~A. Weyn, Haiyu Dong, Jayesh~K. Gupta, Kit Thambiratnam, Alexander~T. Archibald, Chun-Chieh Wu, Elizabeth Heider, Max Welling, Richard~E. Turner, and Paris Perdikaris.
\newblock A {{Foundation Model}} for the {{Earth System}}, November 2024.

\bibitem[Nguyen et~al.(2024)Nguyen, Shah, Bansal, Arcomano, Maulik, Kotamarthi, Foster, Madireddy, and Grover]{nguyenScalingTransformerNeural2024}
Tung Nguyen, Rohan Shah, Hritik Bansal, Troy Arcomano, Romit Maulik, Veerabhadra Kotamarthi, Ian Foster, Sandeep Madireddy, and Aditya Grover.
\newblock Scaling transformer neural networks for skillful and reliable medium-range weather forecasting, October 2024.

\bibitem[Price et~al.(2024)Price, {Sanchez-Gonzalez}, Alet, Andersson, {El-Kadi}, Masters, Ewalds, Stott, Mohamed, Battaglia, Lam, and Willson]{priceGenCastDiffusionbasedEnsemble2024}
Ilan Price, Alvaro {Sanchez-Gonzalez}, Ferran Alet, Tom~R. Andersson, Andrew {El-Kadi}, Dominic Masters, Timo Ewalds, Jacklynn Stott, Shakir Mohamed, Peter Battaglia, Remi Lam, and Matthew Willson.
\newblock {{GenCast}}: {{Diffusion-based}} ensemble forecasting for medium-range weather, May 2024.

\bibitem[Lang et~al.(2024)Lang, Alexe, Clare, Roberts, Adewoyin, Bouall{\`e}gue, Chantry, Dramsch, Dueben, Hahner, Maciel, {Prieto-Nemesio}, O'Brien, Pinault, Polster, Raoult, Tietsche, and Leutbecher]{langAIFSCRPSEnsembleForecasting2024a}
Simon Lang, Mihai Alexe, Mariana C.~A. Clare, Christopher Roberts, Rilwan Adewoyin, Zied~Ben Bouall{\`e}gue, Matthew Chantry, Jesper Dramsch, Peter~D. Dueben, Sara Hahner, Pedro Maciel, Ana {Prieto-Nemesio}, Cathal O'Brien, Florian Pinault, Jan Polster, Baudouin Raoult, Steffen Tietsche, and Martin Leutbecher.
\newblock {{AIFS-CRPS}}: {{Ensemble}} forecasting using a model trained with a loss function based on the {{Continuous Ranked Probability Score}}, December 2024.

\bibitem[Bonev et~al.(2025)Bonev, Kurth, Mahesh, Bisson, Kossaifi, Kashinath, Anandkumar, Collins, Pritchard, and Keller]{bonevFourCastNet3Geometric2025a}
Boris Bonev, Thorsten Kurth, Ankur Mahesh, Mauro Bisson, Jean Kossaifi, Karthik Kashinath, Anima Anandkumar, William~D. Collins, Michael~S. Pritchard, and Alexander Keller.
\newblock {{FourCastNet}} 3: {{A}} geometric approach to probabilistic machine-learning weather forecasting at scale, July 2025.

\bibitem[Kossaifi et~al.(2026)Kossaifi, Kovachki, Mardani, Leibovici, Ravuri, Shokar, Calvello, Abbas, Harrington, Subramaniam, Brenowitz, Bonev, Byeon, Kreis, Durran, Vahdat, Pritchard, and Kautz]{kossaifiDemystifyingDataDrivenProbabilistic2026}
Jean Kossaifi, Nikola Kovachki, Morteza Mardani, Daniel Leibovici, Suman Ravuri, Ira Shokar, Edoardo Calvello, Mohammad~Shoaib Abbas, Peter Harrington, Akshay Subramaniam, Noah Brenowitz, Boris Bonev, Wonmin Byeon, Karsten Kreis, Dale Durran, Arash Vahdat, Mike Pritchard, and Jan Kautz.
\newblock Demystifying {{Data-Driven Probabilistic Medium-Range Weather Forecasting}}, January 2026.

\bibitem[Yu et~al.(2026)Yu, Huang, Calotoiu, and Hoefler]{yuScalingLawsGlobal2026}
Yuejiang Yu, Langwen Huang, Alexandru Calotoiu, and Torsten Hoefler.
\newblock Scaling {{Laws}} of {{Global Weather Models}}, February 2026.

\bibitem[Kochkov et~al.(2024)Kochkov, Yuval, Langmore, Norgaard, Smith, Mooers, Kl{\"o}wer, Lottes, Rasp, D{\"u}ben, Hatfield, Battaglia, {Sanchez-Gonzalez}, Willson, Brenner, and Hoyer]{kochkovNeuralGeneralCirculation2024}
Dmitrii Kochkov, Janni Yuval, Ian Langmore, Peter Norgaard, Jamie Smith, Griffin Mooers, Milan Kl{\"o}wer, James Lottes, Stephan Rasp, Peter D{\"u}ben, Sam Hatfield, Peter Battaglia, Alvaro {Sanchez-Gonzalez}, Matthew Willson, Michael~P. Brenner, and Stephan Hoyer.
\newblock Neural {{General Circulation Models}} for {{Weather}} and {{Climate}}.
\newblock \emph{Nature}, July 2024.
\newblock ISSN 0028-0836, 1476-4687.
\newblock \doi{10.1038/s41586-024-07744-y}.

\bibitem[{Cresswell-Clay} et~al.(2025){Cresswell-Clay}, Liu, Durran, Liu, Espinosa, Moreno, and Karlbauer]{cresswell-clayDeepLearningEarth2025}
Nathaniel {Cresswell-Clay}, Bowen Liu, Dale~R. Durran, Zihui Liu, Zachary~I. Espinosa, Raul~A. Moreno, and Matthias Karlbauer.
\newblock A {{Deep Learning Earth System Model}} for {{Efficient Simulation}} of the {{Observed Climate}}.
\newblock \emph{AGU Advances}, 6\penalty0 (4):\penalty0 e2025AV001706, 2025.
\newblock ISSN 2576-604X.
\newblock \doi{10.1029/2025AV001706}.

\bibitem[{Watt-Meyer} et~al.(2024){Watt-Meyer}, Henn, McGibbon, Clark, Kwa, Perkins, Wu, Harris, and Bretherton]{watt-meyerACE2AccuratelyLearning2024}
Oliver {Watt-Meyer}, Brian Henn, Jeremy McGibbon, Spencer~K. Clark, Anna Kwa, W.~Andre Perkins, Elynn Wu, Lucas Harris, and Christopher~S. Bretherton.
\newblock {{ACE2}}: {{Accurately}} learning subseasonal to decadal atmospheric variability and forced responses, November 2024.

\bibitem[Cachay et~al.(2024)Cachay, Henn, {Watt-Meyer}, Bretherton, and Yu]{cachayProbabilisticEmulationGlobal2024}
Salva~R{\"u}hling Cachay, Brian Henn, Oliver {Watt-Meyer}, Christopher~S. Bretherton, and Rose Yu.
\newblock Probabilistic {{Emulation}} of a {{Global Climate Model}} with {{Spherical DYffusion}}, November 2024.

\bibitem[Chapman et~al.(2025)Chapman, Schreck, Sha, II, Kimpara, Zanna, Mayer, and Berner]{chapmanCAMulatorFastEmulation2025}
William~E. Chapman, John~S. Schreck, Yingkai Sha, David John~Gagne II, Dhamma Kimpara, Laure Zanna, Kirsten~J. Mayer, and Judith Berner.
\newblock {{CAMulator}}: {{Fast Emulation}} of the {{Community Atmosphere Model}}, April 2025.

\bibitem[Flora and Potvin(2025)]{floraWoFSCastMachineLearning2025}
Montgomery~L. Flora and Corey Potvin.
\newblock {{WoFSCast}}: {{A Machine Learning Model}} for {{Predicting Thunderstorms}} at {{Watch-to-Warning Scales}}.
\newblock \emph{Geophysical Research Letters}, 52\penalty0 (10):\penalty0 e2024GL112383, 2025.
\newblock ISSN 1944-8007.
\newblock \doi{10.1029/2024GL112383}.

\bibitem[Abdi et~al.(2025)Abdi, Jankov, Madden, Vargas, Smith, Frolov, Flora, and Potvin]{abdiHRRRCastDatadrivenEmulator2025}
Daniel Abdi, Isidora Jankov, Paul Madden, Vanderlei Vargas, Timothy~A. Smith, Sergey Frolov, Montgomery Flora, and Corey Potvin.
\newblock {{HRRRCast}}: A data-driven emulator for regional weather forecasting at convection allowing scales, July 2025.

\bibitem[Pathak et~al.(2024)Pathak, Cohen, Garg, Harrington, Brenowitz, Durran, Mardani, Vahdat, Xu, Kashinath, and Pritchard]{pathakKilometerScaleConvectionAllowing2024}
Jaideep Pathak, Yair Cohen, Piyush Garg, Peter Harrington, Noah Brenowitz, Dale Durran, Morteza Mardani, Arash Vahdat, Shaoming Xu, Karthik Kashinath, and Michael Pritchard.
\newblock Kilometer-{{Scale Convection Allowing Model Emulation}} using {{Generative Diffusion Modeling}}, August 2024.

\bibitem[Mardani et~al.(2023)Mardani, Brenowitz, Cohen, Pathak, Chen, Liu, Vahdat, Nabian, Ge, Subramaniam, Kashinath, Kautz, and Pritchard]{mardaniResidualCorrectiveDiffusion2023}
Morteza Mardani, Noah Brenowitz, Yair Cohen, Jaideep Pathak, Chieh-Yu Chen, Cheng-Chin Liu, Arash Vahdat, Mohammad~Amin Nabian, Tao Ge, Akshay Subramaniam, Karthik Kashinath, Jan Kautz, and Mike Pritchard.
\newblock Residual {{Corrective Diffusion Modeling}} for {{Km-scale Atmospheric Downscaling}}, 2023.

\bibitem[Brenowitz et~al.(2025)Brenowitz, Ge, Subramaniam, Gupta, Hall, Mardani, Vahdat, Kashinath, and Pritchard]{brenowitzClimateBottleGenerative2025}
Noah~D. Brenowitz, Tao Ge, Akshay Subramaniam, Aayush Gupta, David~M. Hall, Morteza Mardani, Arash Vahdat, Karthik Kashinath, and Michael~S. Pritchard.
\newblock Climate in a {{Bottle}}: {{Towards}} a {{Generative Foundation Model}} for the {{Kilometer-Scale Global Atmosphere}}, 2025.

\bibitem[Perkins et~al.(2026)Perkins, Kwa, McGibbon, Arcomano, Clark, {Watt-Meyer}, Bretherton, and Harris]{perkinsHiROACEFastSkillful2026}
W.~Andre Perkins, Anna Kwa, Jeremy McGibbon, Troy Arcomano, Spencer~K. Clark, Oliver {Watt-Meyer}, Christopher~S. Bretherton, and Lucas~M. Harris.
\newblock {{HiRO-ACE}}: {{Fast}} and skillful {{AI}} emulation and downscaling trained on a 3 km global storm-resolving model, February 2026.

\bibitem[Bar-Tal et~al.(2023)Bar-Tal, Yariv, Lipman, and Dekel]{Bar-Tal2023-sg}
Omer Bar-Tal, Lior Yariv, Yaron Lipman, and Tali Dekel.
\newblock {MultiDiffusion}: Fusing diffusion paths for controlled image generation.
\newblock \emph{arXiv [cs.CV]}, February 2023.

\bibitem[Bonev et~al.(2023)Bonev, Kurth, Hundt, Pathak, Baust, Kashinath, and Anandkumar]{bonevSphericalFourierNeural2023}
Boris Bonev, Thorsten Kurth, Christian Hundt, Jaideep Pathak, Maximilian Baust, Karthik Kashinath, and Anima Anandkumar.
\newblock Spherical {{Fourier Neural Operators}}: {{Learning Stable Dynamics}} on the {{Sphere}}, June 2023.

\bibitem[Keisler(2022)]{keislerForecastingGlobalWeather2022}
Ryan Keisler.
\newblock Forecasting {{Global Weather}} with {{Graph Neural Networks}}, February 2022.

\bibitem[Mouallem et~al.(2023)Mouallem, Harris, and Chen]{mouallemImplementationNovelDuoGrid2023}
Joseph Mouallem, Lucas Harris, and Xi~Chen.
\newblock Implementation of the {{Novel Duo-Grid}} in {{GFDL}}'s {{FV3 Dynamical Core}}.
\newblock \emph{Journal of Advances in Modeling Earth Systems}, 15\penalty0 (12):\penalty0 e2023MS003712, 2023.
\newblock ISSN 1942-2466.
\newblock \doi{10.1029/2023MS003712}.

\bibitem[Peebles and Xie(2023)]{peeblesScalableDiffusionModels2023}
William Peebles and Saining Xie.
\newblock Scalable {{Diffusion Models}} with {{Transformers}}, March 2023.

\bibitem[Hassani et~al.(2023)Hassani, Walton, Li, Li, and Shi]{hassaniNeighborhoodAttentionTransformer2023a}
Ali Hassani, Steven Walton, Jiachen Li, Shen Li, and Humphrey Shi.
\newblock Neighborhood {{Attention Transformer}}, May 2023.

\bibitem[Hassani et~al.(2025)Hassani, Zhou, Kane, Huang, Chen, Shi, Walton, Hoehnerbach, Thakkar, Isaev, Zhang, Xu, Wu, Hwu, Liu, and Shi]{hassaniGeneralizedNeighborhoodAttention2025}
Ali Hassani, Fengzhe Zhou, Aditya Kane, Jiannan Huang, Chieh-Yun Chen, Min Shi, Steven Walton, Markus Hoehnerbach, Vijay Thakkar, Michael Isaev, Qinsheng Zhang, Bing Xu, Haicheng Wu, Wen-mei Hwu, Ming-Yu Liu, and Humphrey Shi.
\newblock Generalized {{Neighborhood Attention}}: {{Multi-dimensional Sparse Attention}} at the {{Speed}} of {{Light}}, April 2025.

\bibitem[Hoffmann et~al.(2022)Hoffmann, Borgeaud, Mensch, Buchatskaya, Cai, Rutherford, Casas, Hendricks, Welbl, Clark, Hennigan, Noland, Millican, van~den Driessche, Damoc, Guy, Osindero, Simonyan, Elsen, Rae, Vinyals, and Sifre]{hoffmannTrainingComputeOptimalLarge2022}
Jordan Hoffmann, Sebastian Borgeaud, Arthur Mensch, Elena Buchatskaya, Trevor Cai, Eliza Rutherford, Diego de~Las Casas, Lisa~Anne Hendricks, Johannes Welbl, Aidan Clark, Tom Hennigan, Eric Noland, Katie Millican, George van~den Driessche, Bogdan Damoc, Aurelia Guy, Simon Osindero, Karen Simonyan, Erich Elsen, Jack~W. Rae, Oriol Vinyals, and Laurent Sifre.
\newblock Training {{Compute-Optimal Large Language Models}}, March 2022.

\bibitem[Kaplan et~al.(2020)Kaplan, McCandlish, Henighan, Brown, Chess, Child, Gray, Radford, Wu, and Amodei]{kaplanScalingLawsNeural2020}
Jared Kaplan, Sam McCandlish, Tom Henighan, Tom~B. Brown, Benjamin Chess, Rewon Child, Scott Gray, Alec Radford, Jeffrey Wu, and Dario Amodei.
\newblock Scaling {{Laws}} for {{Neural Language Models}}, January 2020.

\bibitem[Zhang(2019)]{zhangMakingConvolutionalNetworks2019}
Richard Zhang.
\newblock Making {{Convolutional Networks Shift-Invariant Again}}, June 2019.

\bibitem[Yu et~al.(2025)Yu, Xiong, Nie, Sheng, Liu, and Luo]{yuPixelDiTPixelDiffusion2025}
Yongsheng Yu, Wei Xiong, Weili Nie, Yichen Sheng, Shiqiu Liu, and Jiebo Luo.
\newblock {{PixelDiT}}: {{Pixel Diffusion Transformers}} for {{Image Generation}}, November 2025.

\bibitem[He et~al.(2026)He, Ma, Guo, Shan, Fu, Chen, Huang, and Li]{he2026hyperdit}
Yu~He, Lichen Ma, Zipeng Guo, Xinyuan Shan, Jingling Fu, Dong Chen, Junshi Huang, and Yan Li.
\newblock {HyperDiT}: Hyper-connected transformers for high-fidelity pixel-space diffusion.
\newblock \emph{arXiv preprint arXiv:2605.15741}, 2026.

\bibitem[Su et~al.(2023)Su, Lu, Pan, Murtadha, Wen, and Liu]{suRoFormerEnhancedTransformer2023}
Jianlin Su, Yu~Lu, Shengfeng Pan, Ahmed Murtadha, Bo~Wen, and Yunfeng Liu.
\newblock {{RoFormer}}: {{Enhanced Transformer}} with {{Rotary Position Embedding}}, November 2023.

\bibitem[Heo et~al.(2024)Heo, Park, Han, and Yun]{heoRotaryPositionEmbedding2024}
Byeongho Heo, Song Park, Dongyoon Han, and Sangdoo Yun.
\newblock Rotary {{Position Embedding}} for {{Vision Transformer}}.
\newblock https://arxiv.org/abs/2403.13298v2, March 2024.

\bibitem[Surcel et~al.(2015)Surcel, Zawadzki, and Yau]{surcelStudyScaleDependence2015}
Madalina Surcel, Isztar Zawadzki, and M.~K. Yau.
\newblock A {{Study}} on the {{Scale Dependence}} of the {{Predictability}} of {{Precipitation Patterns}}.
\newblock \emph{Journal of the Atmospheric Sciences}, January 2015.
\newblock \doi{10.1175/JAS-D-14-0071.1}.

\bibitem[Roberts and Lean(2008)]{robertsScaleSelectiveVerificationRainfall2008}
Nigel~M. Roberts and Humphrey~W. Lean.
\newblock Scale-{{Selective Verification}} of {{Rainfall Accumulations}} from {{High-Resolution Forecasts}} of {{Convective Events}}.
\newblock \emph{Monthly Weather Review}, January 2008.
\newblock \doi{10.1175/2007MWR2123.1}.

\bibitem[Parishani et~al.(2018)Parishani, Pritchard, Bretherton, Terai, Wyant, Khairoutdinov, and Singh]{parishaniInsensitivityCloudResponse2018}
Hossein Parishani, Michael~S. Pritchard, Christopher~S. Bretherton, Christopher~R. Terai, Matthew~C. Wyant, Marat Khairoutdinov, and Balwinder Singh.
\newblock Insensitivity of the {{Cloud Response}} to {{Surface Warming Under Radical Changes}} to {{Boundary Layer Turbulence}} and {{Cloud Microphysics}}: {{Results From}} the {{Ultraparameterized CAM}}.
\newblock \emph{Journal of Advances in Modeling Earth Systems}, 10\penalty0 (12):\penalty0 3139--3158, 2018.
\newblock ISSN 1942-2466.
\newblock \doi{10.1029/2018MS001409}.

\bibitem[Terai et~al.(2020)Terai, Pritchard, Blossey, and Bretherton]{teraiImpactResolvingSubkilometer2020}
C.~R. Terai, M.~S. Pritchard, P.~Blossey, and C.~S. Bretherton.
\newblock The {{Impact}} of {{Resolving Subkilometer Processes}} on {{Aerosol-Cloud Interactions}} of {{Low-Level Clouds}} in {{Global Model Simulations}}.
\newblock \emph{Journal of Advances in Modeling Earth Systems}, 12\penalty0 (11):\penalty0 e2020MS002274, 2020.
\newblock ISSN 1942-2466.
\newblock \doi{10.1029/2020MS002274}.

\bibitem[Peng et~al.(2024)Peng, Blossey, Hannah, Bretherton, Terai, Jenney, and Pritchard]{pengImprovingStratocumulusCloud2024}
Liran Peng, Peter~N. Blossey, Walter~M. Hannah, Christopher~S. Bretherton, Christopher~R. Terai, Andrea~M. Jenney, and Michael Pritchard.
\newblock Improving {{Stratocumulus Cloud Amounts}} in a 200-m {{Resolution Multi-Scale Modeling Framework Through Tuning}} of {{Its Interior Physics}}.
\newblock \emph{Journal of Advances in Modeling Earth Systems}, 16\penalty0 (3):\penalty0 e2023MS003632, 2024.
\newblock ISSN 1942-2466.
\newblock \doi{10.1029/2023MS003632}.

\bibitem[Arakawa(2004)]{arakawa2004cumulus}
Akio Arakawa.
\newblock The cumulus parameterization problem: Past, present, and future.
\newblock \emph{Journal of climate}, 17\penalty0 (13):\penalty0 2493--2525, 2004.

\bibitem[Satoh et~al.(2014)Satoh, Tomita, Yashiro, Miura, Kodama, Seiki, Noda, Yamada, Goto, Sawada, et~al.]{satoh2014non}
Masaki Satoh, Hirofumi Tomita, Hisashi Yashiro, Hiroaki Miura, Chihiro Kodama, Tatsuya Seiki, Akira~T Noda, Yohei Yamada, Daisuke Goto, Masahiro Sawada, et~al.
\newblock The non-hydrostatic icosahedral atmospheric model: Description and development.
\newblock \emph{Progress in Earth and Planetary Science}, 1\penalty0 (1):\penalty0 18, 2014.

\bibitem[Ma et~al.(2022)Ma, Klein, Lee, Ahn, Tao, and Gleckler]{maSuperiorDailySubDaily2022}
Hsi-Yen Ma, Stephen~A. Klein, Jiwoo Lee, Min-Seop Ahn, Cheng Tao, and Peter~J. Gleckler.
\newblock Superior {{Daily}} and {{Sub}}-{{Daily Precipitation Statistics}} for {{Intense}} and {{Long}}-{{Lived Storms}} in {{Global Storm}}-{{Resolving Models}}.
\newblock \emph{Geophysical Research Letters}, 49\penalty0 (8):\penalty0 e2021GL096759, April 2022.
\newblock ISSN 0094-8276, 1944-8007.
\newblock \doi{10.1029/2021GL096759}.

\bibitem[Hohenegger et~al.(2023)Hohenegger, Korn, Linardakis, Redler, Schnur, Adamidis, Bao, Bastin, Behravesh, Bergemann, et~al.]{hohenegger2023icon}
Cathy Hohenegger, Peter Korn, Leonidas Linardakis, Ren{\'e} Redler, Reiner Schnur, Panagiotis Adamidis, Jiawei Bao, Swantje Bastin, Milad Behravesh, Martin Bergemann, et~al.
\newblock Icon-sapphire: simulating the components of the earth system and their interactions at kilometer and subkilometer scales.
\newblock \emph{Geoscientific Model Development}, 16\penalty0 (2):\penalty0 779--811, 2023.

\bibitem[Hadade et~al.(2025)Hadade, Klocke, Enkovaara, Lunttila, Rackow, Engels, Frauen, Redler, Kontkanen, Jung, Sein, Sandu, Reuter, Wedi, Milinski, {Doblas-Reyes}, Castrillo, Acosta, Girona, and Manninen]{hadadeDestinationEarthClimate2025}
Ioan Hadade, Daniel Klocke, Jussi Enkovaara, Tuomas Lunttila, Thomas Rackow, Jan~Frederik Engels, Claudia Frauen, Ren{\'e} Redler, Jenni Kontkanen, Thomas Jung, Dmitry Sein, Irina Sandu, Balthasar Reuter, Nils Wedi, Sebastian Milinski, Francisco {Doblas-Reyes}, Miguel Castrillo, Mario Acosta, Sergi Girona, and Pekka Manninen.
\newblock Destination {{Earth}}: {{The Climate Change Adaptation Digital Twin}}.
\newblock In \emph{Proceedings of the {{International Conference}} for {{High Performance Computing}}, {{Networking}}, {{Storage}} and {{Analysis}}}, {{SC}} '25, pages 99--110, New York, NY, USA, November 2025. Association for Computing Machinery.
\newblock ISBN 979-8-4007-1466-5.
\newblock \doi{10.1145/3712285.3771790}.

\bibitem[Henry et~al.(2020)Henry, Dachapally, Pawar, and Chen]{henryQueryKeyNormalizationTransformers2020}
Alex Henry, Prudhvi~Raj Dachapally, Shubham Pawar, and Yuxuan Chen.
\newblock Query-{{Key Normalization}} for {{Transformers}}, October 2020.

\bibitem[Qiu et~al.(2025)Qiu, Wang, Zheng, Huang, Wen, Yang, Men, Yu, Huang, Huang, Liu, Zhou, and Lin]{qiuGatedAttentionLarge2025}
Zihan Qiu, Zekun Wang, Bo~Zheng, Zeyu Huang, Kaiyue Wen, Songlin Yang, Rui Men, Le~Yu, Fei Huang, Suozhi Huang, Dayiheng Liu, Jingren Zhou, and Junyang Lin.
\newblock Gated {{Attention}} for {{Large Language Models}}: {{Non-linearity}}, {{Sparsity}}, and {{Attention-Sink-Free}}.
\newblock https://arxiv.org/abs/2505.06708v1, May 2025.

\bibitem[Rasp et~al.(2024)Rasp, Hoyer, Merose, Langmore, Battaglia, Russel, {Sanchez-Gonzalez}, Yang, Carver, Agrawal, Chantry, Bouallegue, Dueben, Bromberg, Sisk, Barrington, Bell, and Sha]{raspWeatherBench2Benchmark2024a}
Stephan Rasp, Stephan Hoyer, Alexander Merose, Ian Langmore, Peter Battaglia, Tyler Russel, Alvaro {Sanchez-Gonzalez}, Vivian Yang, Rob Carver, Shreya Agrawal, Matthew Chantry, Zied~Ben Bouallegue, Peter Dueben, Carla Bromberg, Jared Sisk, Luke Barrington, Aaron Bell, and Fei Sha.
\newblock {{WeatherBench}} 2: {{A}} benchmark for the next generation of data-driven global weather models, January 2024.

\bibitem[Pendergrass and Hartmann(2014)]{pendergrass2014two}
Angeline~G Pendergrass and Dennis~L Hartmann.
\newblock Two modes of change of the distribution of rain.
\newblock \emph{Journal of Climate}, 27\penalty0 (22):\penalty0 8357--8371, 2014.

\bibitem[Kooperman et~al.(2016)Kooperman, Pritchard, Burt, Branson, and Randall]{koopermanRobustEffectsCloud2016}
Gabriel~J. Kooperman, Michael~S. Pritchard, Melissa~A. Burt, Mark~D. Branson, and David~A. Randall.
\newblock Robust effects of cloud superparameterization on simulated daily rainfall intensity statistics across multiple versions of the {{Community Earth System Model}}.
\newblock \emph{Journal of Advances in Modeling Earth Systems}, 8\penalty0 (1):\penalty0 140--165, 2016.
\newblock ISSN 1942-2466.
\newblock \doi{10.1002/2015MS000574}.

\bibitem[Goyal et~al.(2018)Goyal, Doll{\'a}r, Girshick, Noordhuis, Wesolowski, Kyrola, Tulloch, Jia, and He]{goyalAccurateLargeMinibatch2018}
Priya Goyal, Piotr Doll{\'a}r, Ross Girshick, Pieter Noordhuis, Lukasz Wesolowski, Aapo Kyrola, Andrew Tulloch, Yangqing Jia, and Kaiming He.
\newblock Accurate, {{Large Minibatch SGD}}: {{Training ImageNet}} in 1 {{Hour}}, April 2018.

\bibitem[McCandlish et~al.(2018)McCandlish, Kaplan, Amodei, and Team]{mccandlishEmpiricalModelLargeBatch2018}
Sam McCandlish, Jared Kaplan, Dario Amodei, and OpenAI~Dota Team.
\newblock An {{Empirical Model}} of {{Large-Batch Training}}, December 2018.

\end{thebibliography}
